\documentclass[12pt,preprint]{aastex}

\usepackage{verbatim}
\usepackage{rotating}
\usepackage{lscape}

\def\percc{$\rm cm^{-3}$}
\def\h2{$\rm H_2$}

\def\e#1{$\times 10^{#1}$}

\def\kms{km~s$^{-1}$}

\def\pv{$p$V}

\newcommand{\msun}{M$_{\odot}$}
\newcommand{\zsun}{Z$_{\odot}$}
\newcommand{\halpha}{H$\alpha$}

\newcommand{\hi}{H{\sc I}}

\newcommand{\hoii}{Ho~{\sc II}}

\title{Does Stellar Feedback Create \hi\ Holes? An HST/VLA Study of Holmberg {\sc II}}
\author{Daniel R.\ Weisz, Evan D.\ Skillman}
\affil{Astronomy Department, University of Minnesota,
Minneapolis, MN 55455}
\email{dweisz@astro.umn.edu, skillman@astro.umn.edu}

\author{John M.\ Cannon}
\affil{Department of Physics \& Astronomy, Macalester College, 1600 Grand Avenue, St. Paul, MN 55125}
\email{jcannon@macalester.edu}

\author{Andrew E.\ Dolphin}
\affil{Raytheon Company,
1151 E Hermans Rd, Tucson, AZ 85756}
\email{adolphin@raytheon.com}

\author{Robert C.\ Kennicutt, Jr.}
\affil{Institute of Astronomy, University of Cambridge, Madingley Road, Cambridge CB3 0HA, UK}
\email{robk@ast.cam.ac.uk}

\author{Janice Lee}
\affil{Observatories of the Carnegie Institution of Washington,
813 Santa Barbara Street, Pasadena, CA 91101, Hubble Fellow}
\email{jlee@obs.carnegiescience.edu}

\author{Fabian Walter}
\affil{Max-Planck-Institut f\"{u}r Astronomie, K\"{o}nigstuhl 17, D-69117 Heidelberg, Germany}
\email{walter@mpia.de}

\author{}
\affil{}
\email{}

\clearpage

\begin{abstract}

We use deep HST/ACS F555W and F814W photometry of resolved stars in the M81 Group dwarf irregular galaxy Ho II to study the hypothesis that the holes identified in the neutral ISM (\hi) are created by stellar feedback.  From the deep photometry, we construct color-magnitude diagrams (CMDs) and measure the star formation histories (SFHs) for stars contained in \hi\ holes from two independent holes catalogs, as well as select control fields, i.e., similar sized regions that span a range of \hi\ column densities.  The CMDs reveal young ($<$ 200 Myr) stellar populations inside all \hi\ holes, which contain very few bright OB stars with ages less than 10 Myr, indicating they are not reliable tracers of \hi\ hole locations while the recent SFHs confirm multiple episodes of star formation within most holes. Converting the recent SFHs into stellar feedback energies, we find that enough energy has been generated to have created all holes.  However, the required energy is not always produced over a time scale that is less than the estimated kinematic age of the hole.  A similar analysis of stars in the control fields finds that the stellar populations of the control fields and \hi\ holes are statistically indistinguishable.  However, because we are only sensitive to holes $\sim$ 100 pc in diameter, we cannot tell if there are smaller holes inside the control fields.   The combination of the CMDs, recent SFHs, and locations of young stars shows that the stellar populations inside \hi\ holes are not coherent, single-aged, stellar clusters, as previously suggested, but rather multi-age populations distributed across each hole.   From a comparison of the modeled and observed integrated magnitudes, and the locations and energetics of stars inside of \hi\ holes, we propose a potential new model: a viable mechanism for creating the observed \hi\ holes in \hoii\ is stellar feedback from multiple generations of SF spread out over tens or hundreds of Myr, and thus, the concept of an age for an \hi\ hole is intrinsically ambiguous.  For \hi\ holes in the outer parts of \hoii, located beyond the HST/ACS coverage, we use Monte Carlo simulations of expected stellar populations to show that low level SF could provide the energy necessary to form these holes.  Applying the same method to the SMC, we find that the holes that appear to be void of stars could have formed via stellar feedback from low level SF.  We further find that \halpha\ and  24$\mu$m\ emission, tracers of the most recent star formation, do not correlate well with the positions of the \hi\ holes. However, UV emission, which traces star formation over roughly the last 100 Myr, shows a much better correlation with the locations of the \hi\ holes.

\end{abstract}


\begin{document}

\section{Introduction}

The role of star formation (SF) in creating structure and shaping the interstellar medium (ISM) is an important but unresolved topic.   Studies of the neutral hydrogen (\hi) component in a wide variety of nearby galaxies, including our own, reveal numerous holes, shells, and bubbles (for simplicity, we will refer to all types as holes throughout the paper), which span a great range in size and age \citep[e.g.,][]{hei79, bri86, puc92, oey95, kim98, wal99a, mul03, rel07, chu08, bag09}.  The origin of these ISM features has long been attributed to stellar feedback from massive stars, i.e., stellar winds and Type {\sc II}\ supernovae (SNe) \citep[][and references therein]{wea77, cas80, mcc87, ttb88}.  Typical ages of these ISM features, measured from the hole diameters and \hi\ expansion velocities, range from $\sim$ 10$^{6}$ $-$ 10$^{8}$ yr \citep[e.g.,][]{oey95, wal99a, hat05}, which suggests that even if massive O and B type stars, the primary contributors to stellar feedback, are no longer present inside the \hi\ holes, we should still see remnant stellar populations (assuming a universal IMF).  

While the stellar feedback hypothesis of \hi\ hole creation is appealing due to its intuitive simplicity, observational and theoretical studies do not provide much supporting evidence.  The close proximity of dwarf galaxies in the Local Group (LG) has permitted detailed studies of correlation between stars and \hi\ holes.  In the LMC, \citet{kim98} compared the positions of 103 giant or supergiant \hi\ shells with \halpha, and found a weaker correlation than expected given the young ages of the shells. Similarly, \citet{book08} find that while \halpha\ traces sites of secondary triggered SF around supergiant \hi\ shells in the LMC, it is not exclusively associated with stars interior to the holes. \citet{hat05} cross-correlated the locations of 509 \hi\ holes in the SMC with known catalogs of OB associations, supergiants, Cepheids, Wolf-Rayet (WR) stars, SN remnants, and stars clusters, finding 59 \hi\ shells with no stellar component, and that holes without associated OB stars exceed those with them by factor of $\sim$ 1.5.  Further, in the LG dwarf irregular galaxy (dI) IC~1613, \citet{sil06} find that even the \hi\ structures that do contain young stars do not have the observed properties consistent with the traditional stellar feedback theory.

The studies of  the origins of \hi\ holes in galaxies beyond the LG show similarly mixed results when testing the stellar feedback hypothesis.  In the pioneering study of M31, \citet{bri86} found a strong correlation between OB associations and \hi\ shells smaller than $\sim$ 300 pc.  Correlating high resolution \hi\ imaging of the M81~Group dI Holmberg {\sc{ii}} with \halpha, \citet{puc92} found strong evidence for a stellar feedback origin to the 51 \hi\ holes.  In a follow up study of \hoii, \citet{rho99} used integrated BVR photometry to search for stars within the \hi\ holes and found only 14\% had a stellar component. A third study of \hoii\ concludes that \halpha\ does not adequately trace the progenitor stellar populations; however, detected FUV emission associated with \hi\ holes supports the stellar feedback model \citep{ste00}. In another M81~Group dI, IC~2574, \citet{pas08} use integrated colors from imaging from the Large Binocular Telescope to show that there is not a one to one correspondence between stars and \hi\ holes, while an evaluation of alternative models by \citet{san02} finds that \hi\ holes in IC~2574 are likely formed by stellar feedback.   

Although a number of studies of select individual holes have found associated young stars \citep[e.g.,][]{oey95, vdh96, wei09}, the overall evidence for a stellar feedback origin for all \hi\ holes is not as compelling and a variety of alternate explanations have been proposed.  For example, gamma ray bursts (GRBs) have sufficient energy to create holes in the \hi\ without leaving a remnant stellar cluster \citep[e.g.,][]{loeb98}.  Similarly, collisions between high velocity \hi\ clouds and a galaxy \citep[e.g.,][]{tt86} or ram pressure stripping from the intergalactic medium \citep[e.g.,][]{bur04} could provide the means to create observed ISM features. \citet{rho99} suggest that modified energetics, either by an overestimation in the energies necessary to evacuate \hi\ holes or by a non-standard IMF, could explain the observed discrepancies, although they acknowledge the latter is not likely. An alternate possibility is that gravitational instabilities and/or turbulence naturally give rise to structure in the ISM \citep[e.g.,][]{elm97, dib05}.  The study of Constellation {\sc III} by \citet{har04} has cast doubt upon a number of these alternate theories and favors structure creation by numerous SF events. While it is likely that many of these mechanisms play a role in shaping the ISM at some level, quantifying the impact of stellar feedback on the ISM can give key insights into the feedback processes and self-regulation of SF.

A further alternative is that the traditional theory linking stellar feedback to the creation of \hi\ holes is not complete.  Specifically, the most rigorous observational test of the stellar feedback model \citep{rho99} assumed that single age stellar clusters generated the SNe energy necessary to evacuate the \hi\ holes.  This is a reasonable assumption based on an extrapolation from the theoretical models of the explosion and expansion of a single SN in the ISM \citep[e.g.,][]{che74, wea77}. In this scenario, SNe from a single generation of stars (i.e., a cluster) explode, releasing energy into to an isotropic ISM,  thus creating the observed hole and quenching future SF, by sweeping up and heating of the gas interior to the hole.  However, the study of spatially resolved SFHs in nearby dwarf galaxies has provided evidence that SF within $\sim$ 500 Myr can return to the same locations \citep{dea97, dom02, wei09}, demonstrating the SF does not always suppress subsequent generations of SF.  The creation of \hi\ holes by multiple generations of SNe is a scenario that would likely have complicated dynamics, but preserves the intuitive simplicity linking the energy production of stars with the structures in the ISM.

In this paper, we use HST/ACS imaging of \hoii\ to detect and study the resolved stellar populations within the \hi\ holes independently catalogued by \citet{puc92} and The \hi\ Nearby Galaxy Survey \citep[THINGS;][]{wal08, bag09}.  Nearby dIs, such as \hoii\ \citep[roughly a solid body rotator in the inner 2 kpc;][]{puc92}, are excellent environments for both the study of resolved stars and ISM features.  Because the \hi\ disks of dIs are typically solid body rotators, and thus lack sheer and differential rotation, \hi\ holes and potential progenitor stellar components travel in unison for time scales on order of $\sim$ 10$^{8}$ yr \citep[e.g.,][]{ski88, ski96}.  Further, the low metallicities of nearby dIs minimize internal extinction effects, which allow us to take a precise census of individual stars.  

From the resolved stellar populations of \hoii, we construct color-magnitude diagrams (CMDs) and measure star formation histories (SFHs) corresponding to catalogued \hi\ holes that fall into our ACS field of view.  Using the SFHs we can compute the energy associated with stellar feedback and compare it to the energy required to evacuate an \hi\ hole, allowing us to quantify the efficiency of stellar feedback and compare it to predictions from models.  We further compare the stellar populations of the \hi\ holes to those of select control fields, similar in size to the \hi\ holes, which span a wide range in \hi\ surface density and do not overlap with catalogued \hi\ holes.  The resolved stellar populations also permit us to directly test the conclusions of previous studies of \hoii\ by \citet{puc92} and \citet{rho99}, including comparisons to \halpha, 24$\mu$m, and UV imaging of the same regions.  Using the results of the stellar populations in the \hi\ holes located within the ACS field of view, we use simulated CMDs to demonstrate the plausibility of \hi\ hole creation due to stellar feedback in regions of low gas and stellar densities, such as the outer regions of galaxies.  We then apply this method of analysis to the SMC and discuss how observations of apparently empty holes does not rule out the possibility of a stellar feedback origin.

\section{Observations and Photometry}

As part of a larger HST program aimed at imaging multiple dIs in the M81~Group (GO-10605), we observed \hoii\ on 2006, December 30 using HST/ACS \citep{for98}.  To capture most of the optical galaxy, we used two ACS fields, observing each field for 4660 sec in F555W (V) and 4660 sec in F814W (I), with a CR-split in order to reduce the impact of cosmic rays, and a dithering strategy to cover the chip gap.  The images were processed with the HST pipeline and we performed photometry and artificial star tests using DOLPHOT, a version of HSTphot \citep{dol00} optimized for ACS observations\footnote[1]{Photometry and artificial star tests for the HST/ACS imaging of \hoii\ is available via the ACS Nearby Galaxy Survey Treasury program \citep[ANGST;][]{jdal09}: http://www.nearbygalaxies.org}.  Photometry was carried out as part of a larger program and the specific details, e.g., rejection criteria, completeness, can be found in \citet{wei08}.  Following the photometric cuts, we merged the two fields into a single photometric list, appropriately removing duplicate stars, resulting in a total of 388,945 well measured stars for the combined fields.  The deep photometry that is used in this paper has a limiting magnitude of $M_{V}$ $\sim$ 0, which allows us to see main sequence stars (MS) up to $\sim$ 300 Myr and Blue Helium Burning stars (BHeBs) up to $\sim$ 1 Gyr in age \citep{gza05, wei08}.  To quantify the accuracy of the photometry we ran artificial star tests on the entire galaxy and found 50\% completeness limits of m$_{F555W}$ $=$ 27.8 (M$_{F555W}$ $=$ $+$0.15) and m$_{F814W}$ $=$ 27.3 (M$_{F814W}$ $=$ $-$0.15), assuming a distance modulus of 27.65 \citep{kar02}.  A number of \hi\ holes in nearby galaxies have inferred kinematic ages (from \hi\ observations) as old as a $\sim$ 10$^{8}$ yr \citep[e.g.,][]{puc92, wal99a}.  If some of the older \hi\ holes were created by SF on this time scale, only deep photometry of resolved stellar populations would probe faint enough limits to detect the remaining stars.

\begin{figure}[t]
\begin{center}
\plotone{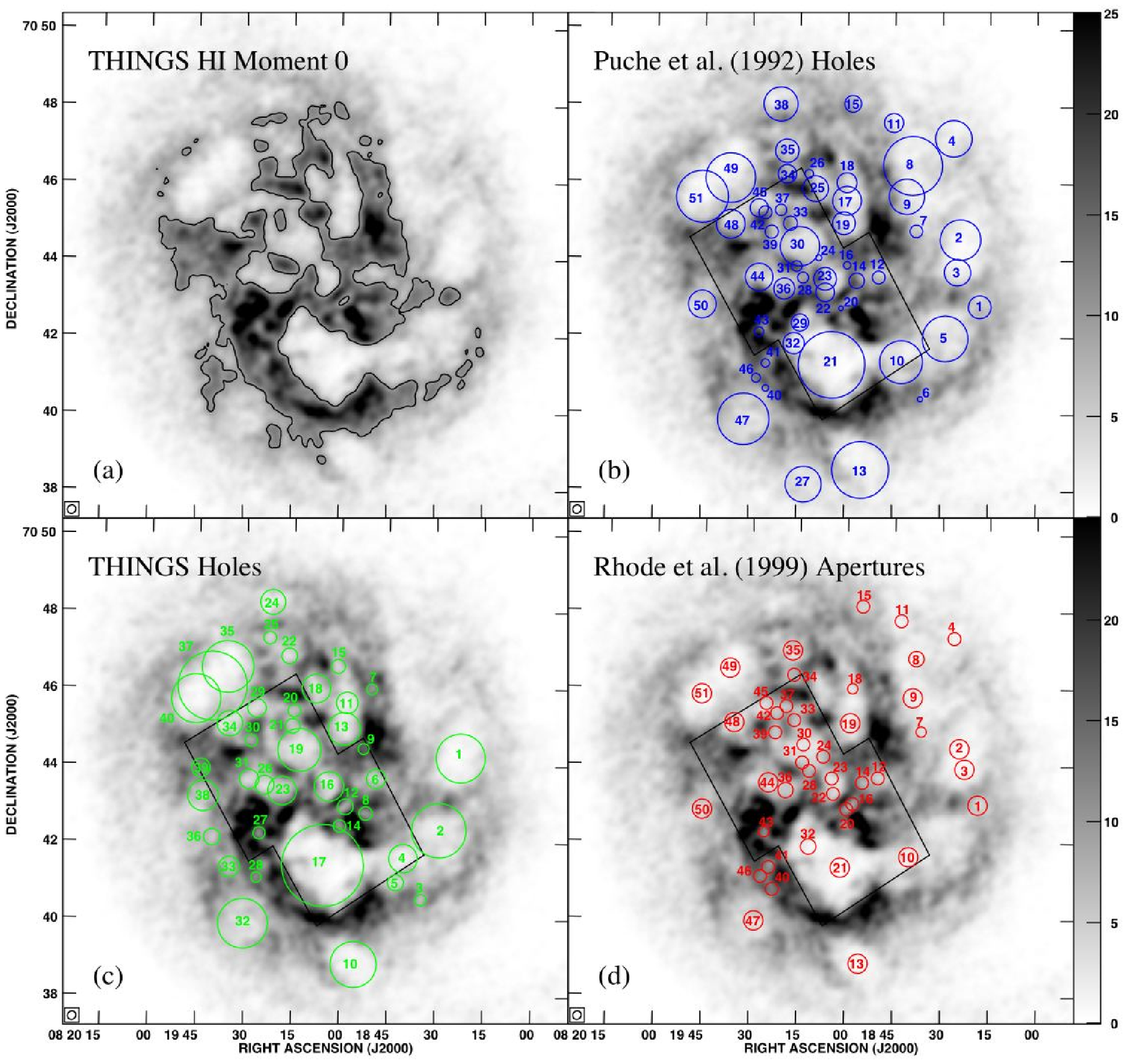}
\caption{ (a) \hi\ surface density image of \hoii\ processed by the THINGS program \citep[THINGS;][]{wal08} with a contour of 10$^{-21}$ cm$^{-2}$ overlaid; (b) \hi\ holes catalogued by \citet{puc92}; (c) \hi\ holes from the THINGS catalog \citep{bag09}; (d) apertures used by \citet{rho99} with the ACS footprint overlaid in black.  Note the `shredded' appearance of the \hi\ distribution as well as the numerous separate holes cataloged in regions of low \hi\ column density.}  
\label{map1}
\end{center}
\end{figure}

The \hi\ image we used in our analysis (panel (a) of Figure \ref{map1}) was observed with the VLA using the B, C, D configurations in 1990 and 1991 (the full details of the observations are available in \citealt{puc92}).  As part of the THINGS program \citep{wal08}, the observations were reprocessed with their software pipeline, achieving a final spatial resolution of $\sim$ 6\arcsec\ and a velocity resolution of 2.6 \kms.  Both P92 and the THINGS team independently constructed catalogs of \hi\ holes in \hoii\ by examining them in the position-velocity (\pv) cuts and radius-velocity space in the \hi\ data cube.  Hole identification was done by eye using the following criteria as a guide: presence of the hole with a stationary center in multiple channel maps, sufficient contrast between the hole and its immediate surroundings, and the shape of the hole in radius-velocity space must be described by an ellipse \citep{puc92, bag09}.  Both catalogs contain the locations, diameters, expansion velocities of each of the holes (see Tables \ref{tab1} $-$ \ref{tab4} and Figure \ref{map1}).  Additionally, the THINGS catalog classifies each hole such that a type 1 hole has completely blown out of the disk of the galaxy, i.e, a break in the \pv\ diagram, a type 2 hole only shows an approaching or receding side, and a type 3 hole has both sides present in \pv\ space.  While the P92 and THINGS holes catalogs are generally similar, the sizes, locations, and measured expansion velocities are not identical.  For example, the THINGS catalog assigns an expansion velocity of 7 \kms to any blown out hole, while P92 lists expansion velocities for each hole, and does not indicate a blow out.  Because of the high stellar density, slight differences in the hole locations and size can lead to different stellar populations.  We explore this further in \S 4.3.

\begin{figure}[t]
\begin{center}
\epsscale{0.8}\plotone{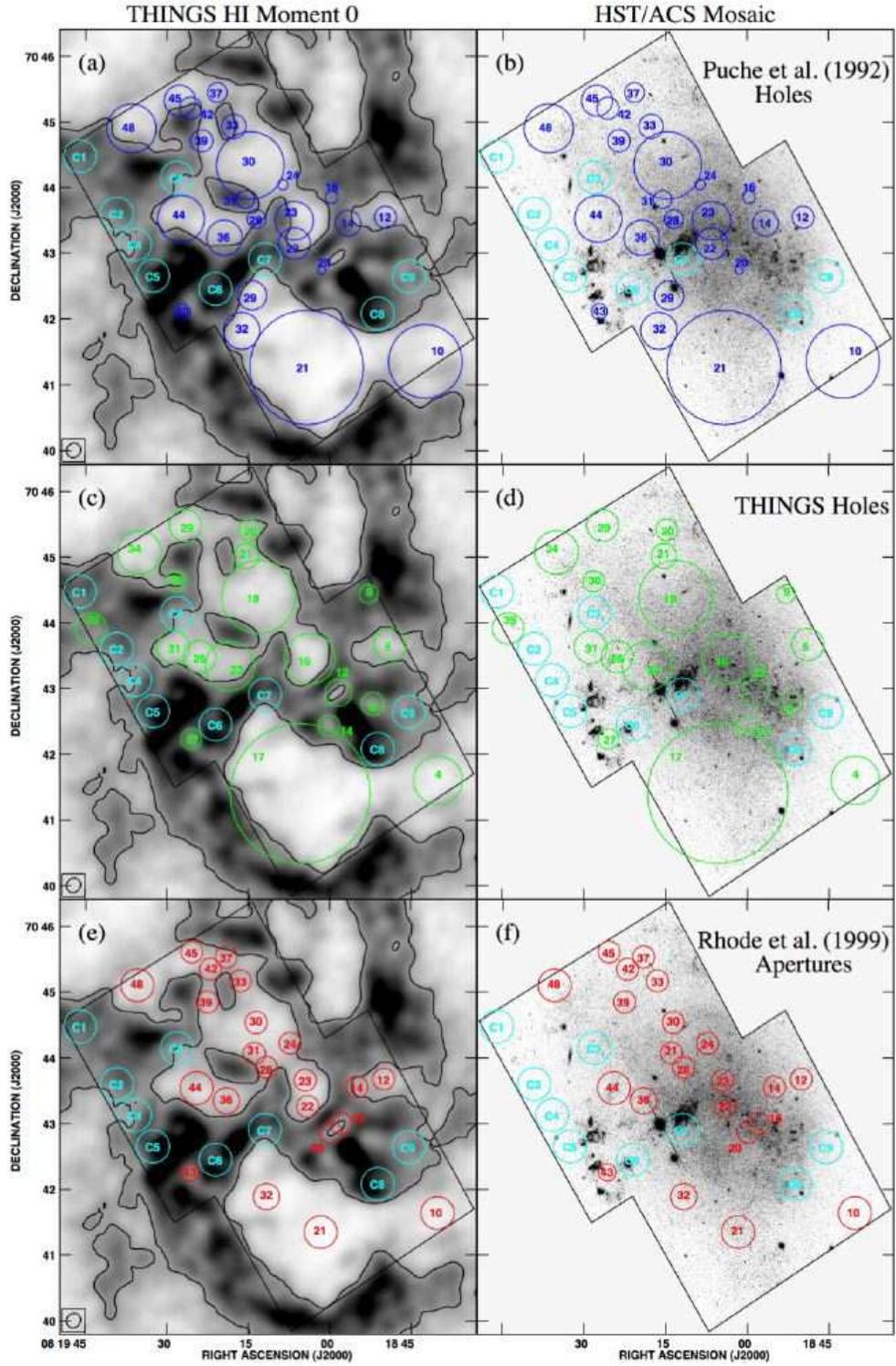}
\caption{(a) THINGS processed \hi\ image of \hoii\ \citep{wal08} with a contour of 10$^{-21}$ cm$^{-2}$ overlaid and (b) ACS drizzled image of \hoii\ with \citet{puc92} holes in blue and control fields in cyan; (c) and (d) THING holes catalog \citep{bag09} in green and control fields in cyan; (e) and (f) \citet{rho99} apertures in red and control fields in cyan.}

\label{map2}
\end{center}
\end{figure}

With a distance of 3.40 Mpc, measured from the tip of the red giant branch \citep{kar02}, the high spatial resolution of HST/ACS (1 pixel $=$ 0.05\arcsec; 1\arcsec\ $\sim$ 16.5 pc) allows us to probe scales much smaller than the minimum detectable \hi\ hole size ($\sim$ 100 pc).  Treating each hole as circular and using the central coordinates and diameters from each hole catalog, we overlaid these apertures over the HST/ACS drizzled image and found that 23 of the 51 holes from P92 and 19 of the 39 THINGS holes were within ACS coverage (Figure \ref{map2}).  In the P92 catalog, a portion of holes 10, 21, 25, 34, and 48 are outside our ACS field of view.  We included those that primarily overlapped with the ACS data (10, 21, and 48), but excluded the others from the sample (25 and 34).  From the THINGS catalog, we excluded holes 18 and 38 for the same reason.  For consistency we used the \hi\ hole numbering systems presented in the P92 and THINGS catalogs.  Additionally, we selected nine control fields, regions in comparable size to the \hi\ holes that do not overlap cataloged \hi\ holes and span a range in \hi\ column densities, and number them c1 $-$ c9 (Figures \ref{map1} and \ref{map2}).

\section{Stellar Components of \hi\ holes in \hoii}

\subsection{Luminous Main Sequence Stars Within the \hi\ Holes}

\begin{figure}[t]
\begin{center}
\plotone{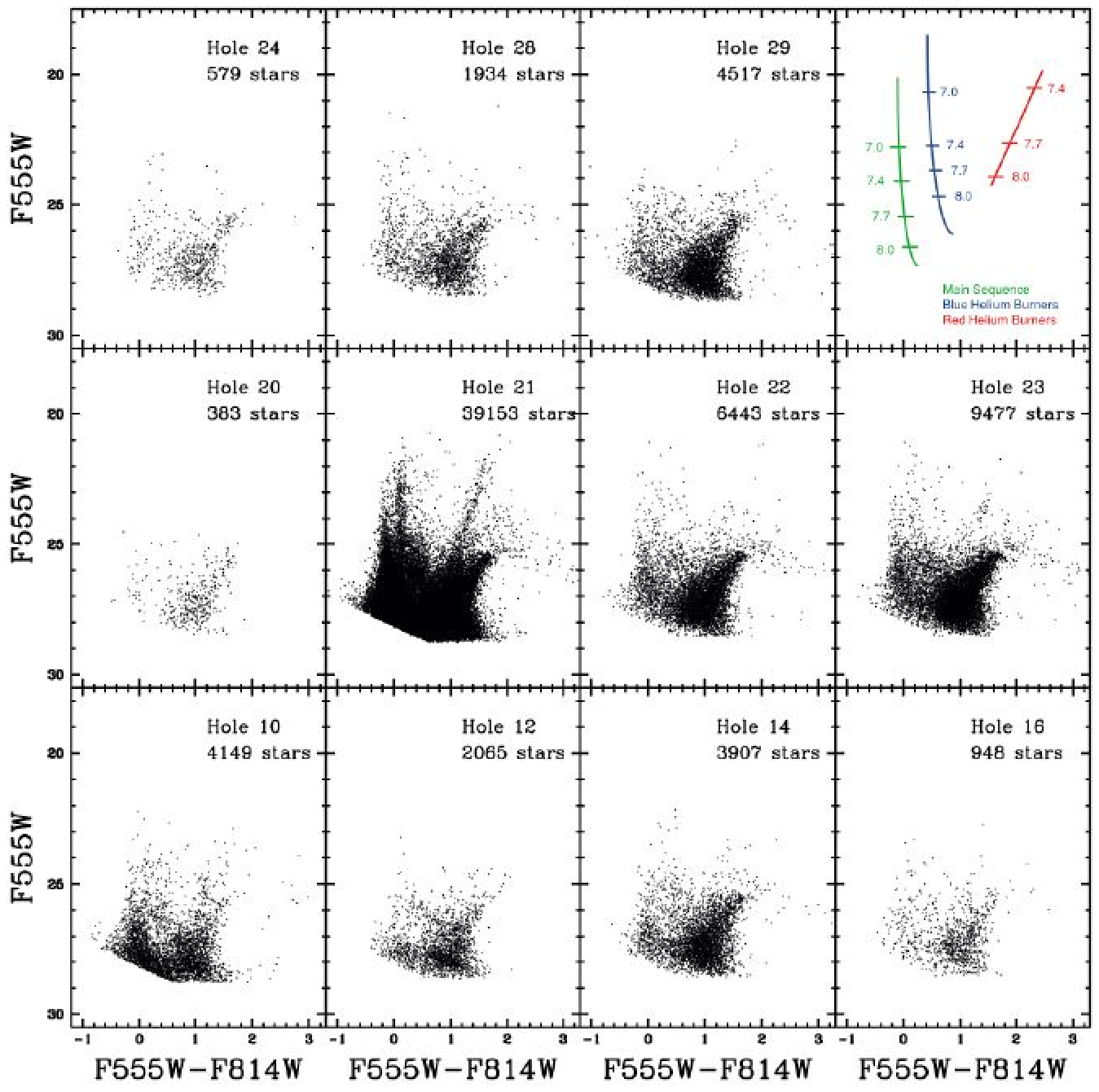}
\caption{HST/ACS CMDs of the stars inside the \hi\ holes cataloged by \citet{puc92}, corrected for foreground reddening, A$_{V}$ $=$ 0.11 \citep{sch98}.  The schematic in the upper right hand corner shows the ages of each type of star, MS (green), BHeB (green), and RHeB (red), with the logarithm of the  ages (turn off age for the MS) shown as a function of magnitude and color on the CMD.  Although sparse in some fields, note the presence of young MS and BHeB stars in all the CMDs.  The fact the the BHeBs span a range in magnitudes indicate that multiple episodes of recent SF must have taken place, as different age BHeBs do not overlap on the CMD. In contrast, BHeBs from a single age cluster would have an overdensity at only one magnitude.}  
\label{p92_cmd1}
\end{center}
\end{figure}

\begin{figure}[t]
\begin{center}
\plotone{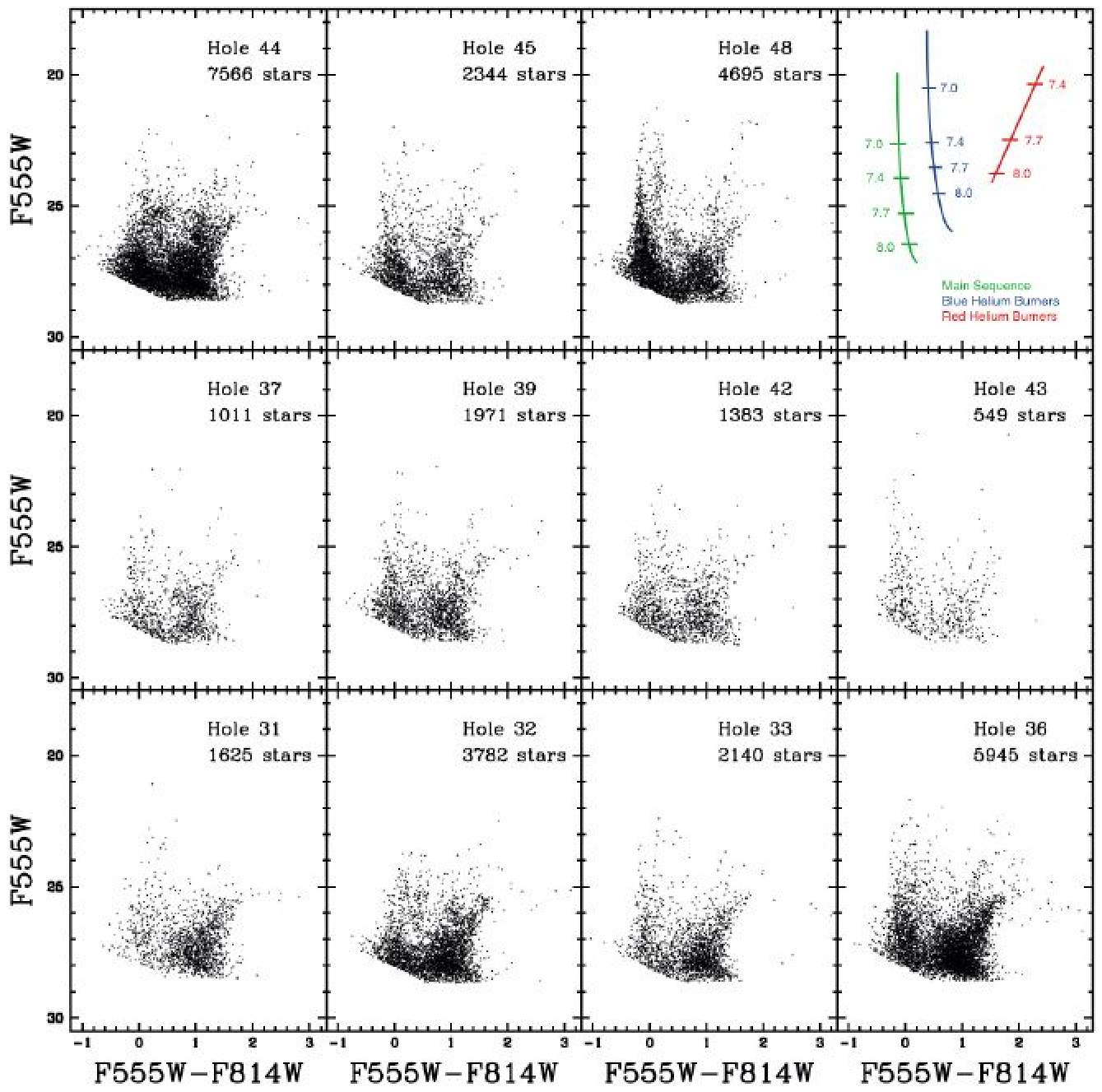}
\caption{HST/ACS CMDs of the stars inside the \hi\ holes cataloged by \citet{puc92}, corrected for foreground reddening, A$_{V}$ $=$ 0.11 \citep{sch98}.  The schematic in the upper right hand corner shows the ages of each type of star, MS (green), BHeB (green), and RHeB (red), with the logarithm of the ages (turn off age for the MS) shown as a function of magnitude and color on the CMD. Although sparse in some fields, note the presence of young MS and BHeB stars in all the CMDs.  The fact the the BHeBs span a range in magnitudes indicate that multiple episodes of recent SF must have taken place, as different age BHeBs do not overlap on the CMD. In contrast, BHeBs from a single age cluster would have an overdensity at only one magnitude.}  
\label{p92_cmd2}
\end{center}
\end{figure}

\begin{figure}[t]
\begin{center}
\plotone{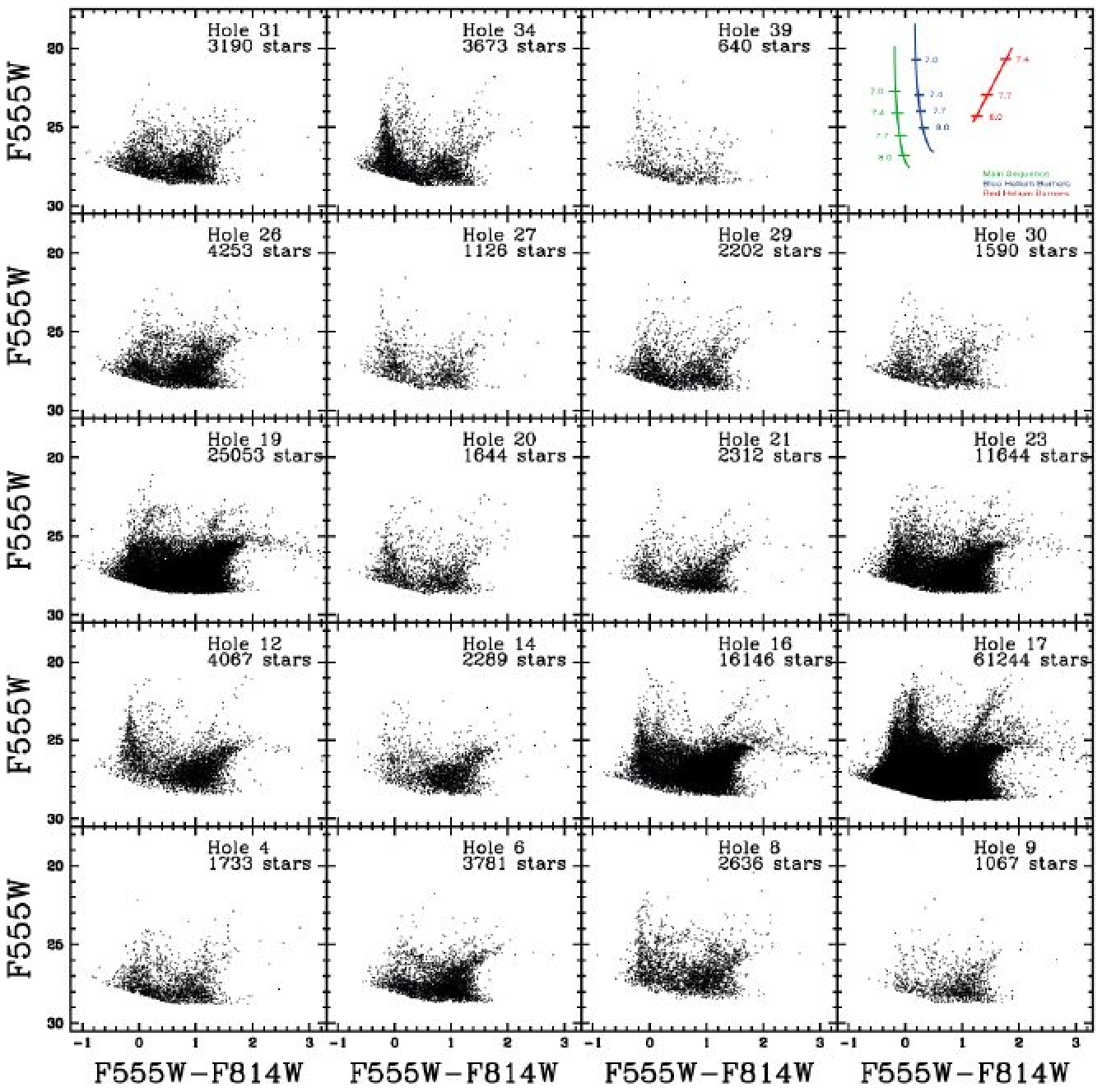}
\caption{HST/ACS CMDs of the stars inside the \hi\ holes cataloged by \citet{bag09}, corrected for foreground reddening, A$_{V}$ $=$ 0.11 \citep{sch98}.  The schematic in the upper right hand corner shows the ages of each type of star, MS (green), BHeB (green), and RHeB (red), with the logarithm of the  ages (turn off age for the MS) shown as a function of magnitude and color on the CMD. Although sparse in some fields, note the presence of young MS and BHeB stars in all the CMDs.  The fact the the BHeBs span a range in magnitudes indicate that multiple episodes of recent SF must have taken place, as different age BHeBs do not overlap on the CMD. In contrast, BHeBs from a single age cluster would have an overdensity at only one magnitude.}  
\label{THINGS_cmd}
\end{center}
\end{figure}

\begin{figure}[t]
\begin{center}
\plotone{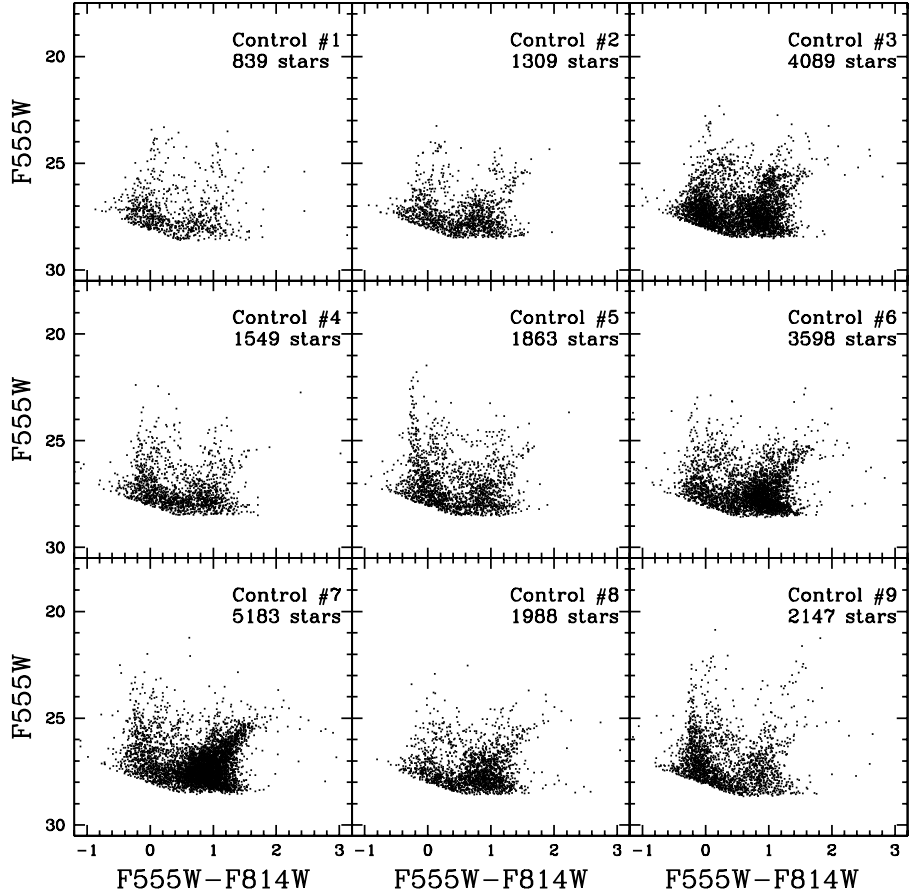}
\caption{HST/ACS CMDs of the nine control fields in \hoii\ (Table \ref{tab3} and Figure \ref{map2}), corrected for foreground reddening, A$_{V}$ $=$ 0.11 \citep{sch98}. Although sparse in some fields, note the presence of young MS and BHeB stars in all the CMDs.  The fact the the BHeBs span a range in magnitudes indicate that multiple episodes of recent SF must have taken place, as different age BHeBs do not overlap on the CMD. In contrast, BHeBs from a single age cluster would have an overdensity at only one magnitude. Comparing the CMDs of the control fields to those associated with the \hi\ holes (Figure \ref{p92_cmd1} $-$ \ref{THINGS_cmd}), there is not a clear difference in the composition of the two stellar populations despite having vastly different \hi\ column densities.} 
\label{ccmd1}
\end{center}
\end{figure}

\begin{figure}[t]
\begin{center}
\plotone{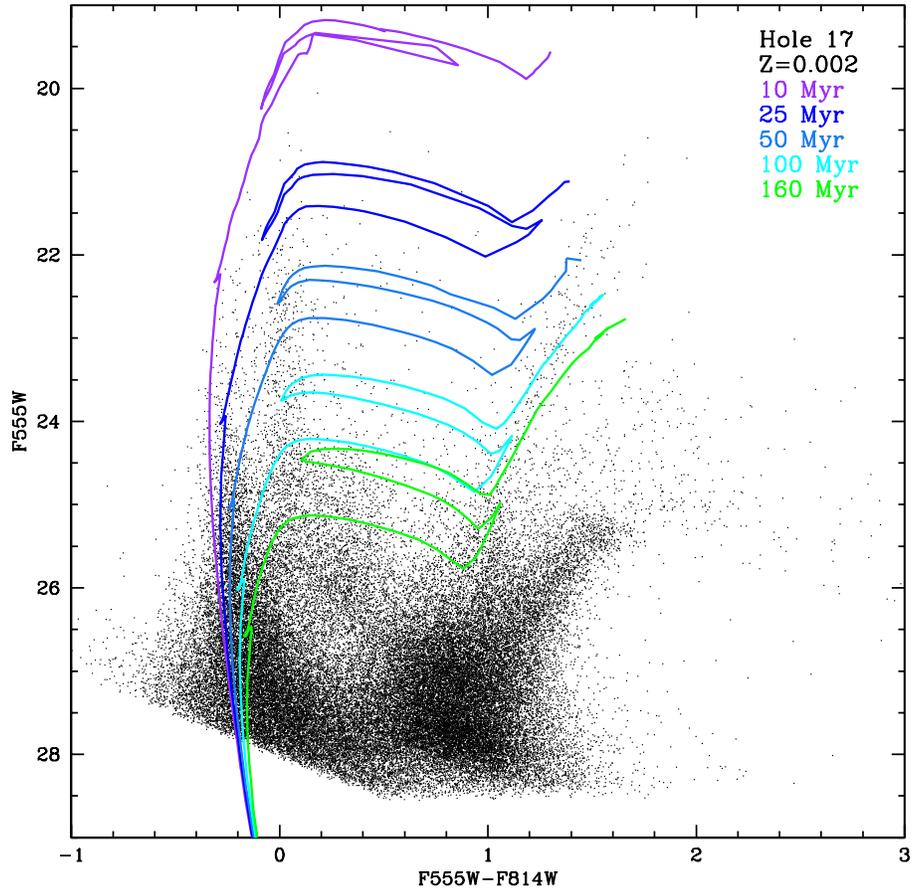}
\caption{HST/ACS CMD of THINGS hole 17 with the Z=0.002 isochrones from \citet{mar08} overlaid.  The observed CMD has been corrected for foreground reddening.  Notice that although the observed MS and isochrones show excellent agreement, the BHeB models become increasingly redder with age, when compared with the data.  This is not an effect of metallicity, as we matched the isochrones to the measured nebular abundance \citep{mil96}.  Instead, this is an example of the color mismatch for BHeB models at metallicites greater than $\sim$ 10\% \zsun.}
\label{iso_cmd}
\end{center}
\end{figure}

From our photometry we constructed CMDs  of the stars in each of the 23 holes in the P92 catalog, 19 holes in the THINGS catalog, and nine control fields (Figures \ref{p92_cmd1} $-$ \ref{ccmd1} and Tables \ref{tab1}, \ref{tab2}, \& \ref{tab5}).  Notably, we found that all holes in both catalogs contain a significant number of young stars; enough to easily identify CMD features by visual inspection and compare with the models of \citet{mar08} (see Figure \ref{iso_cmd}).  However, these CMDs are not consistent with the expected CMD for a single age stellar cluster. The holes also have red giant (RGB) and asymptotic giant branch (AGB) stars with ages $<$ 1 Gyr, as expected given that the older stars are fairly smoothly distributed throughout the optical galaxy (see Figure 11 in \citealt{wei08}).

Young stellar clusters (e.g., OB associations) are traditionally associated with the formation of \hi\ holes \citep[e.g.,][]{bri86, hat05}, and we tested this notion by making a census of the youngest and brightest stars.  We identified young stars based on the stellar evolution models of \citet{mar08}, which suggest that the brightest 10 Myr old MS star at its turnoff has M$_{F555W}$ $\sim$ $-$5, or m$_{F555W}$ $\sim$ 22.65 in \hoii.  We found 302 MS stars with m$_{F555W}$ $<$ 22.65 that also lay within holes in the P92 catalog.  Further, restricting the color to be less than 0.05, i.e., m$_{F555W}$$-$m$_{F814W}$ $<$ 0.05, (to exclude Red Helium Burning stars (RHeBs), bright foreground stars, and most BHeBs) we find 109 stars that match our criteria.  Of these 109 stars, 21 of them belong to hole 21, which is the largest and most populated hole with a diameter of $\sim$ 2 kpc and $\sim$ 40,000 stars in the CMD. Applying the above criteria for luminous MS stars to the remaining 21 holes, we found that all the holes have fewer than 10 bright MS stars and 20 holes have fewer than two bright MS stars.  Similarly for the \hi\ holes in the THINGS catalog, we found 347 stars with m$_{F555W}$ $<$ 22.65, and 104 stars with m$_{F555W}$ $<$ 22.65 and m$_{F555W}$$-$m$_{F814W}$ $<$ 0.05.  Of these 104 stars, 30 belong to the largest hole (number 17). The other 18 holes all have fewer than 10 young MS stars, and 13 have fewer than two young MS stars.  For the stars located in the control fields (Figure \ref{ccmd1}), we found that 32 match the bright MS star criteria, with c7 containing 21 of these stars, and six of the nine control fields containing two or less bright MS stars.  The control field have more bright MS stars by a factor of $\sim$ 1.5 when compared to the \hi\ holes.  We explore the impact of small number statistics, stochastic effects on the SFR, and production of massive stars in \S 5.

From these basic calculations, it seems that the location of young OB type stars are not a reliable tracer of the locations of \hi\ holes, nor do they accurately represent the presence of young remnant stellar populations inside \hi\ holes.  We will discuss the spatial distribution of these stars further in \S 5.2.

\subsection{The SFHs of Stars Within the \hi\ Holes}

Upper MS stars and BHeBs offer the opportunity to age date stellar populations more than 10 Myr old.  To extract a more complete picture of the recent SF within the \hi\ holes, we measured the SFHs for the stars inside each \hi\ hole and compared the timing of SF activity to the inferred kinematic ages, i.e., the radius of the hole divided by its expansion velocity.  We derived the SFHs, from the CMDs, using SFH recovery code of \citet{dol02}.  This method constructs synthetic CMDs, using the stellar evolution models of \citet{mar08}, and compares them to the observed CMD using a maximum likelihood technique. To obtain this solution, we used a combination of fixed (e.g, binary fraction, IMF) and searchable (e.g., distance, extinction) parameters.  We allowed the program to search for the best fit metallicity per time bin, with the condition that the metallicity must monotonically increase with time toward the present. We found the mean metallicity in the most recent time bins from the SFHs to be consistent with the observed value of $\sim$ 10\% \zsun\ \citep{mil96}, which is best fit by Z=0.002 isochrones (Figure \ref{iso_cmd}). Throughout this paper, we use the following values when measuring SFHs or simulating CMDs: a standard power law IMF with x $=$ $-$2.3 from 0.1 to 100 \msun, a binary fraction $=$  0.35, the stellar evolution models of \citet{mar08}, a distance of 3.40 Mpc, a foreground extinction of A$_{F555W}$ $=$ 0.11 \citep{sch98}, and 50\% completeness limits of m$_{F555W}$ $=$ 27.9 and m$_{F814W}$ $=$ 27.4 \citep{wei08}. To quantify the errors in the SFHs, we added the systematic uncertainties from the isochrones and the statistical uncertainties from Monte Carlo tests in quadrature.  Quantifying the best fit solution and associated uncertainties from synthetic CMD matching has been subject to a number of extensive studies \citep[e.g.,][]{tgf89, ts96, gabc96, m97, hol99, hvg99, dol02, ia02, yl07}, which explore the nuances of this method beyond the scope of this paper.  Full details on the method we used in this paper to measure the SFHs and quantify the associated errors can be found in \citet{dol02}.

\begin{figure}[t]
\begin{center}
\plotone{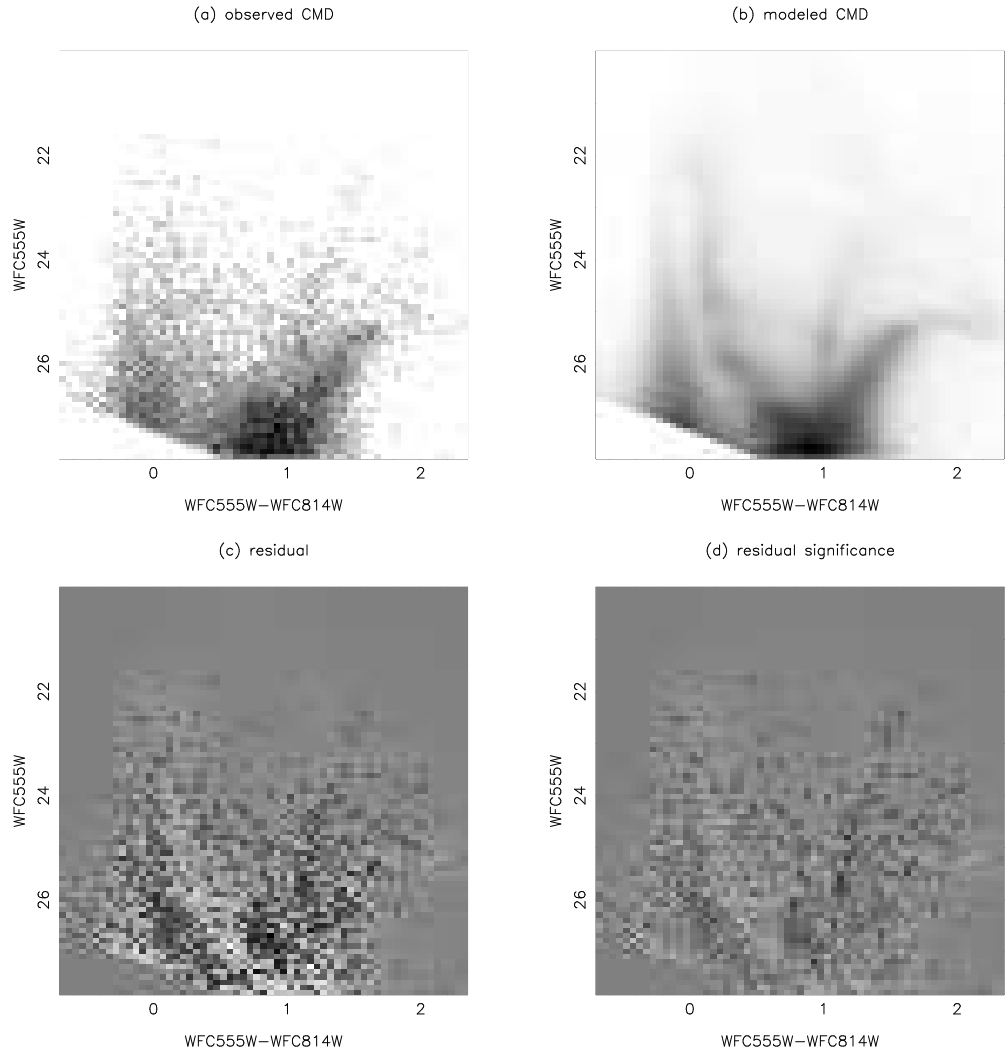}
\caption{Comparison of the observed (panel (a)) and model CMDs (panel (b)) for THINGS hole 23 as measured using the code of \citet{dol02}.  The bottom panels show the residual CMD (panel (c)) and the significance of the residuals (panel (d)).  In the bottom panels black points indicate many more observed stars than synthetic star, while white points have more synthetic than observed stars.  From the residual significance diagram we do not see any particularly poorly matched areas, indicating the synthetic CMD matches the data quite well.}
\label{out_cmd}
\end{center}
\end{figure}

As an example we present the observed, model, residual, and significance of the residual CMDs from the code of \citet{dol02} for hole 23 in the THINGS catalog (Figure \ref{out_cmd}).  Visual inspection of the observed and model CMD show broad agreement of features such as the young MS, BHeBs, RHeBs, and the RGB.  The difference between the model and the observed CMD can been seen in the lower two panels, where black points indicate more real stars than synthetic stars, and white points indicating more synthetic than real stars.  The residual significance diagram (panel (d)), reveals no distinct patterns of black or white points, indicating that the model CMD describes the data quite well. 

\begin{figure}[t]
\begin{center}
\plotone{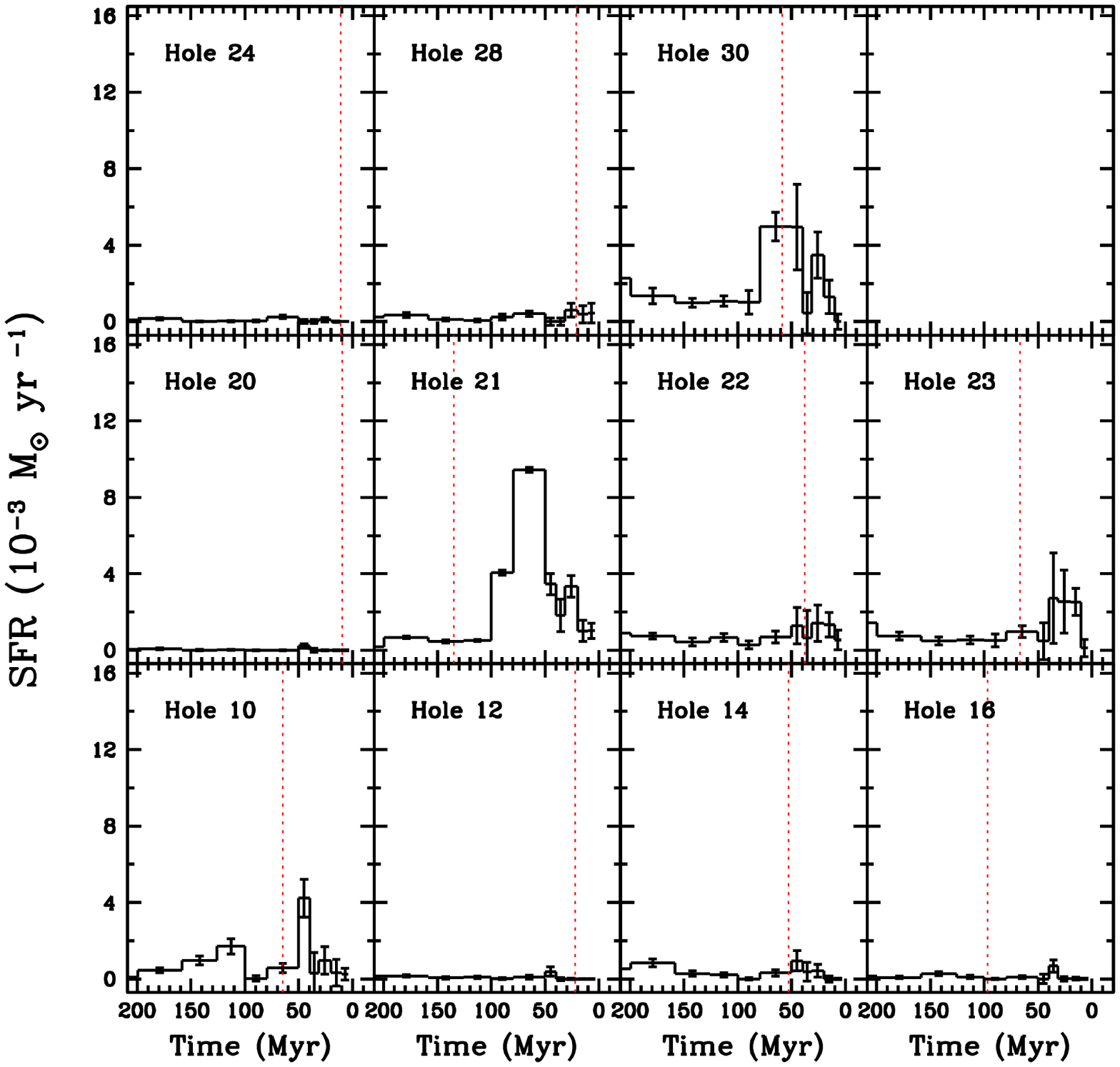}
\caption{The recent SFHs of the stars inside the \hi\ holes cataloged by \citet{puc92} over the last 200 Myr.  The time resolution is $\sim$ 10 Myr until 50 Myr ago, and then $\sim$ 25 Myr there after.  The red dotted line indicates the inferred kinematic age of each hole (Table \ref{tab3}). Note that the SFH of hole 21 has been scaled down by a factor of 10.  In most cases, the majority of SF has taken place more recently than the inferred kinematic age, however, this is not always the case.}  
\label{p92_sfh1}
\end{center}
\end{figure}

\begin{figure}[t]
\begin{center}
\plotone{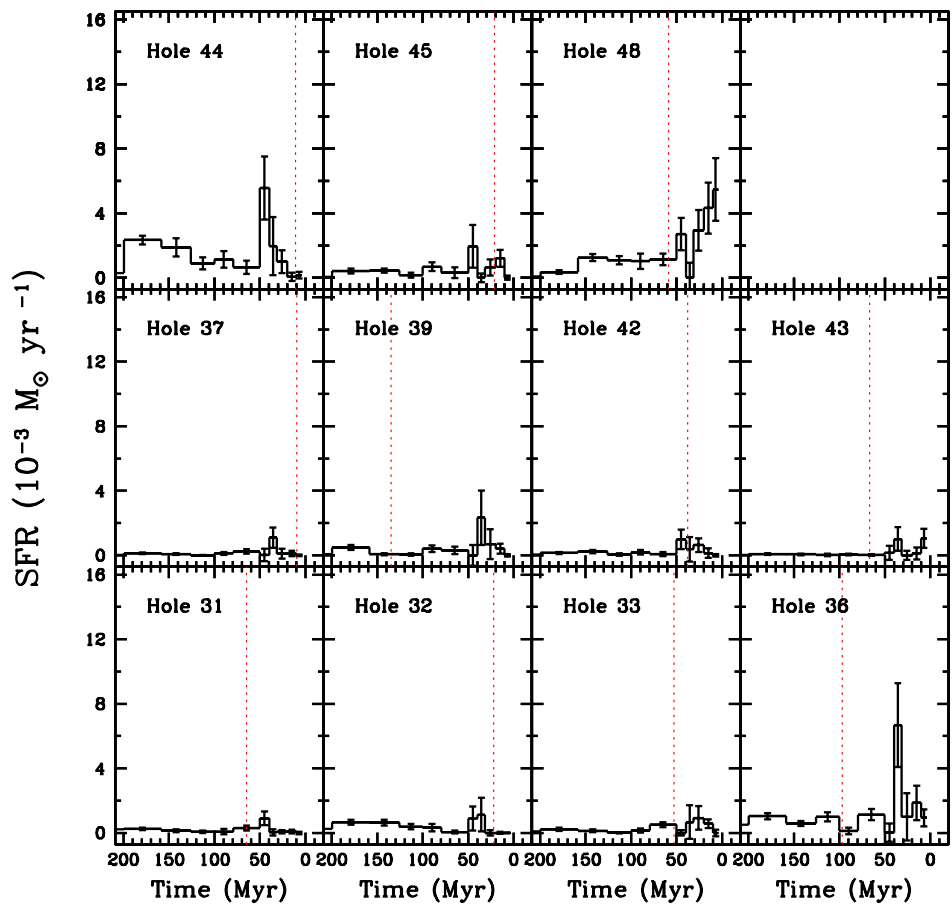}
\caption{The recent SFHs of the stars inside the \hi\ holes cataloged by \citet{puc92} over the last 200 Myr.  The time resolution is $\sim$ 10 Myr until 50 Myr ago, and then $\sim$ 25 Myr there after.  The red dotted line indicates the inferred kinematic age of each hole (Table \ref{tab3}). In most cases, the majority of SF has taken place more recently than the inferred kinematic age, however, this is not always the case.}  
\label{p92_sfh2}
\end{center}
\end{figure}

\begin{figure}[t]
\begin{center}
\plotone{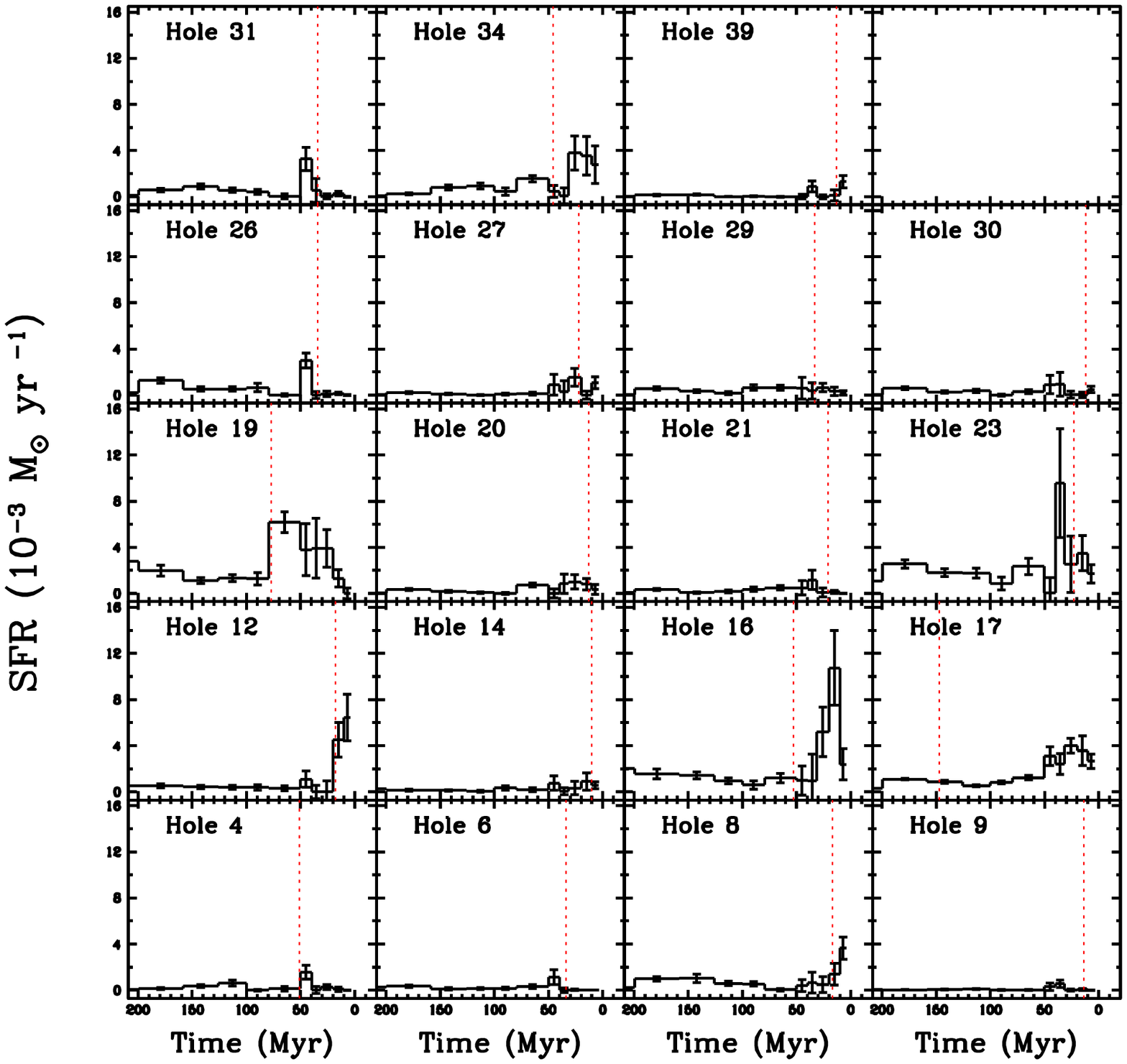}
\caption{The recent SFHs of the stars inside the \hi\ holes cataloged by \citet{bag09} over the last 200 Myr.  The time resolution is $\sim$ 10 Myr until 50 Myr ago, and then $\sim$ 25 Myr there after.  The red dotted line indicates the kinematic age of each hole (Table \ref{tab4}).  Note that the SFH of hole 17 has been scaled down by a factor of 10. In most cases, the majority of SF has taken place more recently than the inferred kinematic age, however, this is not always the case.}  
\label{THINGS_sfh}
\end{center}
\end{figure}

\begin{figure}[t]
\begin{center}
\plotone{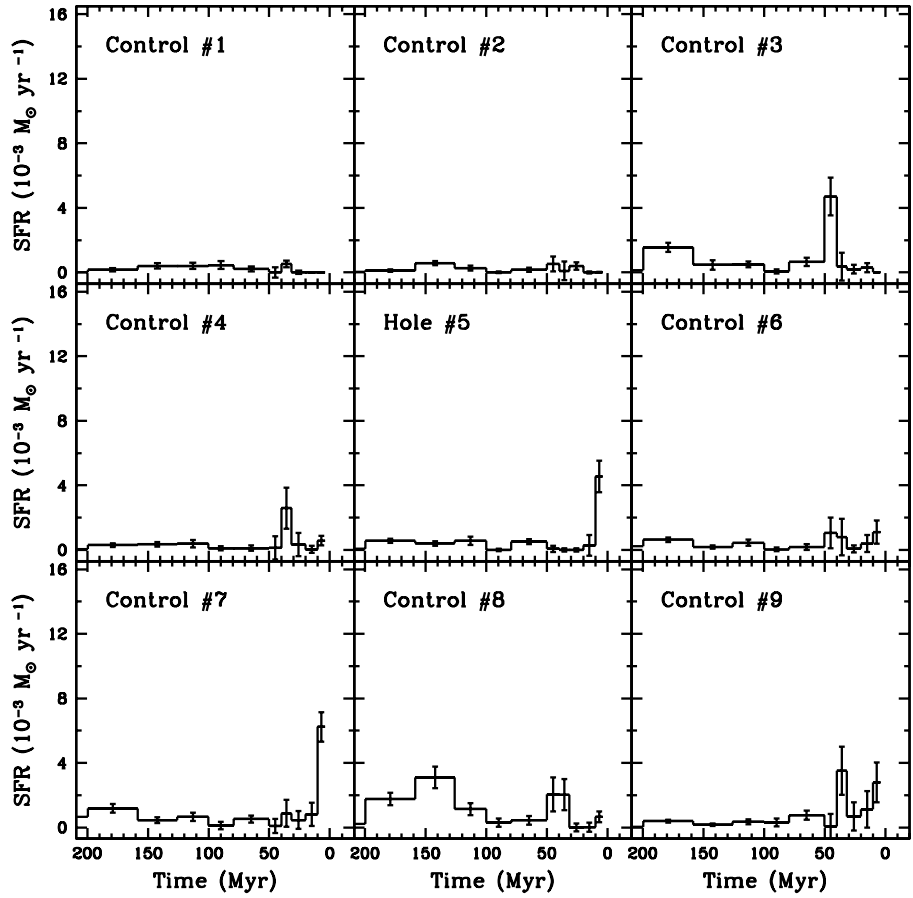}
\caption{The recent SFHs of each control field in \hoii\ over the last 200 Myr.  The time resolution is $\sim$10 Myr ber bin until 50 Myr and $\sim$ 25 Myr there after. }  
\label{csfh1}
\end{center}
\end{figure}

We chose to focus on SF over the past 200 Myr in each of the holes (Figures \ref{p92_sfh1} $-$ \ref{THINGS_sfh}) and control fields (Figure \ref{csfh1}) based on the inferred kinematic ages of the holes derived from both hole catalogs.  The time resolution of our SFHs is $\sim$ 10 Myr up to a look back time of 50 Myr and then $\sim$ 25 Myr from 50 $-$ 200 Myr. These choices provide an adequate look back time and time resolution to see potential SF events responsible for the creation of the \hi\ holes.  Note that kinematic ages for \hi\ holes are traditionally considered an upper limit to the age of the \hi\ holes, as the expansion velocities were most likely higher in the past \citep[e.g.,][]{ttb88}.  We discuss the accuracy of inferred kinematic ages in more detail in \S 5.

As a check for accuracy of SFHs between the two hole catalogs, we compared the SFHs and CMDs from overlapping holes.  The CMD of P92 hole 44 significantly overlaps with THINGS holes 26 and 31.  The CMD of P92 hole 44 has 7566 stars, where as the two THINGS holes have 4253 and 3190, respectively.  In terms of the SFHs, the peak SFR for P92 hole 44 occurs in the 40 $-$ 50 Myr bin with a peak values of 5.6\e{-3} \msun\ yr$^{-1}$.  Both the THINGS holes how peak SFRs in the same time bin, with values of 3.0\e{-3} \msun\ yr$^{-1}$ and 3.2\e{-3} \msun\ yr$^{-1}$, or co-adding them, we find a peak SFR of $\sim$ 6.2\e{-3} \msun\ yr$^{-1}$, consistent with the values for the single P92 hole. Similar tests were carried out for several other overlapping holes, and all were found to have consistent SFRs where expected. It is important to note that because the stellar density is not uniform in \hoii, even small differences in the location or radius of a hole, may result in significant differences in the stellar populations and SFHs.

\begin{figure}[t]
\begin{center}
\plotone{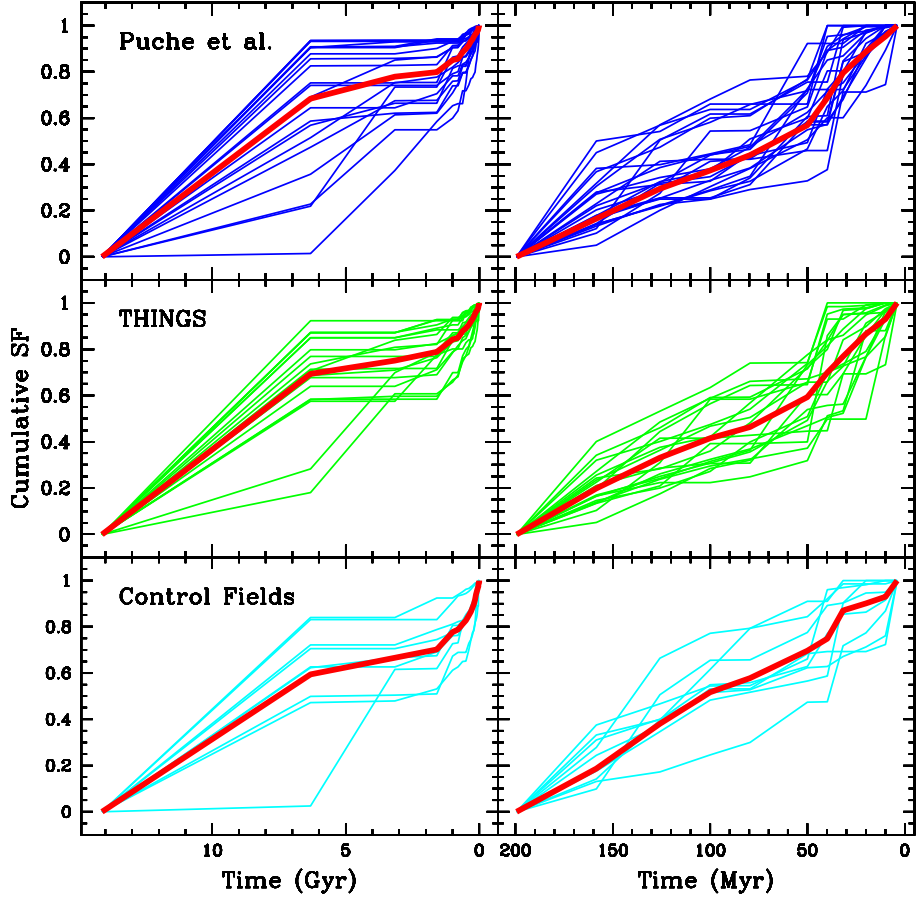}
\caption{Comparison of the cumulative SF of stars within the holes from the \hoii\ for the \citet{puc92}  catalog (blue) and the THINGS catalog \citep{bag09} (green) with the control fields cyan.  The left panels show the cumulative SF over the the past 14.1 Gyr, while the panels on the right show the cumulative SF over the past 200 Myr.  Although any given cumulative SFH is unique, a comparison of the \hi\ holes and control fields suggest that the SFHs of the stars inside \hi\ holes and the control fields are indistinguishable.  For reference, the globally averaged SFR for the galaxy is 3.6\e{-2} \msun\ yr$^{-1}$, and the full SFH of the entire galaxy was measured by \citet{wei08}.}  
\label{cumulative}
\end{center}
\end{figure}

We searched for differences between the stellar populations of the control fields and the \hi\ holes by comparing their cumulative SFHs (i.e., fraction of stars formed as a function of time) over the past 14.1 Gyr (lifetime) and 200 Myr (recent) by performing a KS test comparing the average cumulative SFHs for the holes and control fields (red lines in Figure \ref{cumulative}).  As expected, the majority of the holes and control fields share a common old stellar population, i.e., at least 50\% of the stars were formed greater than 6 Gyrs ago), and the fraction of stars formed in the last 6 Gyrs is also consistent between the hole and control field samples. Although the past 200 Myr is likely more influential when it comes to hole formation, again we see no discernible difference between the recent cumulative SFHs of the two samples.  The KS test confirmed these impressions, indicating that, on both the lifetime and recent time scales, the stellar populations of the holes and control fields have the same parent distribution.  From the perspective of the traditional stellar feedback hypothesis, this is an unanticipated result, as the theory would expect young stars associated with remnant clusters to be predominantly located inside the \hi\ holes, and not control fields.  This is discussed in more detail in \S 5.



\section{Comparing the Energy from SF to the Energy Needed to Create an \hi\ Hole }

Quantifying the energy generated from SF and the energy necessary to create an \hi\ hole is of particular importance in understanding the role stellar feedback plays in creating \hi\ holes.  Fortunately, the interaction of a single SN and the surrounding ISM has been modeled in great detail \citep[e.g.,][]{che74, wea77, hei79}.  In addition, energy associated with SF is computed by the galaxy evolution modeling code STARBURST99 \citep{lei99}.  STARBURST99 takes a SFH as input and uses stellar evolution models to produce predictions for spectrophotometric, chemical, and stellar energy output properties of galaxies that have had episodes of SF within the last 1 Gyr.  For the particular case of stellar feedback, STARBURST99 can compute the energy profiles (i.e., energy produced by stellar winds and SNe) using the SFHs for individual holes, which allows us to directly connect stellar feedback to the creation and evolution \hi\ holes.  In this section, we apply each of these methods to the \hi\ holes in \hoii\ and their stellar populations, and compare them to expected stellar feedback efficiencies from a variety of hydrodynamical simulations.

\subsection{Energy Required to Evacuate an \hi\ Hole}

The minimum amount of energy required to evacuate an \hi\ hole is given by the single blast model of \citet{che74}:

\begin{equation}
E_{Hole} = 5.3\times10^{43}~n_{0}^{1.12}~(\frac{d}{2})^{3.12}~v^{1.4}
\end{equation}

where $E_{Hole}$ is in the units of 10$^{50}$ ergs, $n_{0}$ is the \hi\ volume density (cm$^{-3}$) at the mid-plane of the galaxy, d is the diameter of the \hi\ hole in pc, and v is the expansion velocity in \kms.  While the expansion velocity and diameter of the hole are independently measured quantities,  $n_{0}$ is dependent upon the \hi\ column density, $N_{\hi}$, and scale height of the gas, h, in the following way:

\begin{equation}
n_{0} = \frac{N_{\hi}}{\sqrt{2\pi}h}
\end{equation}

The denominator, $\sqrt{2\pi}h$ is the effective thickness of the gas assuming a Gaussian distribution, with h being the 1$\sigma$ scale height.  


One of the challenges in determining $E_{Hole}$ is accurately measuring $n_{0}$.  Both P92 and \citet{bag09} use a rotation curve to estimate the total mass in a gravitationally bound spherical system.  This value was averaged over the volume of a sphere with a radius, $R_{V_{max}}$ and multiplied by a factor of two.  Measuring $\sigma_{v}$ $=$ 6.8 \kms (\hi\ velocity dispersion), P92 found $h$ $=$ 625 pc, and \citet{bag09} find $h$ $=$ 600 pc.  P92 used this method to compute the scale height of M31, and found it to be in agreement with the mass-model based method of \citet{bri84}.  However,  P92 acknowledged that wide-field, deep photometry of \hoii, would produce an accurate mass model of \hoii, and a more precise knowledge of $h$, $n_{0}$, and $E_{Hole}$.  

\begin{figure}[t]
\begin{center}
\plotone{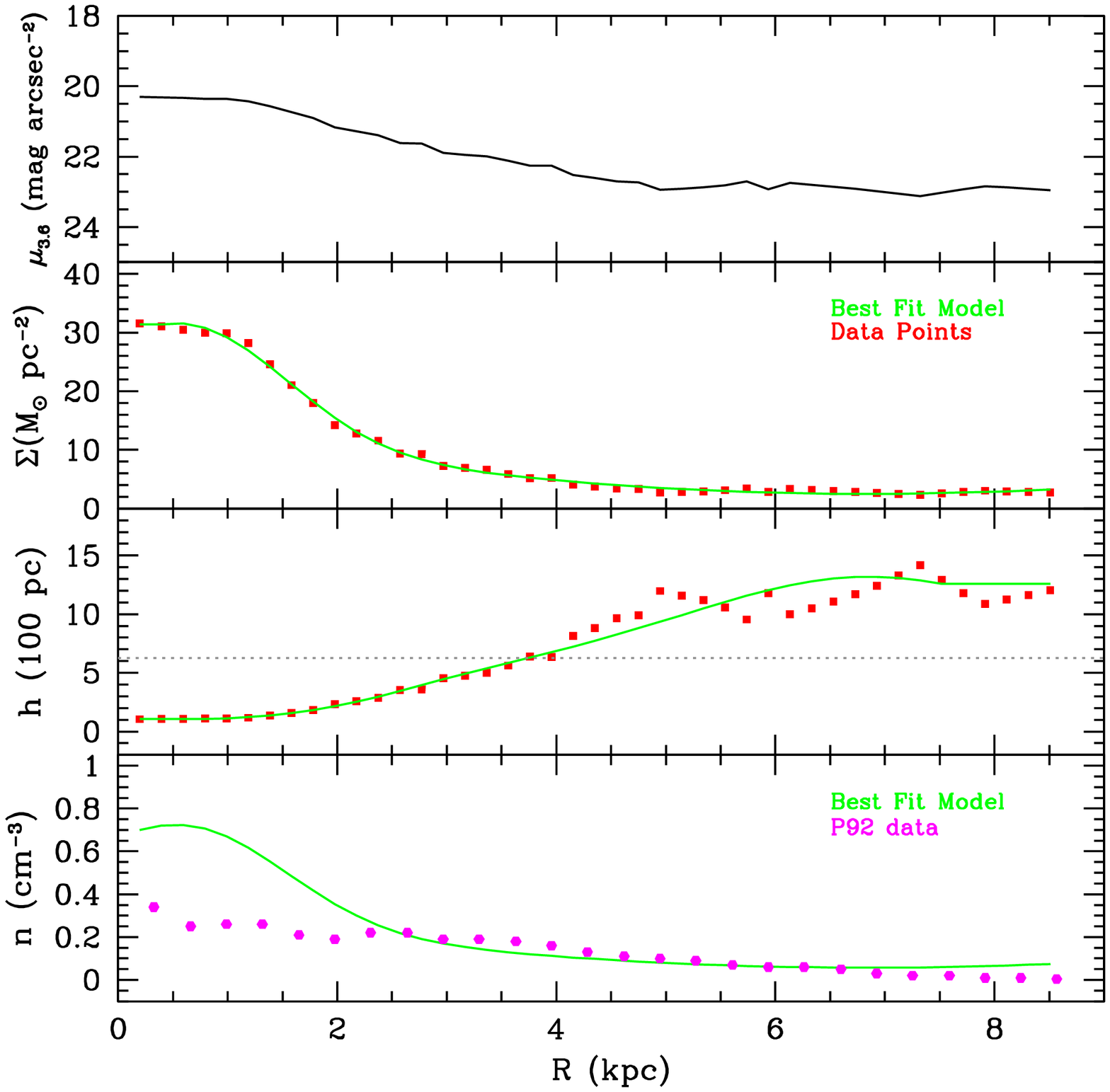}
\caption{\emph{Top panel}: Surface brightness profile of \hoii\ as measured from 3.6$\mu$m imaging from the Local Volume Legacy program \citep[LVL;][]{lee08, dal09} as a function of radius; \emph{Second Panel}: The mass surface density derived from the surface brightness profile following the method of \citet{oh08}.  The red points are the data points and the green line is the best fit analytic function; \emph{Third panel}: The gas scale height of \hoii\ as a function of radius assuming an isothermal disk derived using the method of \citep{kel70, kim98}.  Data points are in red and the best fit analytic function is the green line.  The grey dotted line is the constant value gas scale height used by \citep{puc92}; \emph{Bottom Panel}: The present day \hi\ volume density, $n$, as a function of radius (green line) with the values used by \citet{puc92} in magenta.  Note the difference in the values of n between the best fit model obtained via surface photometry and the values used by \citet{puc92} in the inner $\sim$ 2 kpc.  Because $E_{Hole}$ $\propto$ $n_{0}^{1.12}$ (where $n_{0}$ is the average volume density prior to hole formation), an increase in the volume density leads to an increase in the minimum energy necessary to create an \hi\ hole.  In this case $n_{0}$ measures the average value of the \hi\ volume density, however, SF is more likely to occur at \hi\ peaks, thus $E_{Hole}$ is likely a lower bound (See \S 4.1 for more details).}
\label{scale}
\end{center}
\end{figure}

Fortunately, we had access to high quality 3.6$\mu$m imaging of \hoii\ through the Local Volume Legacy Survey \citep[LVL;][]{lee08, dal09} taken with the \textit{Spitzer Space Telescope}, which allowed us to construct such a mass model.  The LVL team provided a calibrated (including foreground star subtraction) 3.6$\mu$m image of \hoii.  From this image, we used the ISOPHOT package in IRAF\footnote[2]{IRAF is distributed by the National Optical Astronomy Observatory, which is operated by the Association of Universities for Research in Astronomy (AURA) under cooperative agreement with the National Science Foundation.} to perform surface photometry, accounting for inclination and position angle effects \citep[$i$ $=$ 47$^{\circ}$, PA $=$ 177$^{\circ}$; ][]{puc92}  starting at the dynamical center of the galaxy ($\alpha$ $=$ 08:19:18, $\delta$ $=$ $+$70:42:37; J2000), which yielded a surface brightness profile as a function of galactocentric radius (Figure \ref{scale}).  Following the method described in \S\ 4 of \citet{oh08} and assuming the 3.6$\mu$m mass to light ratio, $\Gamma_{\star}$ $=$ 0.5 \citep[comparable to other dwarf irregulars;][]{wal08}, we converted the surface brightness profile into a surface mass density profile, $\Sigma$.  

The isothermal gas scale height equation (Equation 3) derived by \citet{kel70} (and used by \citealt{kim98} for the LMC) relates the surface mass density, $\Sigma$ and the \hi\ velocity dispersion, $\sigma_{v}$, to the \hi\ scale height, $h$.  For \hoii, \citet{bur02} derived $\sigma_{v}$ as a function of galactrocentric radius, allowing us to compute the \hi\ scale height as a function of radius.

\begin{equation}
h = \frac{\langle\ \sigma_{v}^{2} \rangle}{\pi G \Sigma}
\end{equation}

We now have two independent estimates of the gas scale height, namely the \hi\ scale height of P92 and that inferred from surface photometry. We found the \hi\ scale height derived by P92 to be larger by a factor of $\sim$ 3 $-$ 5 in the innermost 2 kpc of the galaxy, and in agreement at $\sim$ 4 kpc.  At radii larger than 2 kpc, $h$ is in better agreement between the different methods.  


We computed new values of $n_{0}$, as a function of radius, using Equation (2) along with the newly calculated \hi\ scale height and the values of $N_{\hi}$ measured by P92.  Because the P92 $N_{\hi}$ values showed multiple large fluctuations in the inner region of the galaxy, likely due to the presence a number of \hi\ holes,  we interpolated over the regions of large fluctuations to ensure a smooth, continuous distribution.  Effectively, the largest difference in the P92 and new estimated values are due to the differences in the \hi\ scales heights.  The \hi\ scale height derived from surface photometry effectively increases the values of  $n_{0}$ by a factor of  $\sim$ 3 compared to the P92 and \citet{bag09} values, which in turn increases  $E_{Hole}$ by approximately the same amount.

It is important to note that the values of $E_{Hole}$ are likely lower limits on the energy necessary to evacuate an \hi\ hole.  The values of $n_{0}$ are the \emph{average} \hi\ volume density at a given radius.  However, observationally, SF occurs at higher than average values of $n_{0}$, suggesting that to get a true sense of $E_{Hole}$ we would need to measure a peak value of $N_{\hi}$ inside the \hi\ holes.  Although this is not observationally plausible, we can get a sense of how a larger values of $N_{\hi}$ would affect $E_{Hole}$.  As an example, we examined control field 5, which is located near one of the highest regions of current SF in \hoii.  From Equation (3) and the peak $N_{\hi}$ measured inside c5, we find an increase by a factor of $\sim$ 8 in the the peak value of $n_{0}$ versus the average value. This translates into an increase in $E_{Hole}$ by a factor of $\sim$ 10, suggesting that $E_{Hole}$ values present here are likely lower limits.

\subsection{Energy Associated with Star Formation}

A basic criteria of the stellar feedback model is that the energy associated with SF must exceed the energy necessary to evacuate a given \hi\ hole.  To calculate the available energy due to SF, we used the galaxy evolution modeling software STARBURST99 \citep{lei99}. We simulated a single instantaneous burst of SF with a fixed mass of 10$^{6}$ \msun using STARBURST99 parameters that matched those described in \S 3.2, including a metallicity of Z $=$ 0.002, a SN cutoff mass of 8 \msun, and a time sampling of 5 Myr.  

Because the energy output from STARBURST99 scales linearly with the input mass, we ran one simulation (with a stellar mass of 10$^{6}$ \msun), and then subsequently scaled it by the integrated stellar mass from the SFHs of stars inside each hole, sampled at 5 Myr intervals.  The cumulative distribution of the energy generated from SF from the present (t $=$ 0) to 200 Myr for the stars inside each \hi\ hole are shown in Figures \ref{p92_energy1} $-$ \ref{THINGS_energy3}.  The inferred kinematic ages are denoted by the grey dashed line, and efficiencies (see \S 4.3) of 100\%, 10\%, and 1\% are indicated by the red, green, and blue dashed lines, respectively.  

We have examined the efficiency over the inferred kinematic age for both catalogs as a function of galactrocentric radius, hole radius, expansion velocity, and inferred kinematic age (Figure \ref{efficiency}).  For both hole catalogs the clearest trend we found is that efficiency seems to increase with expansion velocity. However, this result is likely misleading, because of the differences in assigning expansion velocities.  In the THINGS, catalog, if a hole was blown out, it was assigned an expansion velocity of 7 \kms, whereas P92 attempted to measure the best fit expansion velocity for each hole.  Thus, if we assigned each P92 hole with an expansion velocity lower than 7 \kms, the value of 7 \kms, there would be no trend in the P92 sample, consistent with the THINGS sample. Based on the THINGS catalog classification (see \S 2) of hole types, we found that holes that have not blown out (type 2 or 3), tend to have efficiencies higher than expected from models (see \S 4.3 for a description of the models), in some cases exceeding 100\%. In contrast, holes that have blown out generally fall within the expected range. Again, this trend is likely due to the measurement techniques of the \hi\ expansion velocities.  In general, a blown out hole has stalled expansion, and its expansion velocity is not detectable above the random \hi\ motions.  Thus, the measurements of E$_{Hole}$ are highly uncertain for blown out holes as are the efficiencies.  The underlying implication is that the creation of an \hi\ hole is a complex event that cannot be accurately represented by a simple kinematic age or efficiency, rather a more complex model is necessary (see \S 5 for more discussion).

\subsection{Comparing Energies and Feedback Efficiencies}

We then compared the inferred energy from SF over the kinematic age of each hole, $E_{SFKA}$, to the energy needed to produce the observed properties of each hole, $E_{Hole}$. The ratio of $E_{Hole}$ to $E_{SFKA}$ allows us to calculate the a `putative' stellar feedback efficiency, $\epsilon$, over the inferred kinematic age of each hole (see Tables \ref{tab3} and \ref{tab4}).  Models of SF including feedback and interactions with the ISM expect such efficiencies to range from $\sim$ 1\% $-$ 20\% \citep{the92, cole94, pad97, thor98}. 15 of the 23 P92 holes and 6 of the 19 THINGS holes have putative efficiencies between 1 $-$ 20\%, in agreement with the models.  Overall, a higher fraction of P92 holes are within the expected efficiency range than THINGS holes.  While this may seem to be an unexpected result, this actually underlines the complexity of the connection between stars and \hi\ holes.  For example, P92 hole 42 and THINGS hole 27 are roughly the same size, but slightly different central coordinates, which results in the the THINGS hole having more stars in its CMD by a factor of $\sim$ 2, in turn influencing the SFH and feedback efficiency.  Of the holes that are not consistent with the expected range, one P92 hole (number 48) has an efficiency greater than 100\%, while five THINGS holes (9, 12, 14, 23, 39) also have efficiencies greater than 100\% (Figure \ref{efficiency}).  These outliers suggest that perhaps energy from stellar feedback is not sufficient to have created these holes.   

\begin{figure}[t]
\begin{center}
\plotone{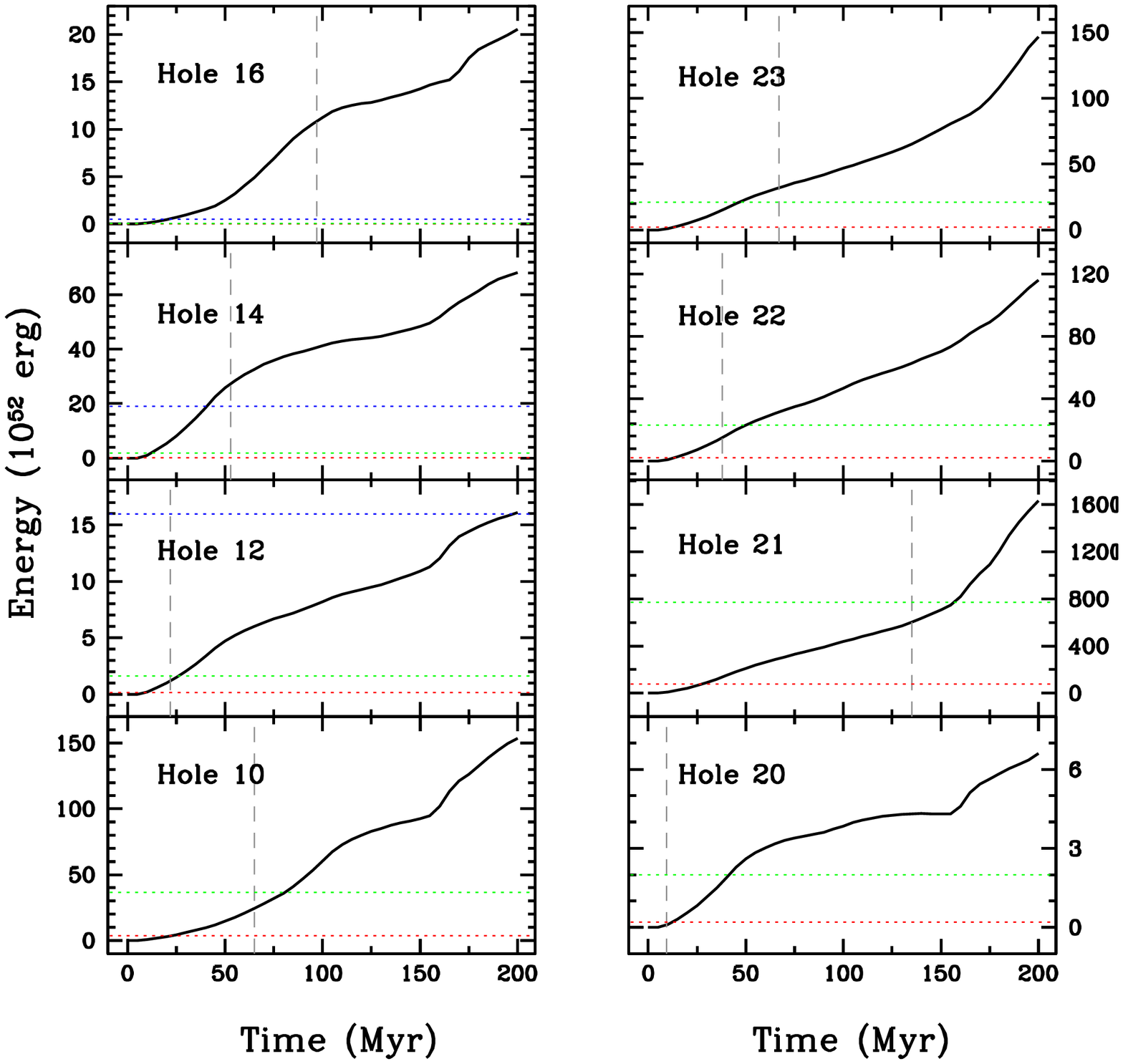}
\caption{The cumulative energies of the stars inside the \hi\ holes of \citet{puc92} computed using STARBURST99 \citep{lei99} with the measured SFHs as input.  The energies are integrated from the present (t $=$ 0) over the past 200 Myr.  The vertical grey dashed line is the kinematic age of each hole.  The horizontal dotted lines are $E_{Hole}$ assuming 100\% (red), 10\% (green), and 1\% (blue) efficiencies.  The intersection of the grey line with the energy profile indicates the putative feedback efficiency over the inferred kinematic age.  In several cases, this falls outside the expected range, and may even be greater than 100\%. This suggests that inferred kinematic age may not represent the true age of the hole.  If instead, we use the intersection 10\% efficiency line (green) and the energy profile as a guide, we can get a rough sense of how old the hole may be, if it formed from multiple generations of SNe.}  
\label{p92_energy1}
\end{center}
\end{figure}

\begin{figure}[t]
\begin{center}
\plotone{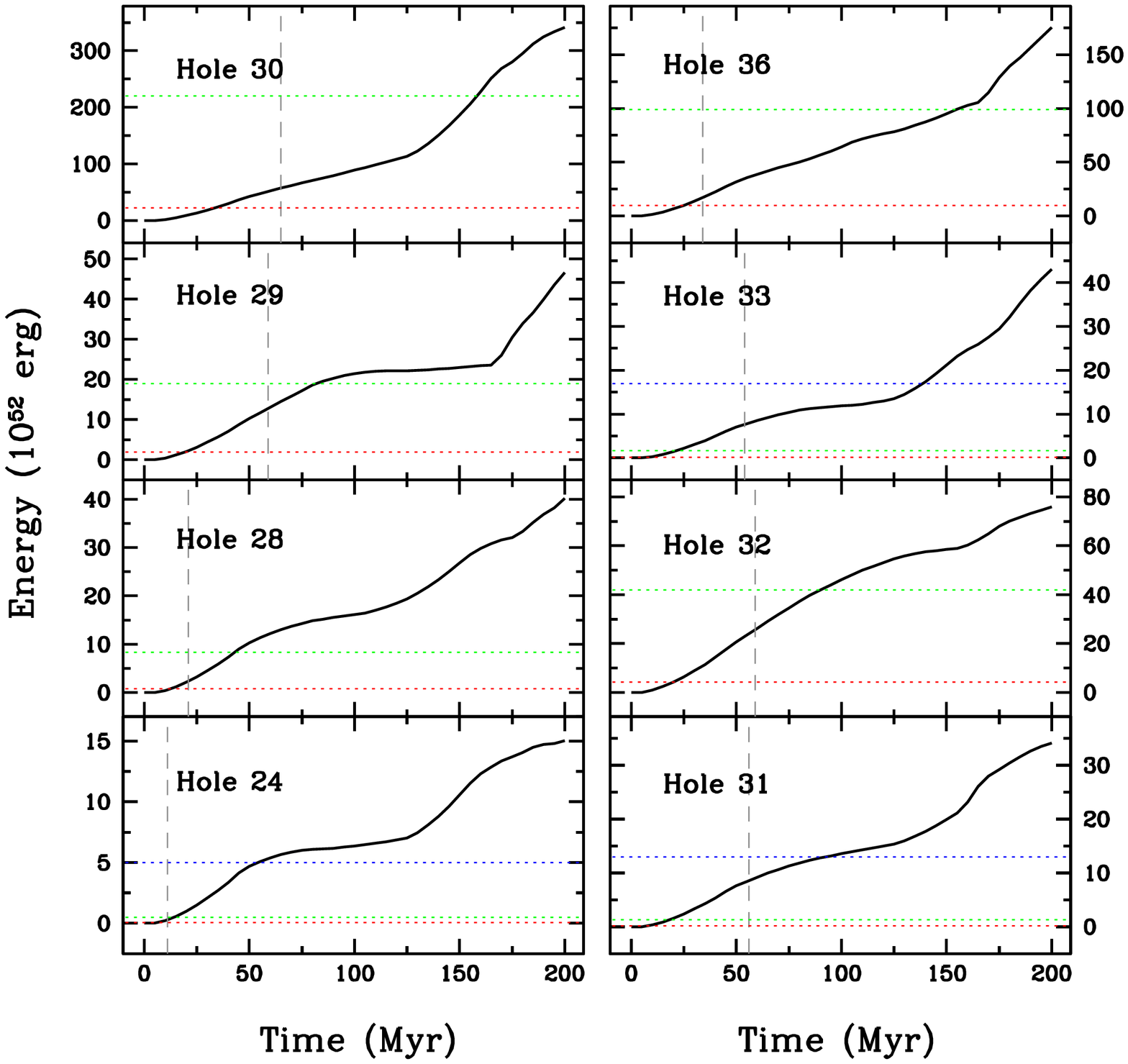}
\caption{The cumulative energies of the stars inside the \hi\ holes of \citet{puc92} computed using STARBURST99 \citep{lei99} with the measured SFHs as input.  The energies are integrated from the present (t $=$ 0) over the past 200 Myr.  The vertical grey dashed line is the kinematic age of each hole.  The horizontal dotted lines are $E_{Hole}$ assuming 100\% (red), 10\% (green), and 1\% (blue) efficiencies. The intersection of the grey line with the energy profile indicates the putative feedback efficiency over the inferred kinematic age.  In several cases, this falls outside the expected range, and may even be greater than 100\%. This suggests that inferred kinematic age may not represent the true age of the hole.  If instead, we use the intersection 10\% efficiency line (green) and the energy profile as a guide, we can get a rough sense of how old the hole may be, if it formed from multiple generations of SNe.}  
\label{p92_energy2}
\end{center}
\end{figure}

\begin{figure}[t]
\begin{center}
\plotone{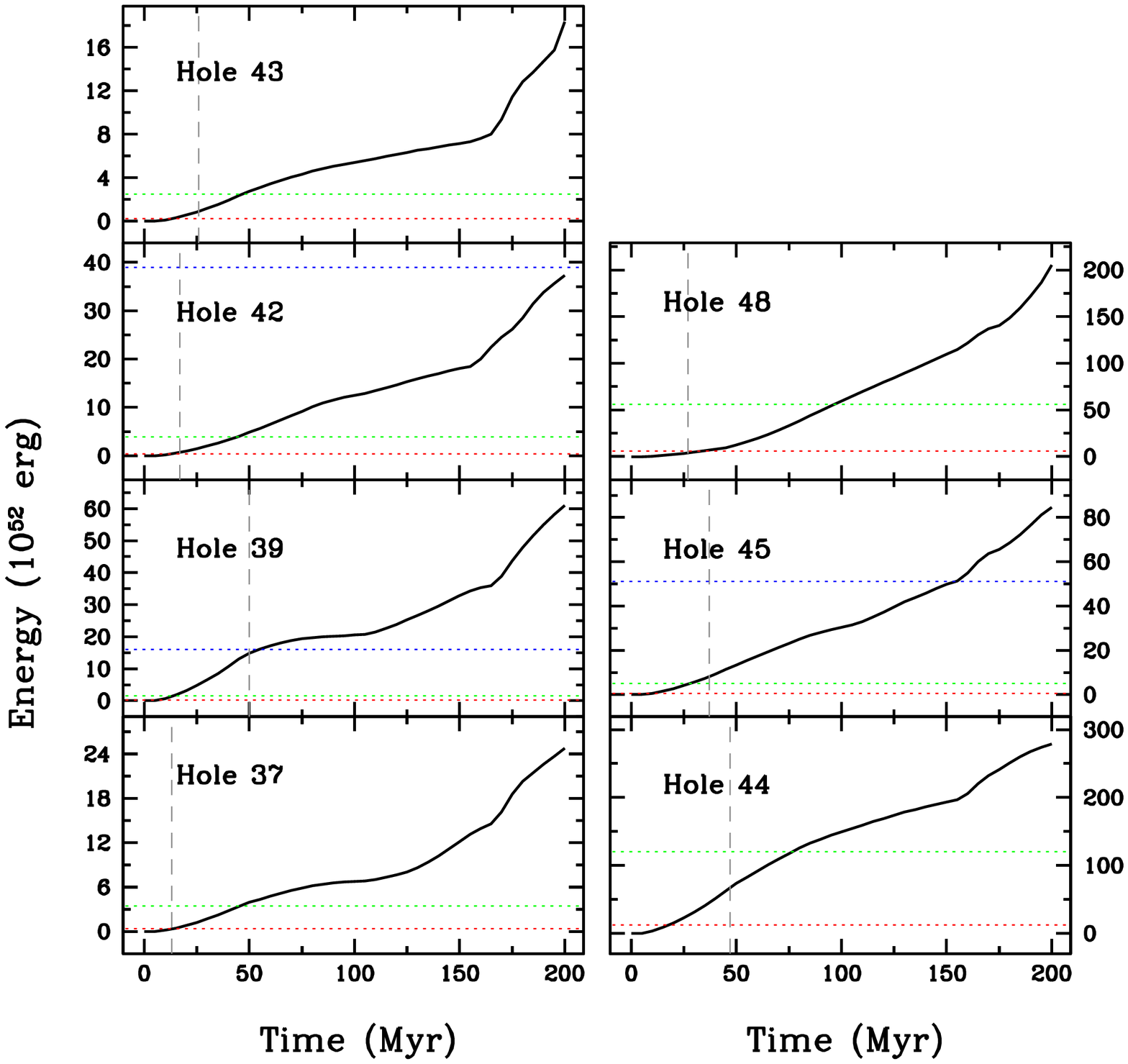}
\caption{The cumulative energies of the stars inside the \hi\ holes of \citet{puc92} computed using STARBURST99 \citep{lei99} with the measured SFHs as input.  The energies are integrated from the present (t $=$ 0) over the past 200 Myr.  The vertical grey dashed line is the kinematic age of each hole.  The horizontal dotted lines are $E_{Hole}$ assuming 100\% (red), 10\% (green), and 1\% (blue) efficiencies. The intersection of the grey line with the energy profile indicates the putative feedback efficiency over the inferred kinematic age.  In several cases, this falls outside the expected range, and may even be greater than 100\%. This suggests that inferred kinematic age may not represent the true age of the hole.  If instead, we use the intersection 10\% efficiency line (green) and the energy profile as a guide, we can get a rough sense of how old the hole may be, if it formed from multiple generations of SNe.}  
\label{p92_energy3}
\end{center}
\end{figure}

\begin{figure}[t]
\begin{center}
\plotone{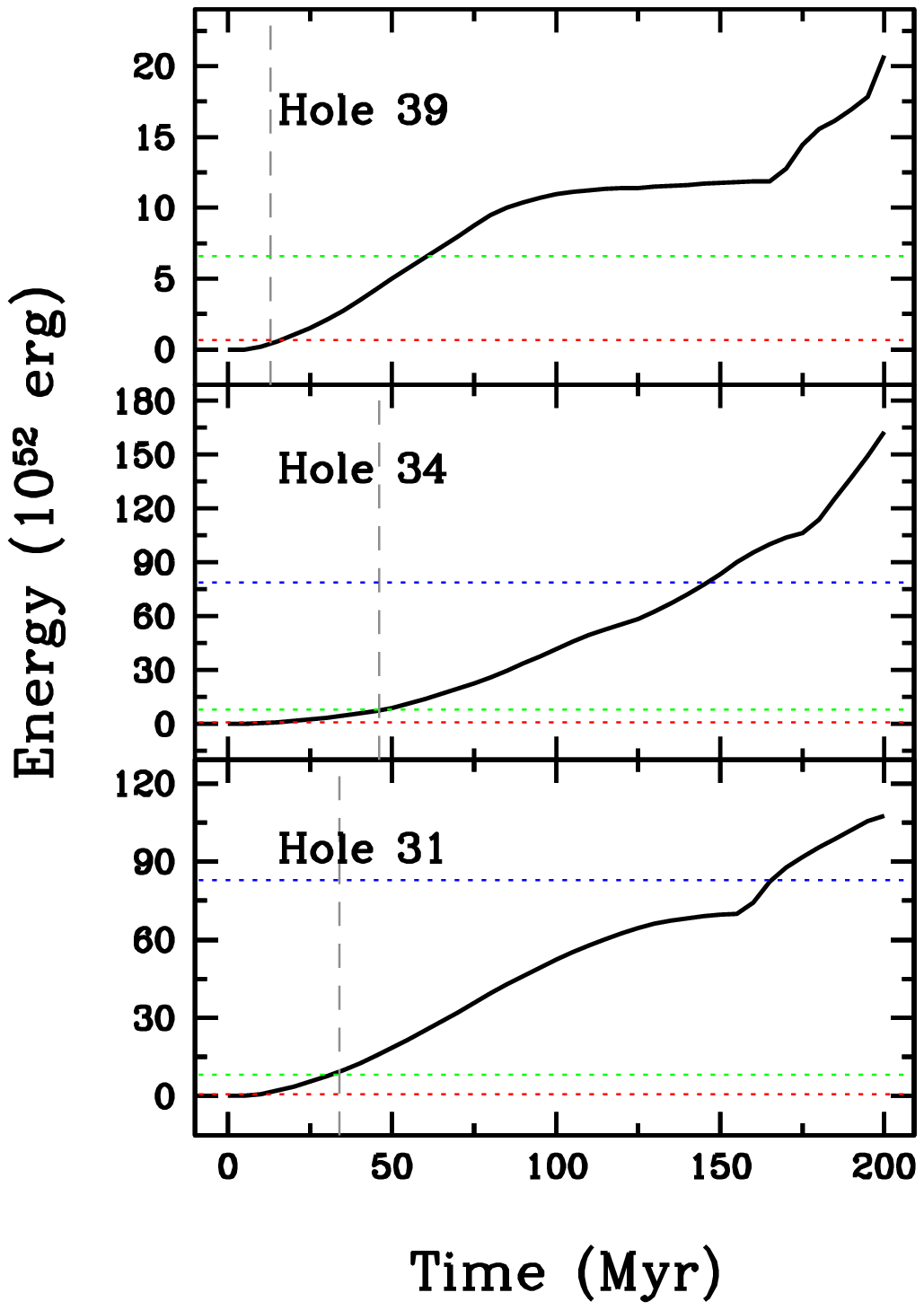}
\caption{The cumulative energies of the stars inside the \hi\ holes of \citet{bag09} computed using STARBURST99 \citep{lei99} with the measured SFHs as input.  The energies are integrated from the present (t $=$ 0) over the past 200 Myr.  The vertical grey dashed line is the kinematic age of each hole.  The horizontal dotted lines are $E_{Hole}$ assuming 100\% (red), 10\% (green), and 1\% (blue) efficiencies. The intersection of the grey line with the energy profile indicates the putative feedback efficiency over the inferred kinematic age.  In several cases, this falls outside the expected range, and may even be greater than 100\%. This suggests that inferred kinematic age may not represent the true age of the hole.  If instead, we use the intersection 10\% efficiency line (green) and the energy profile as a guide, we can get a rough sense of how old the hole may be, if it formed from multiple generations of SNe.}  
\label{THINGS_energy1}
\end{center}
\end{figure}

\begin{figure}[t]
\begin{center}
\plotone{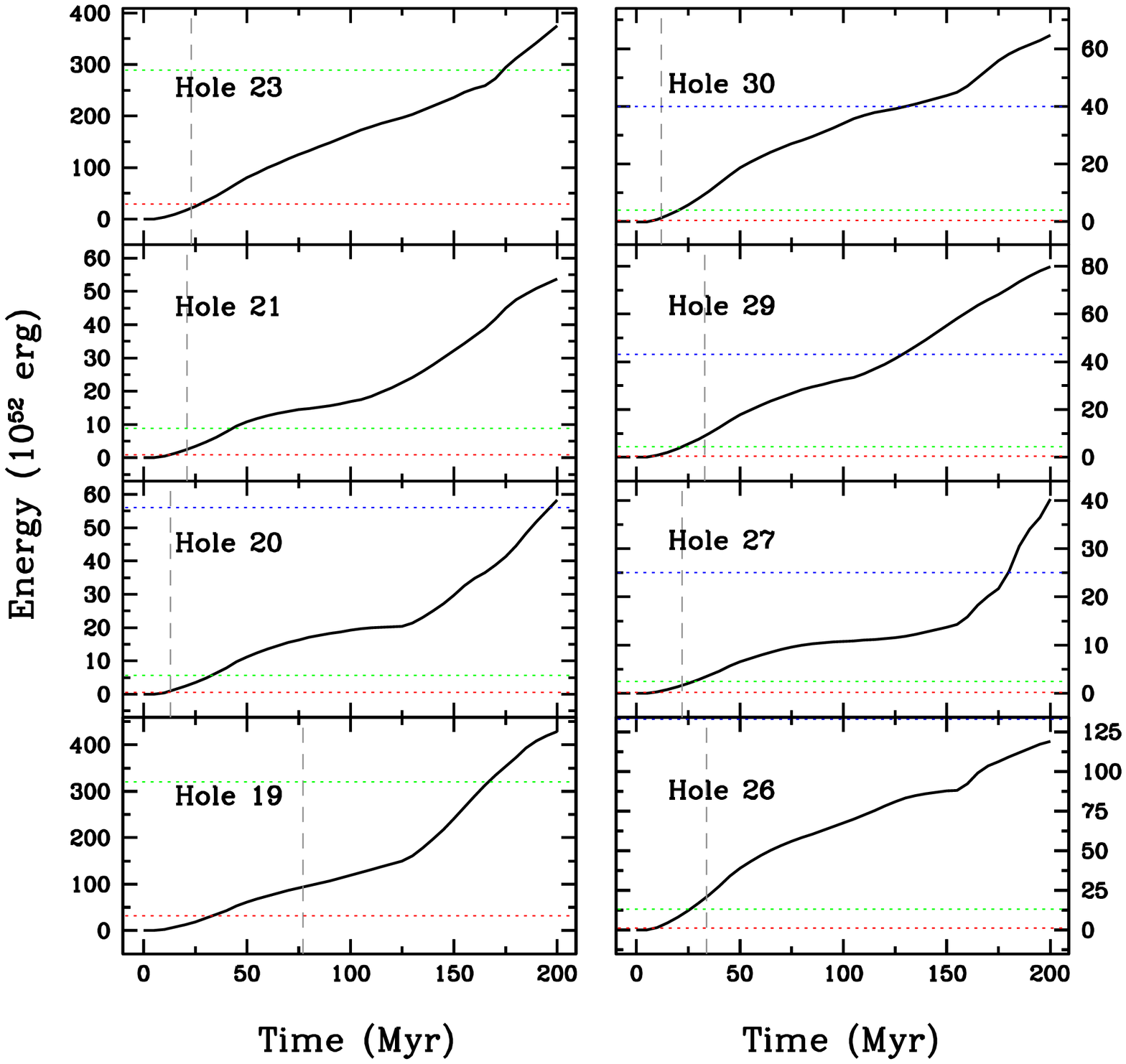}
\caption{The cumulative energies of the stars inside the \hi\ holes of \citet{bag09} computed using STARBURST99 \citep{lei99} with the measured SFHs as input.  The energies are integrated from the present (t $=$ 0) over the past 200 Myr.  The vertical grey dashed line is the kinematic age of each hole.  The horizontal dotted lines are $E_{Hole}$ assuming 100\% (red), 10\% (green), and 1\% (blue) efficiencies. The intersection of the grey line with the energy profile indicates the putative feedback efficiency over the inferred kinematic age.  In several cases, this falls outside the expected range, and may even be greater than 100\%. This suggests that inferred kinematic age may not represent the true age of the hole.  If instead, we use the intersection 10\% efficiency line (green) and the energy profile as a guide, we can get a rough sense of how old the hole may be, if it formed from multiple generations of SNe.}  
\label{THINGS_energy2}
\end{center}
\end{figure}

\begin{figure}[t]
\begin{center}
\plotone{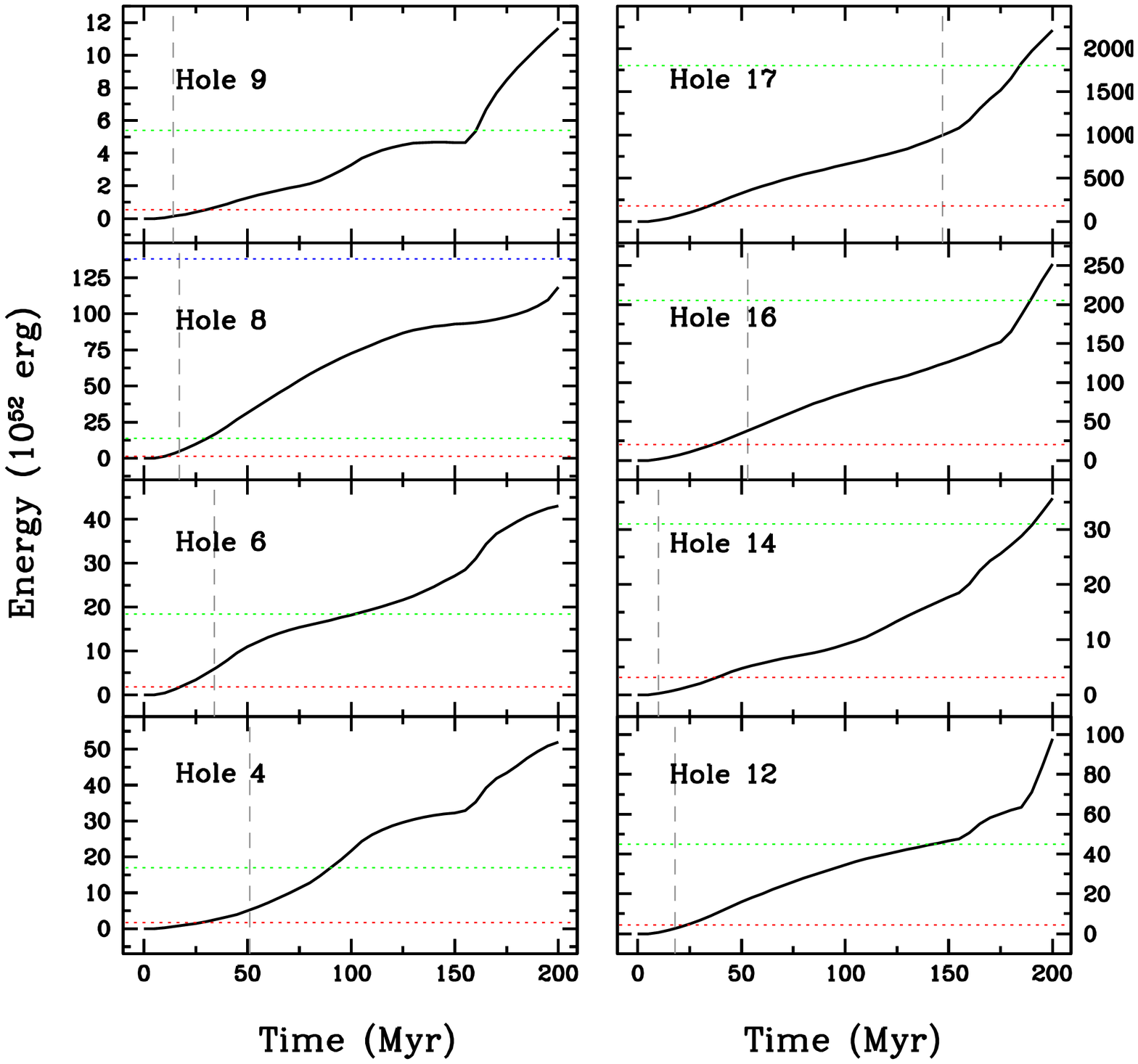}
\caption{The cumulative energies of the stars inside the \hi\ holes of \citet{bag09} computed using STARBURST99 \citep{lei99} with the measured SFHs as input.  The energies are integrated from the present (t $=$ 0) over the past 200 Myr.  The vertical grey dashed line is the kinematic age of each hole.  The horizontal dotted lines are $E_{Hole}$ assuming 100\% (red), 10\% (green), and 1\% (blue) efficiencies. The intersection of the grey line with the energy profile indicates the putative feedback efficiency over the inferred kinematic age.  In several cases, this falls outside the expected range, and may even be greater than 100\%. This suggests that inferred kinematic age may not represent the true age of the hole.  If instead, we use the intersection 10\% efficiency line (green) and the energy profile as a guide, we can get a rough sense of how old the hole may be, if it formed from multiple generations of SNe.}  
\label{THINGS_energy3}
\end{center}
\end{figure}

\begin{figure}[t]
\begin{center}
\plotone{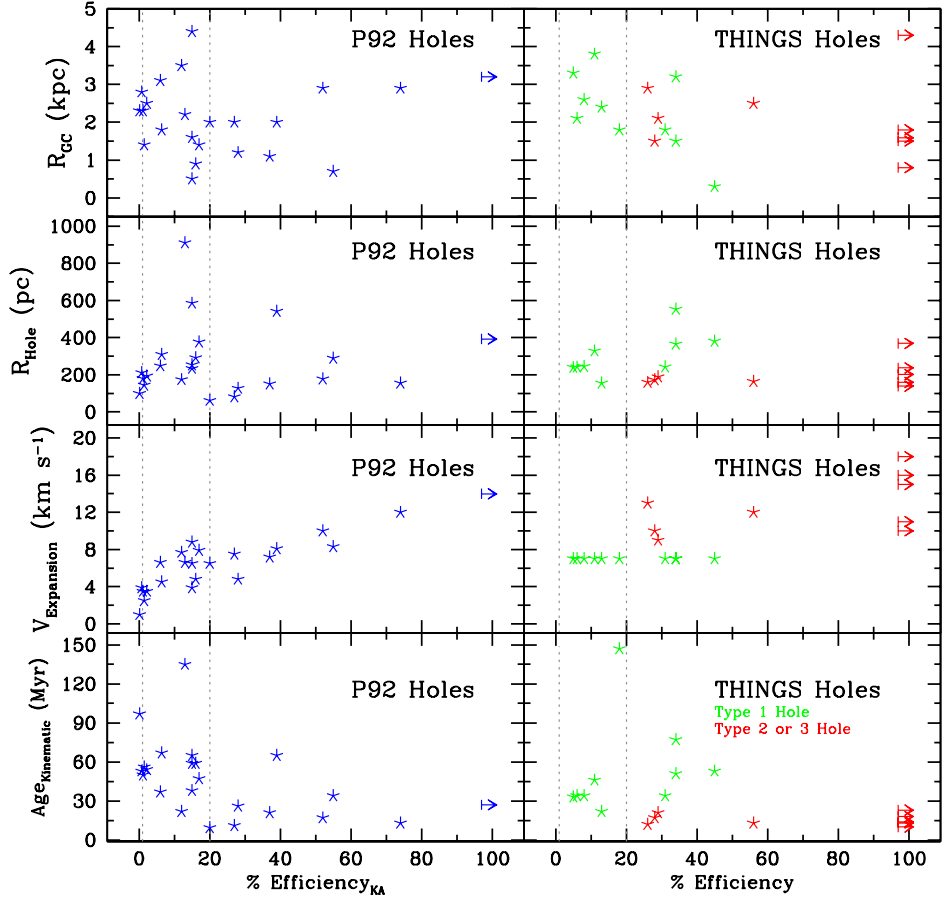}
\caption{Putative stellar feedback efficiencies ($E_{sfka}$/$E_{Hole}$; Tables \ref{tab3} and \ref{tab4}) plotted versus the galactrocentric radius (R$_{GC}$), hole radius (R$_{Hole}$), \hi\ expansion velocity (V$_{Expansion}$), and kinematic age of each hole for the \citet{puc92} and the THINGS \hi\ hole catalogs.  For the THINGS holes, green stars are type 1 holes (blown out), the red stars are type 2 or 3 (not blown out), and the arrow indicate holes that have efficiencies greater than 100\% and have been moved to fit the plot scale.  The vertical grey dashed lines indicate the range of efficiencies predicted by various models, 1\% $-$ 20\% \citep{the92, cole94, pad97, thor98}.  Note that the type 2 and 3 holes in the THINGS hole sample do not fall in to the expected range of putative feedback efficiency.  These holes also tend to be young, expanding quickly, and are generally smaller in size.  Arrows indicate lower limits on the efficiency for both hole catalogs.}  
\label{efficiency}
\end{center}
\end{figure}

For example, P92 hole 14 (Figure \ref{p92_energy1}) has an efficiency of $\sim$ 1\% where the inferred kinematic age intersects the cumulative energy profile.  In comparison, P92 hole 30 (Figure \ref{p92_energy2}) has a kinematic age efficiency of 39\%, outside the theoretical range predicted by models.  Further, ratios that require $\epsilon$ $>$ 100\% indicate that there has not been enough energy for SF during the inferred kinematic age of each hole. However,  Figures \ref{p92_energy1} $-$ \ref{THINGS_energy3} show that there has been sufficient energy available during the most recent 200 Myr even in these extreme cases, suggesting that perhaps the inferred kinematic ages do not truly represent the ages of the holes, or that the concept of an \hi\ hole age is intrinsically ambiguous for holes that contain multiple generations of stars.    

\begin{figure}[t]
\begin{center}
\plotone{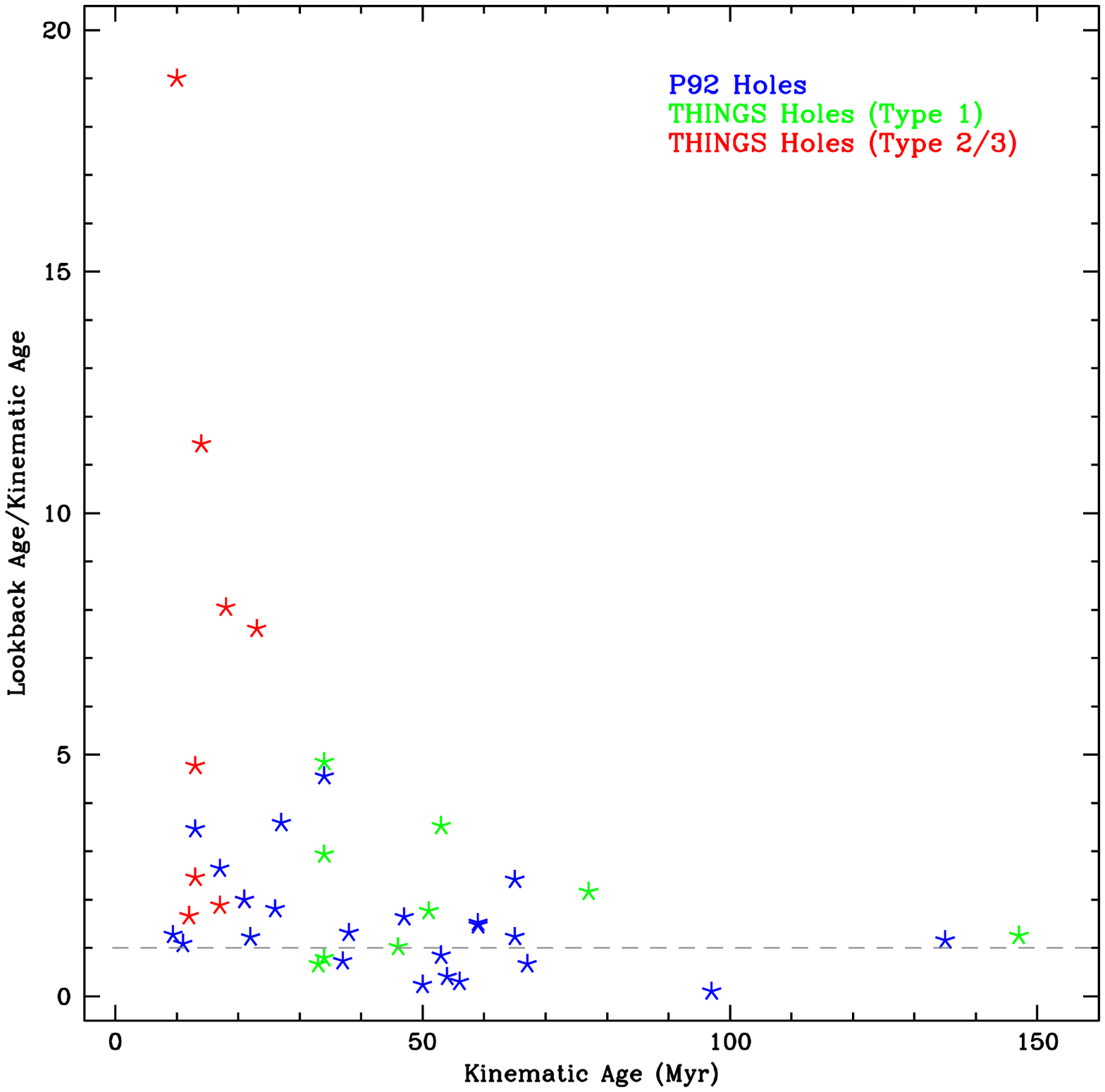}
\caption{The ratio of look back age to inferred kinematic age for the \hi\ holes from the \citet{puc92} and \citet{bag09} as a function of the inferred kinematic age.  The look back age is defined as the time at which the line of 10\% efficiency intersects the energy profiles in Figures \ref{p92_energy1} $-$ \ref{THINGS_energy3}\label{comp_age}.  The spread of values greater than unity (grey dashed line) implies that the inferred kinematic age often underestimates the age of \hi\ holes, suggesting that  the concept of an age for an \hi\ hole created by multiple generations of SF is intrinsically ambiguous.}  
\end{center}
\end{figure}

To explore this possibility, instead of using the inferred kinematic age as a guide, we assumed a nominal 10\% efficiency and recalculated the age of the hole to match the time scale needed to give sufficient energy input from SF.  For example, while P92 hole 48 (Figure \ref{p92_energy3}) has a kinematic age efficiency of 120\%, the energy profile shows that over the past 200 Myr, enough energy was generated from SF to have created the hole, if the hole is $\sim$ 100 Myr old (assuming a nominal 10\% efficiency).  Similarly, for the other  P92 and THINGS holes that have efficiencies outside the theoretical range, the same analysis shows that there is ample energy from SF to have created each of those holes in the past 200 Myr.  We defined a `look back age' as the time at which the line of 10\% efficiency intersects the energy profiles in Figures \ref{p92_energy1} $-$ \ref{THINGS_energy3}.  We then compared the ratio of the look back age to the inferred kinematic age, and found that the inferred kinematic ages are generally lower than the look back times (Figure \ref{comp_age}).  While perhaps not the most rigorous definition of the age of a hole, the look back age suggests that  the inferred kinematic age may not accurately represent the true age of \hi\ holes, even as an upper limit.  We explore the kinematic age discrepancy in greater detail in \S 5.

\subsection{Connecting the Spatially Resolved Star Formation History to the \hi\ Surface Density}


In addition to the measuring the SFR as a function of time, we can use spatial information from the resolved stellar populations to qualitatively compare the location of SF episodes with \hi\ column densities.  Following a method of analysis similar to \citet{dea97}, we used BHeBs from the entire HST/ACS CMD of \hoii\ \citep{wei08} and constructed movies of the spatially resolved recent SFH over the past $\sim$ 200 Myr. The initial masses of BHeBs range from $\sim$ 2 $-$ 20 \msun, giving them lifetimes between $\sim$ 5 $-$ 500 Myr.   More importantly, BHeBs have a one-to-one correspondence between luminosity and age.  BHeBs of a single age therefore occupy a \emph{unique} location on a CMD, making them superb tracers of recent SF.   In contrast, multiple generations of MS stars can occupy the same location on the CMD, making age dating of most individual MS stars from the CMD virtually impossible \citep[see][for a more in depth discussion of age dating stars from CMDs]{dea97, gza05, wei08}.  

By measuring its magnitude, we can assign a unique age to each BHeB, using the models of stellar evolution. We can then trace the location of SF as a function of age based on BHeB positions in the ACS image.  While it is true that many young star clusters dissolve into the field on time scales of $\sim$ 10 Myr \citep[e.g.,][]{lad03}, the stars can maintain spatial coherence for much longer.  Empirically, CMD-based studies of the SMC and LMC find that star forming structures (such as associations) can remain coherent for up to a couple hundred Myr \citep{gie08, bas09}. 

We were able to isolate the BHeBs on the CMD by visual inspection because \hoii\ has a large number of BHeBs, and low foreground and internal reddening values. We constructed a smooth analytic function mapping the magnitudes of the BHeB track to unique ages, using the stellar evolution models of \citet{mar08}.  Applying this function to the observed BHeBs, we assigned each star an individual age, with an uncertainty of $\lesssim$ 20\%.  We created density maps of the BHeBs by placing the stars into 5 Myr bins and converting their $x$ and $y$ coordinates to RA and Dec.  We then smoothed the resulting images with a Gaussian kernel with a FWHM $\sim$ 15\arcsec.  

To convert the BHeB density maps (stars~pixel$^{-1}$) to SFR/area (\msun~yr$^{-1}$~kpc$^{-2}$), we first divided each frame (i.e., 5 Myr binned density maps) by the sum of the pixel values to determine the fractional contribution of each pixel to the total stellar density in each frame.  That is, if a normalized pixel had a value of 0 it did not contribute to the SF in the frame, while a pixel with a value of 0.01 contributed to 1\% of the SF in that frame. The density maps were then connected to the SFH of the field \citep{wei08}. 

\begin{figure}[t]
\begin{center}
\plotone{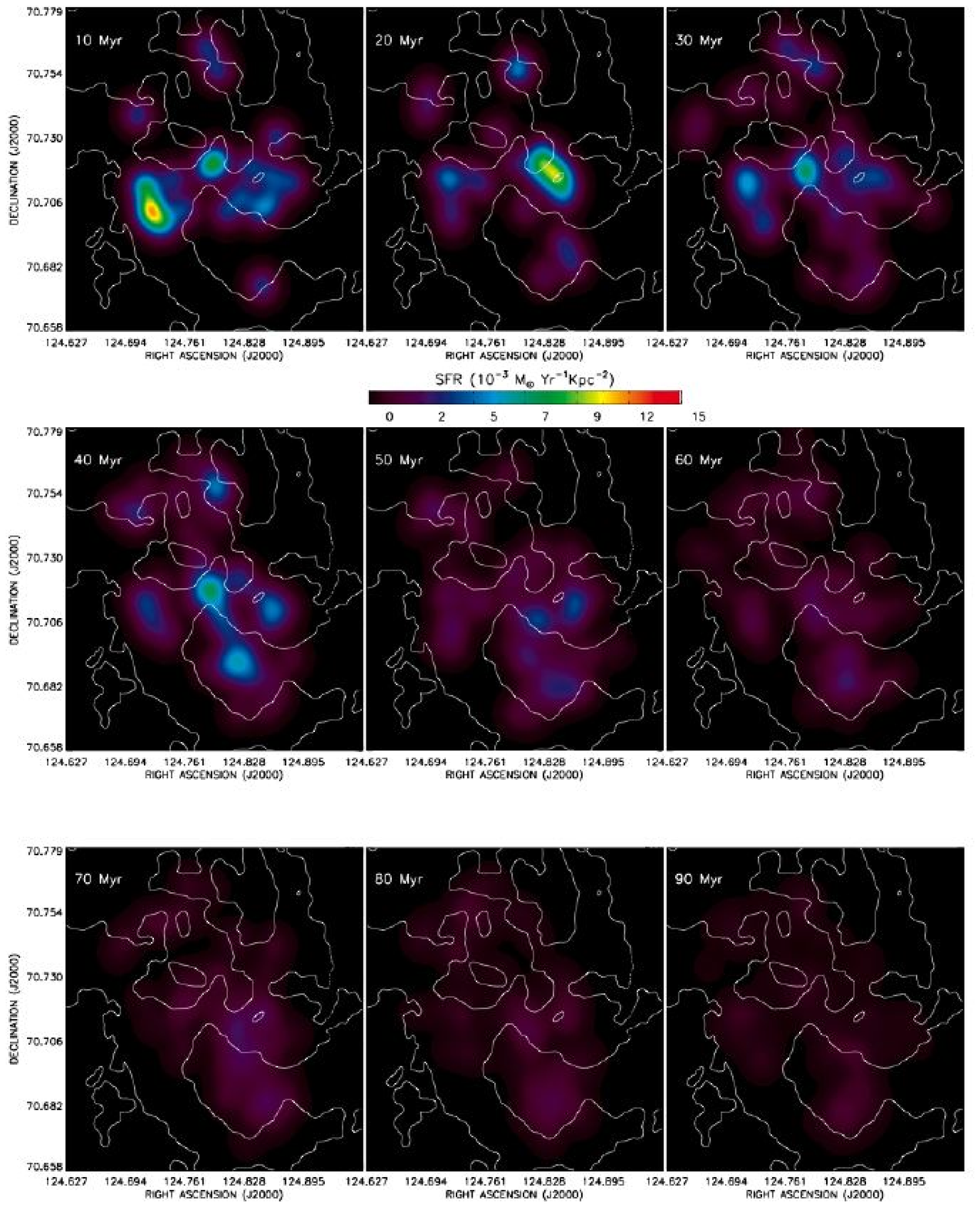}
\caption{The spatially resolved recent SFH of \hoii\ over the last 90 Myr with 10 Myr time resolution as seen on the sky in units of SFR/area.  The \hi\ 10$^{-21}$ cm$^{-2}$ contour is overlaid in white.  Note that majority of the galaxy is dominated by constant low level SF, while only a few regions have experienced elevated episodes of SF.  As expected, the most recent frame shows the peaks of SF coincident with the peaks of the \hi\ distribution.}  
\label{10m_map}
\end{center}
\end{figure}

\begin{figure}[t]
\begin{center}
\plotone{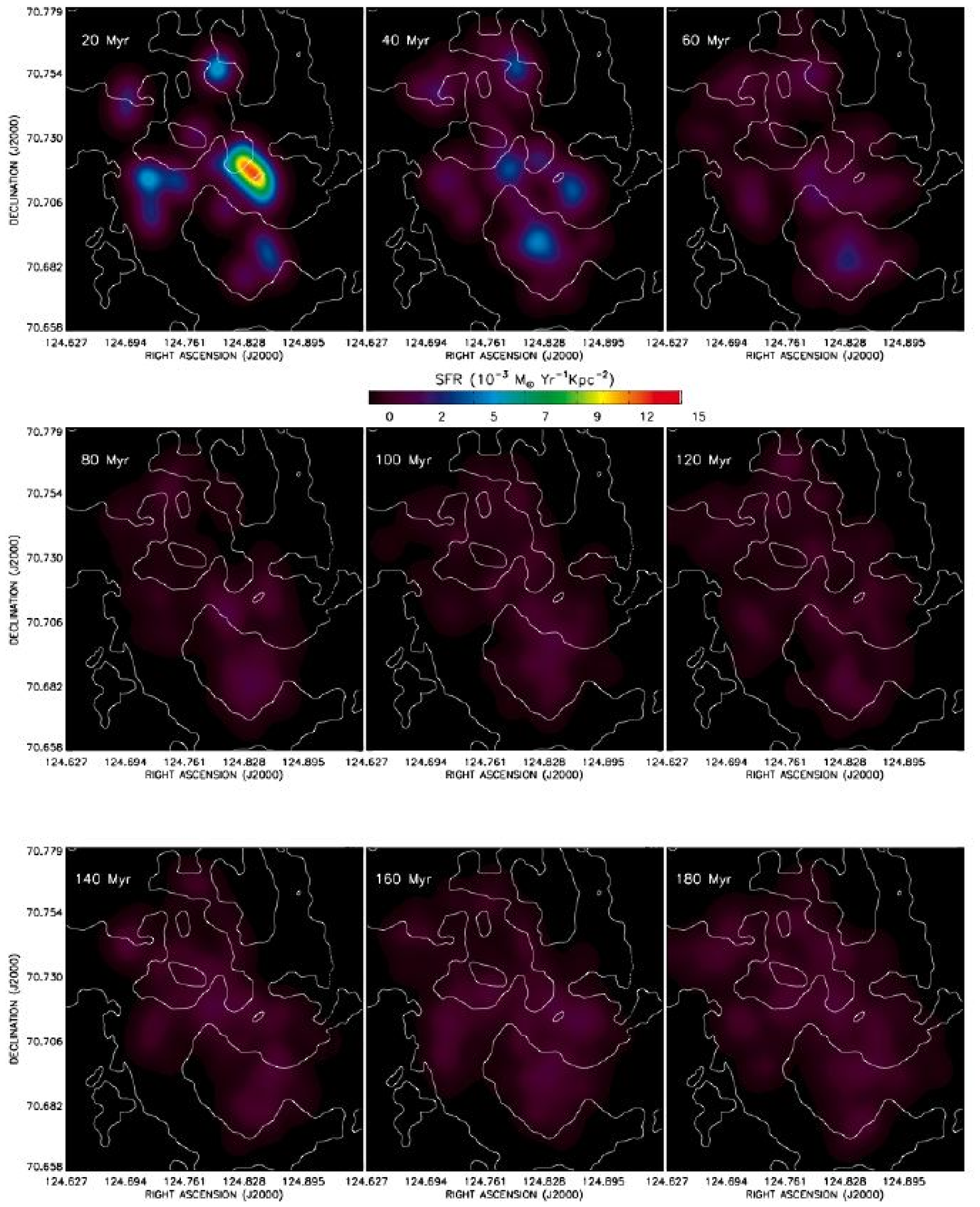}
\caption{The spatially resolved recent SFH of \hoii\ over the last 180 Myr with 20 Myr time resolution as seen on the sky in units of SFR/area.  The \hi\ 10$^{-21}$ cm$^{-2}$ contour is overlaid in white.  Note that majority of the galaxy is dominated by constant low level SF, while only a few regions have experienced elevated episodes of SF.  As expected, the most recent frame shows the peaks of SF coincident with the peaks of the \hi\ distribution.}  
\label{20m_map}
\end{center}
\end{figure}

First, we re-binned the SFH with different time resolutions: 10 Myr for a look back time of $\sim$ 100 Myr and 20 Myr for a look back time of $\sim$ 200 Myr, in order to match the decreasing time resolution at older ages. We then interpolated each to a 5 Myr time sampling to match the temporal resolution of the stellar density maps.  We then multiplied each stellar density map by the corresponding SFR and divided by the physical area of the ACS field of view, thus converting the images from units of stars~pixel$^{-1}$ to SFR/area.  We then performed bilinear interpolation over time and space to ensure smooth and continuous movies.   The result is two movies of the spatially resolved SFHs of \hoii\ with units of SFR/area, with look back times of $\sim$ 100 and $\sim$ 200 Myr, and a spatial resolution of $\sim$ 15\arcsec\ ($\sim$ 250 pc).  Figures \ref{10m_map} and \ref{20m_map} show select still frames of the two movies with \hi\ contours overlaid.

The general trends presented by the spatially resolved SFHs are that low level SF has dominated the galaxy in the most recent $\sim$ 200 Myr, with only a substantial rise in the SF activity in the last $\sim$ 40 Myr.  The smoothness of the SFH implies that the \hi\ holes were not produced by a series of intense episodes of SF, instead the holes were created due to low level SF with some variation above a constant value, which has steadily input energy into the ISM.  As a comparison, we integrated the spatial resolved SFH of the region corresponding to THINGS hole 17 and found general agreement between the two.  For example, both SFHs show a peak at $\sim$ 40 Myr, and from the integration of the spatially resolved SFH we found the SFR at this time to be 3.0\e{-2} \msun\ yr$^{-1}$, whereas the SFH from Figure \ref{THINGS_cmd} shows a SFR of 4.0\e{-2} \msun\ yr$^{-1}$.  While this difference of 30\% may seem large, in fact, the systematic and statistical uncertainties from the two methods are $\sim$ 20\% from the temporal SFH and 50\% from the spatial SFH, which implies these are consistent values.

These movies allow us to investigate the correlation between \hi\ density and SF on different time scales.  Within the last $\sim$ 40 Myr, the SF is well correlated with high \hi\ surface density, as has been seen in \halpha, and UV imaging (see \S 5.2 for a discussion on \halpha\ and UV imaging).  On longer timescales,  we note that the SF patterns do not adhere to the locations of high \hi\ column densities.  In particular, there is a relatively high level of SF activity in the large \hi\ void, corresponding to P92 hole 21 and THINGS hole 17, $\sim$ 40 Myr ago, which has been preceded by low level activity for at least the past $\sim$ 200 Myr. 


\section{Discussion}

\subsection{The Paradoxical Results of \citet{rho99}}

The study of \hoii\ by \citet{rho99} is perhaps one of the most rigorous observational tests of the stellar feedback model of \hi\ hole creation.  Despite the appeal of the stellar feedback theory of \hi\ hole formation, \citet{rho99} found no compelling evidence supporting this hypothesis, and their conclusions seriously challenged the standard model of \hi\ hole formation.  The non-detection of stars in the majority of \hi\ holes contradicted expectations for stellar feedback as a universal mechanism of hole creation, although \citet{rho99} did not rule out stellar feedback as the source of small \hi\ holes in the central regions of \hoii.  


\citet{rho99} searched for single age stellar clusters, but likely did not find them because the single age cluster model can overestimate the surface brightness of the remnant stars by a significant amount, for a variety of reasons we explore in the following sections.  Further, using only integrated light, it can be difficult to distinguish between a single age stellar cluster and a mixed age stellar population, which can make it difficult to age date stars that are detected. As a comparative test, we computed the integrated magnitudes of each of the \citet{rho99} apertures from the HST/ACS CMDs. Our integrated photometry is in agreement with that of \citet{rho99}. A histogram of the resulting integrated magnitudes reveals that the six regions where \citet{rho99} did detect stars are among the brightest in the sample (Figure \ref{rho_hist}).  The rest of the regions were classified as non-detections, despite containing significant numbers of young stars.

\begin{figure}[t]
\begin{center}
\plotone{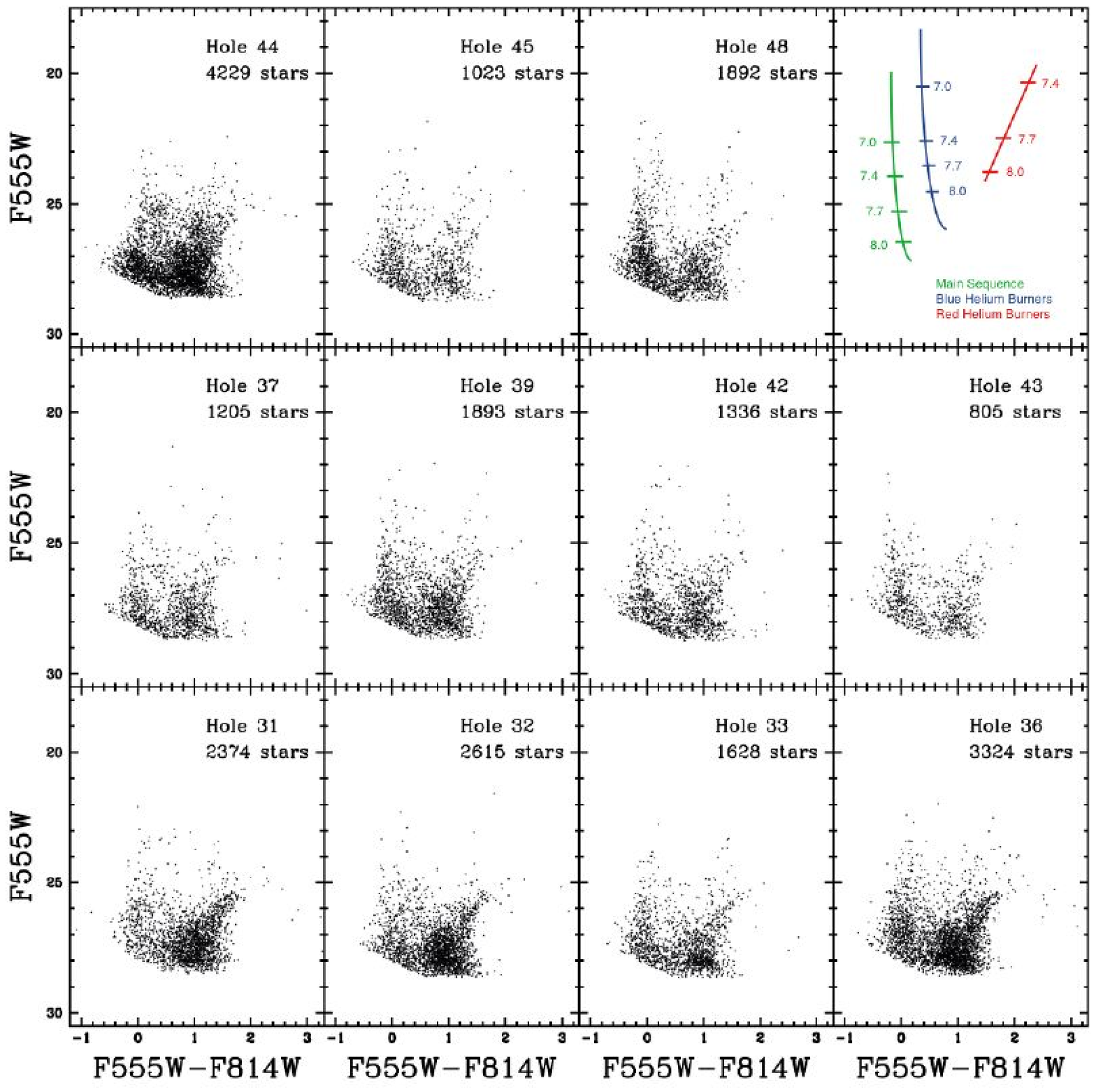}
\caption{HST/ACS CMDs of the photometric apertures used by \citet{rho99}, corrected for foreground reddening, A$_{V}$ $=$ 0.11 \citep{sch98}. The schematic in the upper right hand corner shows the ages of each type of star, MS (green), BHeB (green), and RHeB (red), with the logarithm of the ages shown as a function of magnitude. Although sparse in some fields, note the presence of young MS and BHeB stars in all the CMDs.  The fact the the BHeBs span a range in magnitudes indicate that multiple episodes of recent SF must have taken place, as different age BHeBs do not overlap on the CMD. In contrast, BHeBs from a single age cluster would have an overdensity at only one magnitude. }  
\label{rho_cmd1}
\end{center}
\end{figure}

\begin{figure}[t]
\begin{center}
\plotone{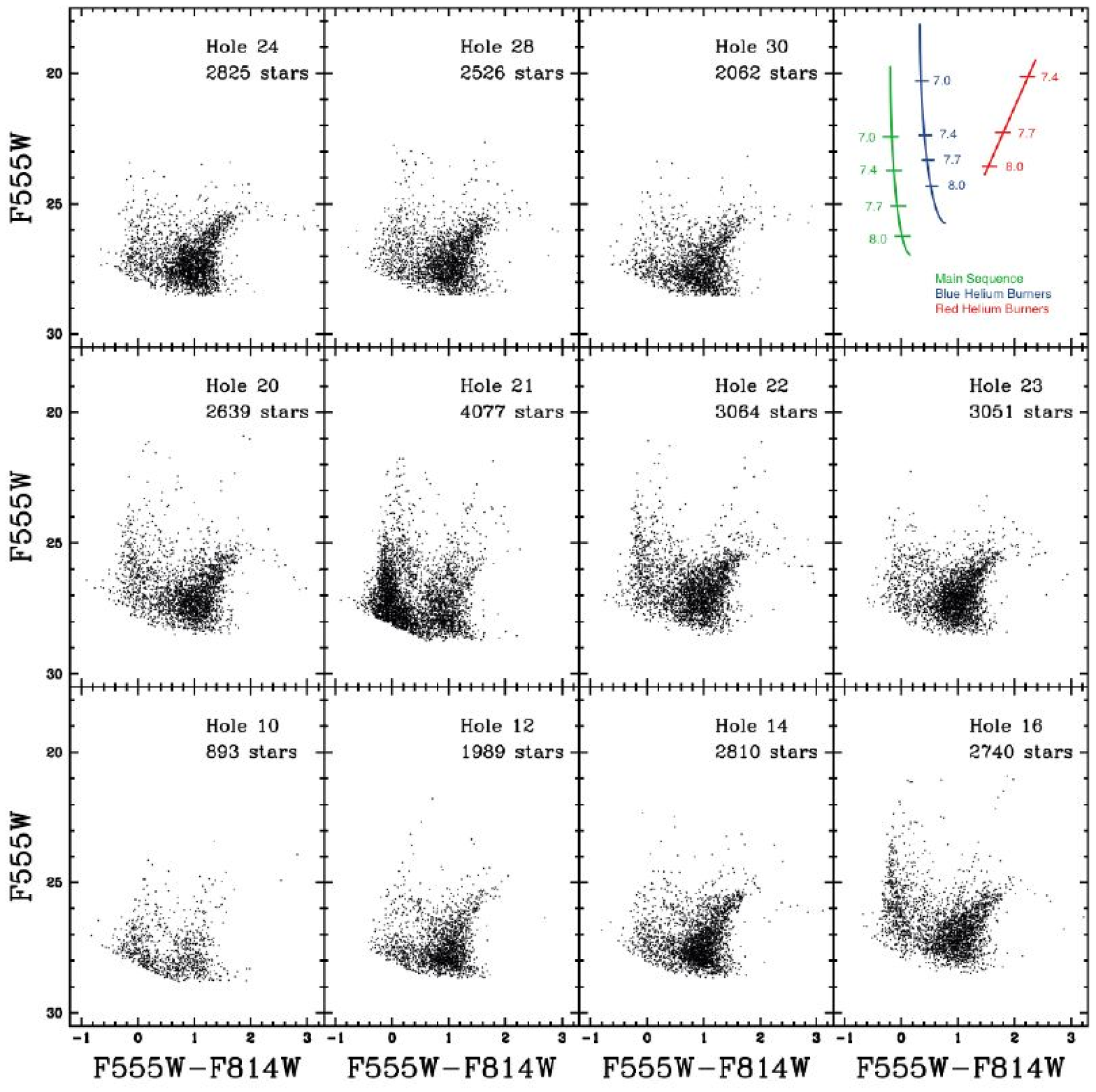}
\caption{HST/ACS CMDs of the photometric apertures used by \citet{rho99}, corrected for foreground reddening, A$_{V}$ $=$ 0.11 \citep{sch98}. The schematic in the upper right hand corner shows the ages of each type of star, MS (green), BHeB (green), and RHeB (red), with the logarithm of the ages shown as a function of magnitude. Although sparse in some fields, note the presence of young MS and BHeB stars in all the CMDs.  The fact the the BHeBs span a range in magnitudes indicate that multiple episodes of recent SF must have taken place, as different age BHeBs do not overlap on the CMD. In contrast, BHeBs from a single age cluster would have an overdensity at only one magnitude.}  
\label{rho_cmd2}
\end{center}
\end{figure}

\begin{figure}[t]
\begin{center}
\plotone{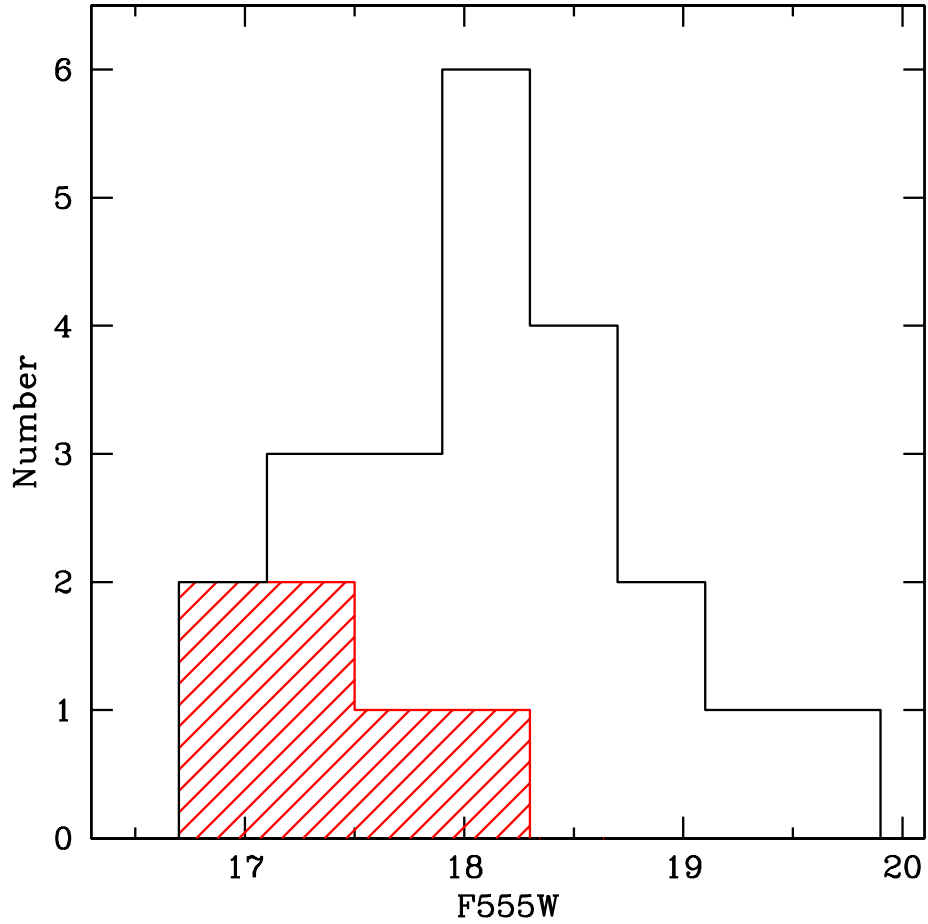}
\caption{A histogram of the integrated ACS m$_{F555W}$ magnitudes from the HST/ACS CMDs corresponding to the photometric apertures used by \citet{rho99}.  The red shaded region represents the apertures that \citet{rho99} classify as containing putative star clusters (Table \ref{rho_tab}).  Note that they regions containing putative stellar clusters are among the brightest in the sample. } 
\label{rho_hist}
\end{center}
\end{figure}

Using the HST/ACS photometry of \hoii\ we constructed CMDs of the apertures (panel (d) of Figure \ref{map1} and Table \ref{rho_tab}) defined by \citet{rho99}.  Figures \ref{rho_cmd1} and \ref{rho_cmd2} unequivocally show well-populated CMDs of mixed age stellar populations, including young MS stars and HeBs, which are likely the remnant stars \citet{rho99} attempted to detect using integrated light.   Our interpretation is that single age stellar populations do not produce a majority of the \hi\ holes in \hoii, confirming the conclusions of \citet{rho99}.


\subsection{A New Model}

Instead of assuming that a single age cluster was responsible for the creation of \hi\ holes, we explored the possibility that a mixed age stellar population producing SNe tens or hundreds of Myr apart was responsible for the creation of \hi\ holes.  Specifically, we asked how a remnant mixed age stellar population would differ from a single age cluster in terms of the integrated light properties, location of stars inside \hi\ holes, and energetics associated with this different mode of SF.

To illustrate this, we generated synthetic CMDs, with the same parameters (e.g., IMF, binary fraction, distance, etc.) used to measure the SFHs in \S 2.2, and computed the mean expected integrated magnitudes of different stellar populations.  To ensure that we were not in a regime where stochastic effects are important, we fixed the integrated mass of the simulations to be 10$^{6}$ \msun, $\sim$ 1\%\ of the stellar mass of \hoii\ and at least a factor of 100 greater than the stellar mass within any given hole.   We limited the simulations to the most recent 100 Myr, the approximate age regime of the majority of \hi\ holes in \hoii.  The simulations were run 100 times, which produced an average statistical error of $\sim$ 0.01 mags.  Effects of completeness and photometric error are similarly $\sim$ 0.01 mags.  Neither source of error has a large impact on the results at magnitudes this bright.  

The first model assumed that all the stellar mass was formed in a $\sim$ 5 Myr period 50 Myr ago, with no subsequent SF.  For this, `burst' model, we found the mean integrated magnitude and color to be m$_{F555W}$ $=$ 15.88 and m$_{F555W}$ $-$ m$_{F814W}$ $=$ 0.48.  For comparison, we distributed the same integrated stellar mass with a constant SFR over the period of 50 $-$ 100 Myr and found m$_{F555W}$ $=$ 16.07 and m$_{F555W}$ $-$ m$_{F814W}$ $=$ 0.51. 

We further considered a two component SFH model.  We distributed 10$^{6}$ \msun\ in stellar mass with a constant SFR from 50 $-$ 90 Myr ago, no SF from 35 $-$ 50 Myr ago, constant SFR from 25 $-$ 35 Myr ago, and quiescence for the most recent 25 Myr.  This multiple episode model produced stars with integrated properties of m$_{F555W}$ $=$ 15.86 and m$_{F555W}$ $-$ m$_{F814W}$ $=$ 0.54. Comparing the single age, constant, and multiple episode models of the same integrated stellar mass, we found they produced results that were indistinguishable in terms of integrated light.  

To illustrate this point empirically, we compared the stellar populations of P92 holes 42 and 43, which both have integrated magnitudes of m$_{F555W}$ $=$ 18.75, but differ in other properties.  For example, hole 42 contains 1383 stars and has an inferred kinematic age of 17 Myr, but hole 43 has 549 stars and an inferred kinematic age of 26 Myr.  The SFHs of stars within the two holes differ significantly in the last 10 Myr.  SF in hole 42 had been more prominent $>$ 10 Myr ago, however within the last 10 Myr, hole 43 has a higher SFR, likely producing young bright stars that dominate the integrated light of the stellar population.  These findings suggest that integrated light can only serve as an indicator of recent SF, but cannot give an accurate account of the SFH.  

\begin{figure}[t]
\begin{center}
\plotone{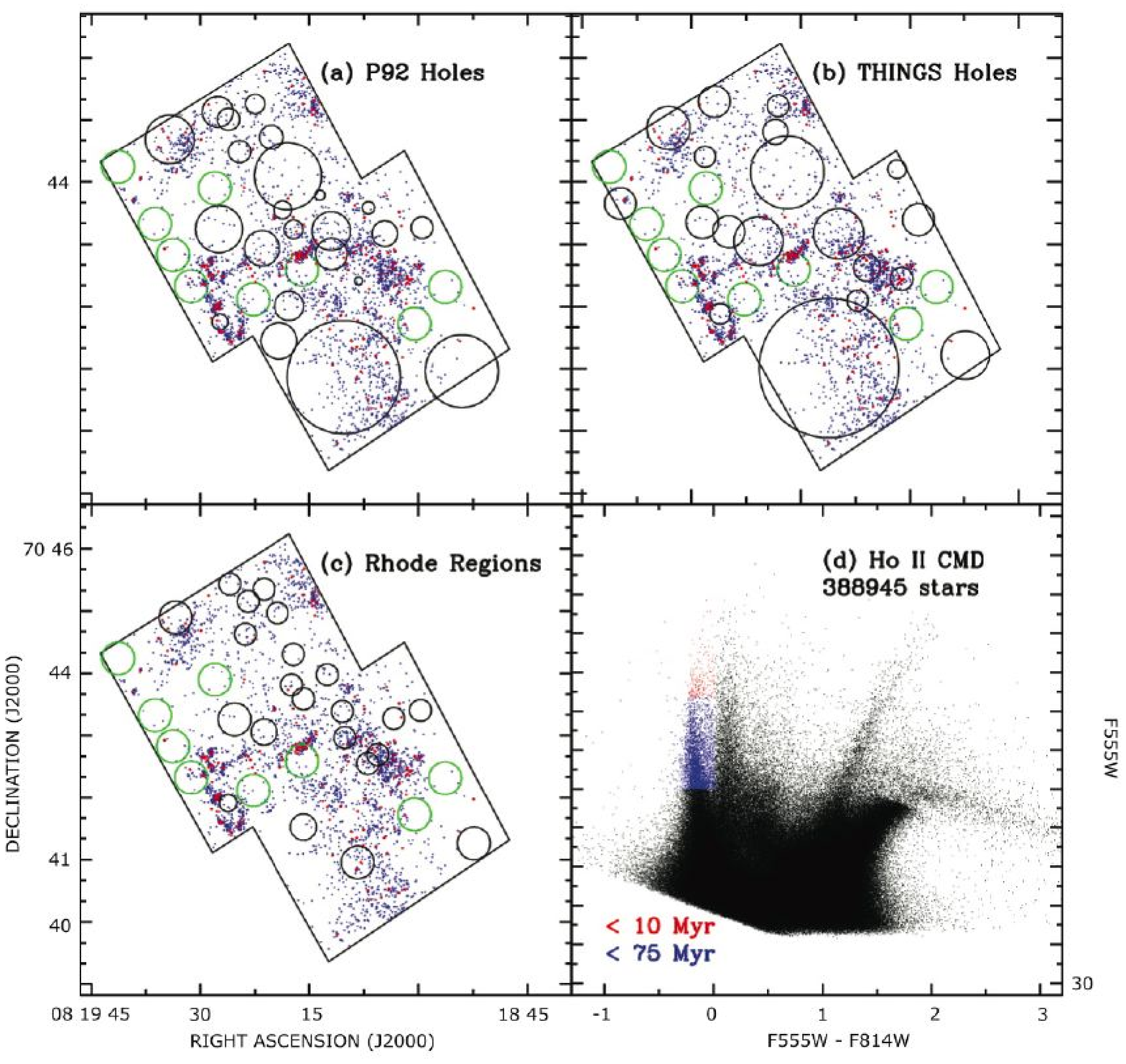}
\caption{The spatial distribution of young MS stars (red $<$ 10 Myr, blue $<$ 75 Myr) plotted over (a) \hi\ holes from the \citet{puc92} catalog; (b) \hi\ holes from the THINGS catalog \citep{bag09}; (c) photometric apertures of \citet{rho99};   Panel (d) is the CMD of the entire galaxy of \hoii\ with the young MS stars color-coded to match the spatial distributions with MS stars $<$ 10 Myr in age in red and $<$ 75 Myr in blue.  Note the tendency of the red points to appear highly clustered, while the blue points are more widely distributed.  Further, note that most of the \hi\ holes do not contain highly clustered regions of young stars, rather these are found near the edges of the holes or near \hi\ peaks.}  
\label{rho_map}
\end{center}
\end{figure}

The dissolution of stellar clusters into the field further complicates integrated light observations and analysis of clusters.  \citet{lad03} note that the most common stellar clusters contain only dozens of stars and disperse into the field within $\sim$ 10 Myr, which is the so-called `infant mortality' scenario of cluster dissolution.  Detailed studies of structure of stellar groups in the SMC and LMC show that only $\sim$ 3 $-$ 7\% of young stars survive in clusters past the age of $\sim$ 10 Myr \citep{gie08, bas09}.  Applying this idea to \hoii, it suggests that the remnant stellar populations would not be highly clustered, but instead would be distributed across the \hi\ hole, resulting in fainter integrated magnitudes than expected for clustered stars.  To look for this effect we selected MS stars from the HST/ACS photometry by applying magnitude and color restrictions that isolated young MS stars ($<$ 10 Myr (red) and $<$ 75 Myr (blue)) and plotted their positions relative to the \hi\ holes catalogs, \citet{rho99} apertures, and control fields (Figure \ref{rho_map}).  While there is some clustering of young stars, particularly those $<$ 10 Myr in age, the clusters of stars are typically on the edges of the \hi\ holes where triggered secondary SF due to the expansion of the hole is likely taking place \citep[e.g.,][]{ttb88}.  The majority of the MS stars $<$ 75 Myr in age are not clustered, which makes the integrated light of these stellar populations appear fainter than had they been bound or appeared as unresolved  stellar clusters.  

One of the most interesting cases presented by \citet{rho99} is that of P92 hole 10.  Based on its \hi\ properties at least several dozen SNe would be necessary to have created hole 10.  The remnant stars  from a burst should be well above the detection criteria used by \citet{rho99} in a single cluster model, but none were detected.  \citet{rho99} assumed a single remnant cluster would be in the center of the hole and used an aperture smaller than the hole to search for the stars.  Comparing the CMD between the actual hole and the aperture shows why the stars likely responsible for creating hole 10 were not detected by \citet{rho99}.  The CMD of the entire hole has $\sim$ 4200 stars and an integrated magnitude of m$_{F555W}$ $=$ 17.68, while the CMD of the \citet{rho99} hole 10 aperture contains 893 stars and has an integrated magnitude of m$_{F555W}$ $=$ 19.75.  Furthermore, the spatial distribution of MS stars $<$ 75 Myr in age (Figure \ref{rho_map}) show that the photometric aperture of hole 10 has no stars, while the hole itself contains a number of young stars.  It appears that the bright stars either did not form centrally in hole 10 or they have rapidly diffused out of the center, if they did form there.  Although this hole was considered among the strongest evidence against the stellar feedback theory of \hi\ hole creation, it now only seems to confirm that the single age cluster model of hole formation is not the appropriate mechanism for \hi\ hole creation.

If \hi\ holes form via multiple generations of SNe, understanding the underlying physics becomes more complicated.  We have to go beyond the simple single blast model of \hi\ hole creation and consider the  complex interplay between the local ISM and stellar feedback, which seems to be the dominant mechanism behind structure formation in the ISM \citep[e.g.,][]{har08}. \citet{hei90} considers the case of `clusters of clusters' of SNe present within an \hi\ hole, noting that sequential explosions would contribute to the mechanical energy expanding the hole as well as significantly alter the gas density inside the hole. \citet{oey97} also conclude that multiple OB associations would increase the mechanical power and time scale of energy input into expanding the shell. \citet{rec06} present a model demonstrating how two separate SF events inside an \hi\ hole would delay the onset of the hole refilling with cold gas with time scales from $\sim$ 100 $-$ 600 Myr depending on local conditions, indicating the multiple episodes of SF spread out over time can help extend the life of an \hi\ hole.  Projection effects could cause generations of stars along a line of site appear to be spatially co-incident within a given \hi\ hole. However, this is likely a small effect because the disk-like nature of \hoii\ implies that the stellar scale height is much less than that of the \hi, suggesting a reasonable assumption is that the young stars are primarily located in the mid-plane of the disk. 

Quantitative modeling of the impact of the complex scenarios presented by multiple generations of SF inside \hi\ holes has yet to be carried out in detail. Presently, the single blast approximation and kinematic age estimations of \hi\ holes are dependent upon quantities (e.g., expansion velocity, volume density) that likely have varied over the history of a given hole. Quantifying the precise impact of these parameters on  \hi\ hole formation and energetics is beyond the scope of this paper.  For example, while why control fields and \hi\ holes contain similar stellar populations, but vastly different \hi\ column densities, is an open question.  However, \hoii\ may not be the best galaxy to answer this question, because we are not sensitive to \hi\ hole smaller than $\sim$ 100 pc.  Thus, if it is the case that larger holes are mergers of multiple smaller holes, we simply may not be sensitive to the small holes forming inside the control fields. Analysis similar to that presented in this paper using galaxies with smaller resolvable \hi\ holes (e.g., LMC or SMC) along with computer modeling that accounts for the effects of an anisotropic ISM and multiple generations of SNe could lend insight into the fundamental physical processes that are responsible for shaping the ISM and driving \hi\ hole formation. 

\subsection{\hi\ Holes in the Outer Regions of \hoii}


Of particular interest to \citet{rho99} are the presence of \hi\ holes outside the the optical body of the galaxy.  While they do not rule out the possibility that the inner holes could have formed from SF, the typical \hi\ densities in the outer regions of \hoii\ are relatively low, which is not very conducive to SF \citep{ken89}.  However, the properties of holes in outer regions of galaxies do not differ from those in inner regions \citep{bri86, puc92, hat05, bag09}, suggesting that the same underlying mechanism, namely stellar feedback, is responsible for \hi\ holes in both environments.  To this end, we used a Monte Carlo method to compute expected stellar populations in the outer holes of \hoii\ to see if their creation via stellar feedback is possible.  

From $E_{Hole}$ we can calculate the SFR over the kinematic age, $SFR_{KA}$, for a typical outer \hi\ hole and then use it to construct simulated CMDs to determine properties of the expected stellar populations.  We find that the 19 holes from \citet{bag09} located outside our ACS coverage have a mean diameter $\sim$ 800 pc, an expansion velocity $\sim$ 8 \kms, a kinematic age $\sim$ 50 Myr, an \hi\ volume density of 0.1 cm$^{-3}$, and $E_{Hole}$ $\sim$ 100 $\times 10^{50}$ erg.  Assuming 10$^{51}$ erg per SN \citep{mcc87} and that 10\% of this energy goes into moving the ISM \citep{the92, cole94, pad97, thor98}, $\sim$ 100 SN are needed to create an \hi\ hole with these properties, equivalent to a SN rate over the kinematic age of $\sim$ 10$^{-6}$ SN~yr$^{-1}$.  Assuming that the SNe come from stars with M $>$ 8 \msun, we find SFR$_{KA}$ $\sim$ 2\e{-5} \msun\ yr$^{-1}$ for all stars, using our standard IMF (\S 3.2). This SFR implies an integrated stellar mass of $\sim$ 10$^{3}$ \msun\ over the kinematic age.

With this SFR as input, we constructed simulated CMDs, via a Monte Carlo process using either a constant or a burst model.  The input for the constant SFR model had a SFR $=$ 2\e{-5} \msun\ yr$^{-1}$ from 50 Myr to 100 Myr.  This focuses on the brightest stars and most recent SF likely to have created the \hi\ hole.  The burst model assumed a burst of SF 50 Myr ago, producing 1000 \msun.  We ran each simulation 100 times to account for the statistical fluctuations.  For the constant model, we find that the average CMD of the remnant stars has $\sim$ 7 stars and an integrated magnitude of m$_{F555W}$ $=$ 24.00, assuming a limiting photometric depth equal to the 50\% completeness limit \citep{wei08}.  For the burst model, the average CMD of the remnant stars had $\sim$ 8 stars and a integrated magnitude of m$_{F555W}$ $=$ 23.75. Both cases results in integrated magnitudes fainter than the predictions by \citet{rho99} as well as the hole and control fields presented in this paper.  The remnant stellar populations in the outer holes are likely only detectable with deep photometry of resolved stars.



Other common methods of searching for stellar populations inside \hi\ holes include the use of \halpha, 24$\mu$m, and UV imaging \citep[e.g.,][]{puc92,ste00,mul03}.  UV traces a stellar population with initial masses $\gtrsim$ 3 \msun\ and time scales of $\sim$ 100 Myr, while \halpha\ (and 24$\mu$m in more dusty environments) traces ionizing photons from MS stars with initial masses of $\gtrsim$ 20 \msun\ and ages $\lesssim$ 10 Myr \citep[see][for a full review of different SFR indicators and their properties]{ken98}.  A number of studies aimed at testing the stellar feedback theory of hole creation via this type of imaging have have done so with a mixed degree of success \citep{puc92, kim98, rho99, ste00, mul03, can05, book08}.  Given the low SFRs we have derived (10$^{-4}$ $-$ 10$^{-5}$ \msun\ yr$^{-1}$) within many of the \hi\ holes, it is not entirely surprising that these imaging techniques have yielded such mixed results.  Simulations have shown that for SFRs below $\sim$ 10$^{-4}$ \msun\ yr$^{-1}$ stochastic processes become important and \halpha\ is no longer a reliable tracer of SF \citep{tre07, thi07, meu09}.  Although a similar study in the UV has yet to be conducted, given the small number of stars expected in any given CMD, it is likely that the UV has a similar cut-off SFR.  Indeed, \citet{ste00} used FUV imaging of \hoii\ to show that there is some agreement between \hi\ hole locations and bright UV regions, but did not find a good one to one correlation.  

\begin{figure}[t]
\begin{center}
\plotone{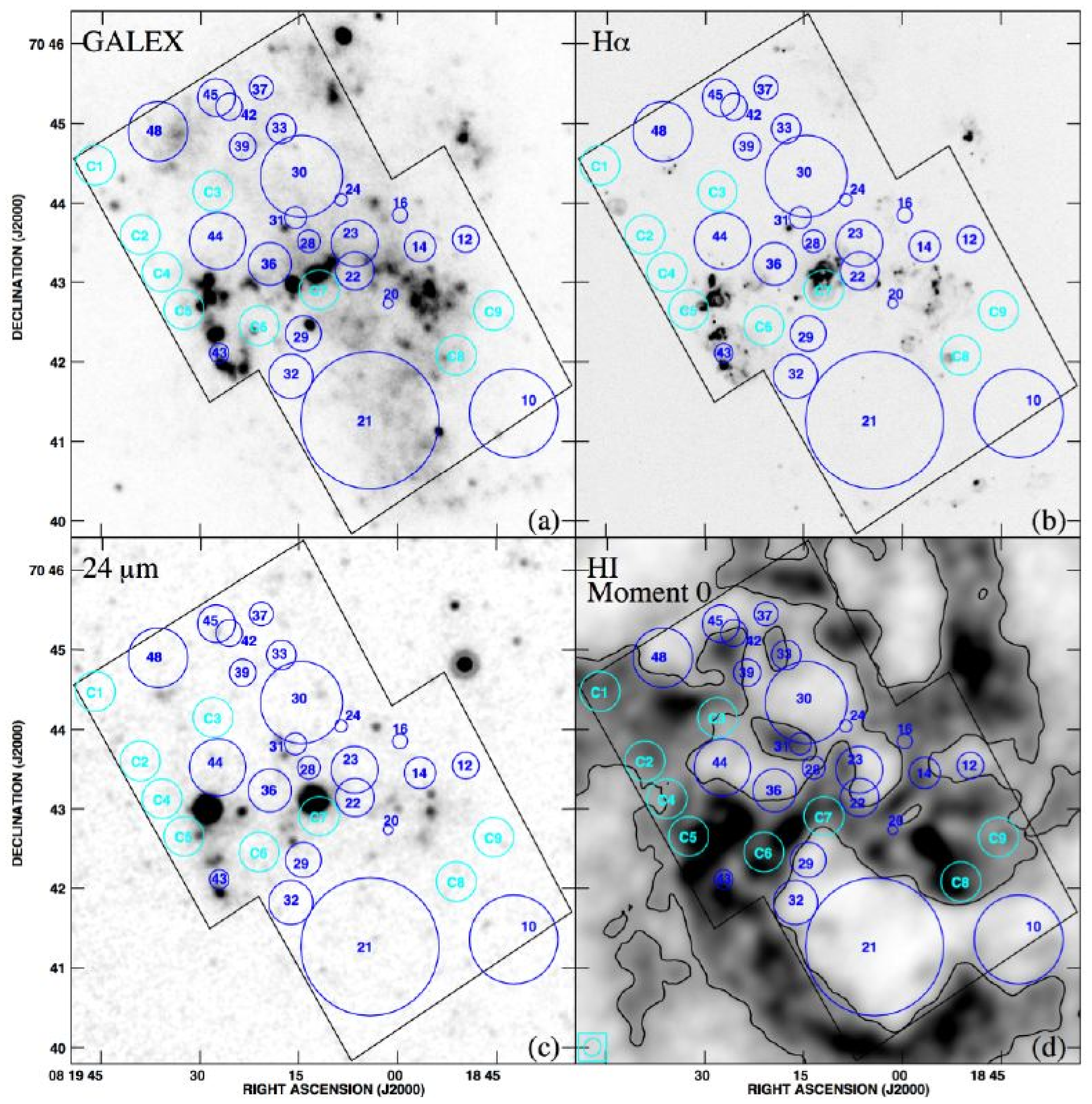}
\caption{\hi\ holes of \citet{puc92} (blue) and control fields (cyan) overlaid on \hoii\ images of (a) GALEX NUV \citep{gil07}; (b) LVL \halpha\ \citep{lee08}; (c) SINGS 24$\mu$m \citet{ken03}; (d) THINGS \hi\ \citep{wal08}. Despite that fact that all the \hi\ holes contain hundreds or thousands of stars, from these images alone, it would seem that the \hi\ holes are void of stars.  At low SFRs ($<$ 10$^{-4}$) \msun\ yr$^{-1}$, it seems that these tracers of recent SF no longer reliably correlate with the underlying stellar populations (see \S 5.3 for further discussion).}
\label{P92_multi}
\end{center}
\end{figure}

\begin{figure}[t]
\begin{center}
\plotone{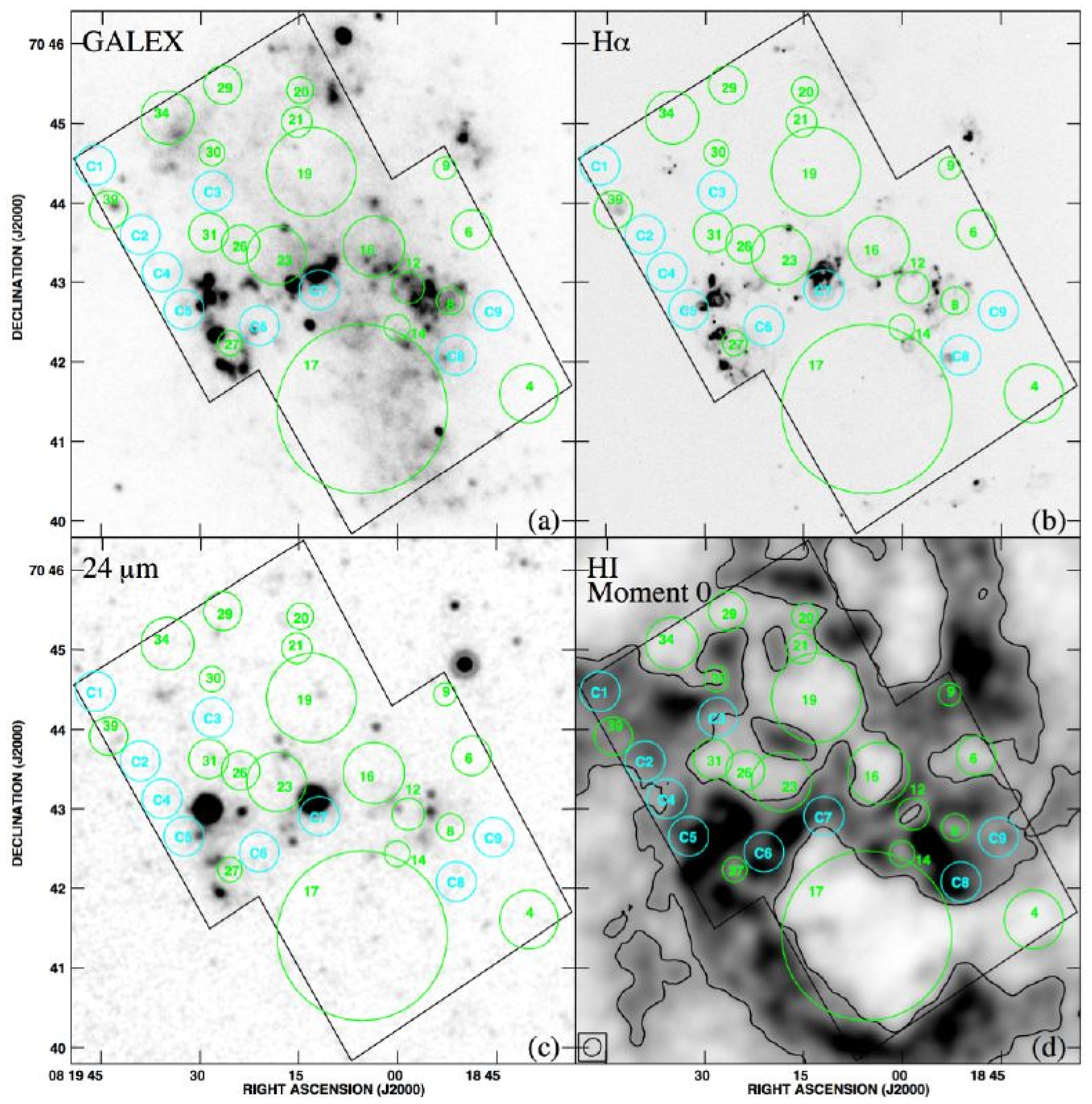}
\caption{\hi\ holes of \citet{bag09} (green) and control fields (cyan) overlaid on \hoii\ images of (a) GALEX NUV \citep{gil07}; (b) LVL \halpha\ \citep{lee08}; (c) SINGS 24$\mu$m \citet{ken03}; (d) THINGS \hi\ \citep{wal08}. Despite that fact that all the \hi\ holes contain hundreds or thousands of stars, from these images alone, it would seem that the \hi\ holes are void of stars.At low SFRs ($<$ 10$^{-4}$) \msun\ yr$^{-1}$, it seems that these tracers of recent SF no longer reliably correlate with the underlying stellar populations (see \S 5.3 for further discussion).}
\label{THINGS_multi}
\end{center}
\end{figure}

\begin{figure}[t]
\begin{center}
\plotone{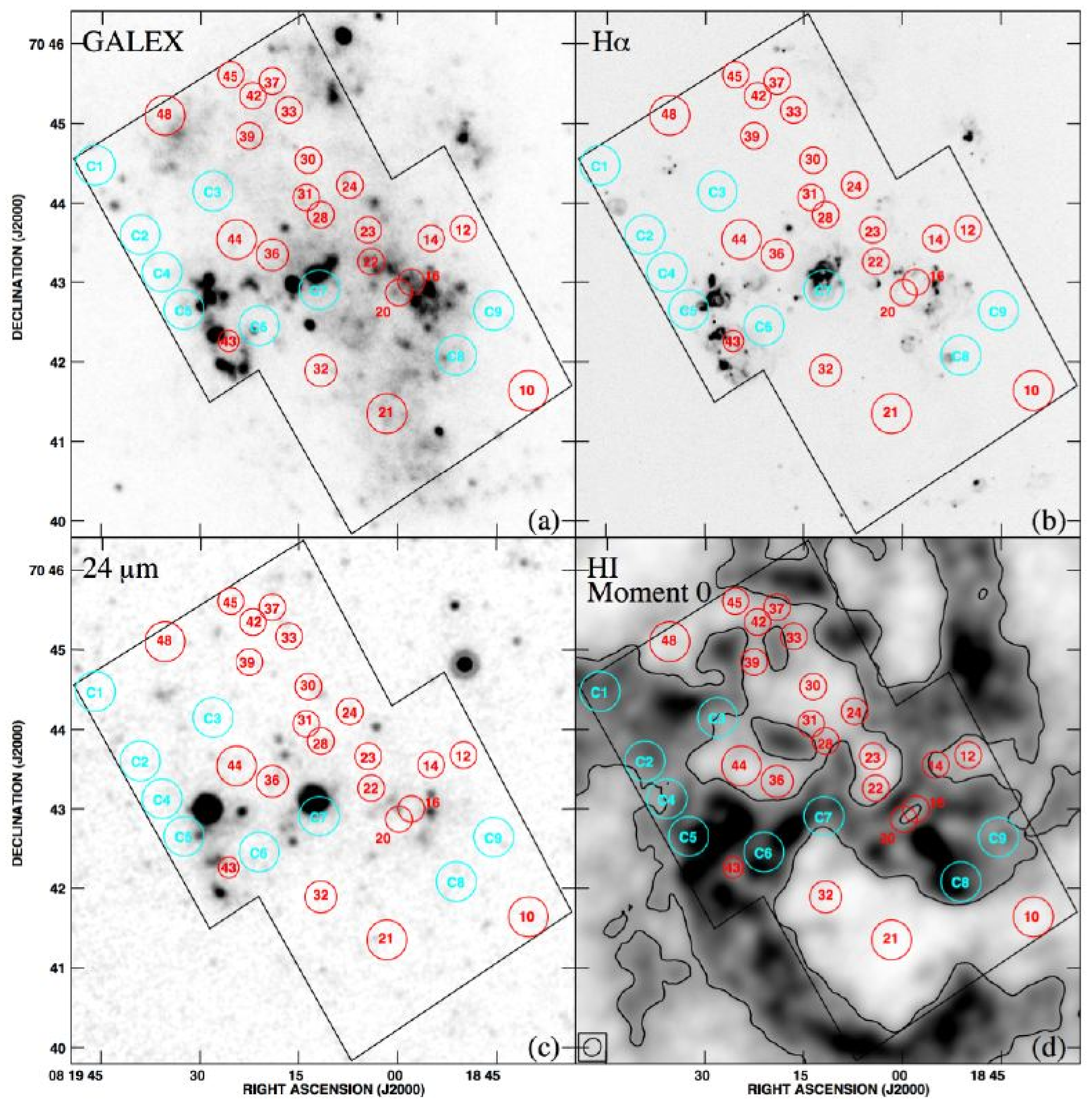}
\caption{Photometric apertures of \citet{rho99} (red) and control fields (cyan) overlaid on \hoii\ images of (a) GALEX NUV \citep{gil07}; (b) LVL \halpha\ \citep{lee08}; (c) SINGS 24$\mu$m \citet{ken03}; (d) THINGS \hi\ \citep{wal08}. Despite that fact that all the \hi\ holes contain hundreds or thousands of stars, from these images alone, it would seem that the \hi\ holes are void of stars.  At low SFRs ($<$ 10$^{-4}$) \msun\ yr$^{-1}$, it seems that these tracers of recent SF no longer reliably correlate with the underlying stellar populations (see \S 5.3 for further discussion).}
\label{rho_multi}
\end{center}
\end{figure}

We can test for UV-bright stellar populations within the outer holes using data from the Local Volume Legacy sample \citep{lee08, dal09}, which contains FUV and NUV imaging originating from the GALEX Nearby Galaxies Survey \citep{gil07}, and the \halpha\ and 24$\mu$m images from the Spitzer Infrared Nearby Galaxies Survey \citep[SINGS][]{ken03}.  Overlaying the \hi\ hole and \citet{rho99} aperture locations over these images, we compared the location of bright regions with known \hi\ holes (Figures \ref{P92_multi} $-$ \ref{rho_multi}).  Because the correlation between low SFRs and \halpha, 24$\mu$m, and UV is not well understood, we are only able to make qualitative conclusions.  From all the images, it is evident that there is not a one to one correlation between any of these fluxes and the stellar populations inside the \hi\ holes as shown in the HST/ACS based CMDs.  This simple test demonstrates that deep photometry is necessary to understand the formation and histories of \hi\ holes.

\subsection{Comparison with the SMC and LMC}

Perhaps the biggest challenge to the stellar feedback model of \hi\ hole creation comes from observations of some of the closest galaxies.  The proximity of the SMC and LMC allow for detailed observations, including high resolution \hi\ imaging, which probes physical scales as small as $\sim$ 15 pc \citep{kim98}.  Matching catalogs of young stellar objects (e.g., OB associations, supergiants, WR stars) and \halpha\ imaging to \hi\ hole catalogs, three recent studies of the LMC \citep{kim98}, the SMC \citep{hat05}, and the Magellanic Bridge \citep{mul03} do not find strong correlations between the locations of young stellar objects and \hi\ holes.  The largest challenge to the stellar feedback creation hypothesis posed by these studies are the presence of young \hi\ holes ($\sim$ 2$-$10 Myr) with no associated young stellar objects \citep{stan07}.  This is particularly problematic in the SMC, where \citet{hat05} identified 59 young holes ($\sim$ 5 $-$ 10 Myr) that appear to be void of young stellar objects.  \citet{hat05} further find that the physical characteristics of these particular holes are no different than those with stellar components.  Further, \citet{hat05} conclude that the properties of the \hi\ holes in the inner and outer region are indistinguishable and that from their analysis there is no evidence to suspect a different creation mechanism for inner and outer holes.

Using the same type of analysis as in \S 5.3, and the assumption that inner and outer holes are indistinguishable in their \hi\ properties \cite{hat05}, we can compute the properties of the expected stellar populations if these young holes were formed due to stellar feedback. For the 59 empty \hi\ holes in the SMC, we find that the mean size is $\sim$ 128 pc, the expansion velocity is 7.5 \kms, and the inferred kinematic age is 8 Myr.  We also compute $E_{Hole}$ $\sim$ 56 \e{50}, assuming an \hi\ volume density of 1.0 \percc, and find that SFR$_{KA}$ $=$ 7\e{-5} \msun\ yr$^{-1}$.  We simulate 100 CMDs assuming a distance modulus of 19, photometric completeness limits of m$_{V}$ $=$ 22 and m$_{I}$ $=$ 21.7 \citep{har04}, and the parameters listed in \S 3.2.   We chose to only consider a burst model of SF at 8 Myr, since the differences between a burst and constant model of SF over such a short time scale are minimal.  We ran 100 simulations, to account for small number statistics, and found that the typical CMD contained $\sim$ 40 stars (within the completeness limits) and the mean magnitude of the brightest star in the CMD was m$_{F555W}$ $\sim$ 14.80.  We found that only 25 out of the 100 simulated CMDs contained at least one star with m$_{F555W}$ $<$ 14, i.e., a bright OB type star.  For m$_{F555W}$ $<$ 15, we found 72 CMDs contained at least one such star.  This test signifies that for a episode of SF 8 Myr ago, only 25\% of the time we would expect to find a star brighter than m$_{F555W}$ $=$ 14, and 72\% of the time for m$_{F555W}$ $=$ 15.  

Applying these statistics to the 509 holes cataloged by \citet{hat05}, we expect 128 \hi\ holes with at least one star brighter than m$_{F555W}$ $=$ 14, or 367 holes for m$_{F555W}$ $=$ 15.  In comparison, \citet{hat05} find 330 \hi\ holes with at least one star cluster,  young star, or WR star.  Initially, they also found 80 holes that appeared void of stars, however follow up observations revealed that 21 of the 80 holes appeared to have A spectral-type stars, leaving 59 holes without detected stellar companions, which is consistent with the 143 of holes we predict will appear void of MS stars at m$_{F555W}$ $=$ 15.  The uncertainties in our calculation are at least 50\%, due to uncertainties in $E_{Hole}$, the stellar evolution models, and statistical error.  Further, we are not able to test out results directly as we cannot reasonably compare our predicted stellar populations to the various catalogs used for correlations by \cite{hat05}.  However, we have demonstrated that even without the detection of a bright object inside a young \hi\ hole, it is possible that the hole was created by stellar feedback.

\section{Conclusions}

In summary:

\begin{itemize}
\item The HST/ACS based CMDs of stars within \hi\ holes from the catalogs of \citet{puc92} and \citet{bag09} revealed significant stellar populations, including young MS and BHeBs with ages less than $\sim$ 50 Myr, in \emph{all} holes.  We found that $\ll$ 1\% of the stars in the composite stellar population of either holes catalog were luminous MS stars less than $\sim$ 10 Myr in age.  This indicates that young OB associations are not reliable tracers of the locations of \hi\ holes.

\item We searched for differences in the stellar populations inside \hi\ holes and those in selected control fields, regions similar in size to holes, but that span a wide range in \hi\ column density.  We found that, on average, the control fields have twice as many luminous (i.e., young) MS stars as the \hi\ holes, however this is consistent with statistical fluctuations from small number statistics tests.  We measured the SFHs and cumulative SFHs of both samples (holes and control fields) and found no significant differences.  However, because we are not sensitive to holes less than $\sim$ 100 pc in size, it could be that smaller holes are already forming in control fields.  Only further similar studies involving an even wider range of hole sizes (e.g., SMC or LMC) can address this question with the requisite level of precision.

\item Using the SFHs for stars in each hole as input into STARBURST99, we calculated the energy associated with SF over the past 200 Myr.  From this we computed the energy input over the inferred kinematic age of each hole and calculated the feedback efficiency.  Compared to theoretical values, 8 of the 23 P92 and 13 of the 19 THINGS holes have efficiencies greater 20\%, with several exceeding 100\%.  Looking at the energy profiles over the last 200 Myr, we found that stellar feedback did produce enough energy, in principle, to have created all the \hi\ holes, just not necessarily within the inferred kinematic age.  This finding suggests that the inferred kinematic age (which is poorly constrained from observations) may not represent the true age of an individual \hi\ hole.  Indeed, the concept of an age for an \hi\ hole created by multiple generations of SF is intrinsically ambiguous.  

\item Using the BHeBs from the full HST/ACS CMD of \hoii, we constructed two spatially resolved recent SFH movies of \hoii\ with look back times of $\sim$ 100 Myr and $\sim$ 200 Myr.  Comparing these movies with the \hi\ distribution, we found that low levels of SF have been common, and only in the past $\sim$ 40 Myr has the SF rate elevated significantly.  Since many of the holes are older than $\sim$ 40 Myr, this suggests that holes are formed through a series of small SF events as opposed to a larger event at a single epoch.

\item From the HST/ACS photometry, we constructed CMDs corresponding to the apertures used by \citet{rho99}.  In all cases, we found mixed age stellar populations of hundreds or thousands of stars.  Using Monte Carlo tests of synthetic CMDs and empirical comparisons from the HST/ACS photometry, we demonstrated that integrated light does not accurately trace the SFHs of stellar populations inside \hi\ holes.  We also show that young MS stars ($<$ 75 Myr) are not highly clustered, which may imply that the single age stellar cluster method of \hi\ formation is not the most likely model. This assumption can account for why \citet{rho99} did not detect a greater number of holes with stars inside. 

\item Using the Monte Carlo technique, we computed expected stellar populations for \hi\ holes outside the optical body of \hoii\ and found that low levels of SF ($\sim$ 10$^{-5}$ \msun) could account for the presence of those holes.  The remnant populations would be quite faint as only a handful of young stars would remain to be seen at the present time.  We also examine the use of other SF tracers, \halpha, 24$\mu$m, and UV, and find that at such low SFRs, they do not reliably correlate with the stellar populations of \hi\ holes.  

\item  We applied similar analysis to the 59 young ($\sim$ 8 Myr) \hi\ holes that appear empty in the SMC \citep{hat05}, finding SFR$_{KA}$ $=$ 7\e{-5} \msun\ yr$^{-1}$.  Using the integrated stellar mass, we assume a burst of SFH at 8 Myr, and construct 100 simulated CMDs.  The resulting synthetic stellar populations show that only 25\% of the CMDs have at least one MS stars brighter than m$_{F555W}$ $=$ 14 and 72\% have at least one MS star brighter than m$_{F555W}$ $=$ 15.  Applying this to the 509 holes in the sample, we expect 128 \hi\ holes with at least one bright star in the 25\% case, 367 holes in the 72\% case, and 143 holes to be without any MS star brighter than m$_{F555W}$ $=$ 15.  \citet{hat05} find 330 holes that have at least one bright object, and 59 \hi\ holes that do not have any objects.  While this comparison is subject to large uncertainties, it demonstrates the plausibility of a stellar feedback origin for young \hi\ holes without a bright stellar companion.

\end{itemize}

The evidence presented by the HST/ACS observations of \hoii\ suggests that stellar feedback provides enough energy, in principle, to create the observed \hi\ holes in \hoii.  The traditional theory that single epoch episodes of SF, and singled aged clusters, are responsible for \hi\ hole creation is unlikely, because of the presence of mixed age stellar populations inside all of the holes, including holes with younger inferred ages.  The conclusions of \citet{rho99} and those in this paper are not contradicting, rather they both seem to reinforce the idea that \hi\ holes are not often created by single aged stellar populations.  The summation of the evidence is that the likely mechanism for the creation of \hi\ holes is stellar feedback from SNe due to multiple generations of SF.  An important caveat is found in the similarity of the stellar populations in control fields and the \hi\ holes.  While this detracts from the potential casual relationship between stellar feedback and \hi\ hole creation, this result could simply be explained by the fact that we are not sensitive to holes smaller than $\sim$ 100 pc.  Applying similar analysis to galaxies with a wider range in \hi\ holes sizes (e.g., SMC or LMC) could reveal the extent of this effect and provide further insight into the potential link between stellar feedback and the formation and growth of \hi\ holes.

\section{Acknowledgments}

Support for this work was provided by NASA through grant GO-10605
from the Space Telescope Science Institute, which is operated by
AURA, Inc., under NASA contract NAS5-26555. DRW is grateful for support 
from a Penrose Fellowship.  EDS is grateful for partial support from the University of Minnesota.
This research has made use of NASA's Astrophysics Data System
Bibliographic Services and the NASA/IPAC Extragalactic Database
(NED), which is operated by the Jet Propulsion Laboratory, California
Institute of Technology, under contract with the National Aeronautics
and Space Administration.  Special thanks to Ioannis Bagetakos and Elias Brinks for sharing their data and results before publication and providing useful guidance.  We would also like to thank Cliff Johnson and Danny Dale for making calibrated \textit{Spitzer} image of \hoii\ available and Julianne Dalcanton for her insightful comments.

\clearpage

\begin{deluxetable}{cccccccc}
\tablecolumns{6}
\tabletypesize{\footnotesize}
\small
\tablewidth{0pt}
\tablecaption{Photometric Properties of Puche et al. \hi\ Holes}
\tablehead{
    \colhead{Hole} &
     \colhead{$\alpha$} &
      \colhead{$\delta$} &
      \colhead{Diameter} &
      \colhead{No. of Stars} &
        \colhead{Integrated} \\
      
       \colhead{Number} &
     \colhead{(J2000)} &
      \colhead{(J2000)} &
      \colhead{(arcsec)}  &
       \colhead{in CMD} &
       \colhead{m$_{F555W}$}\\
       
        \colhead{(1)} &
     \colhead{(2)} &
      \colhead{(3)} &
      \colhead{(4)}  &
       \colhead{(5)} &
              \colhead{(6)}   \\
      
}
\startdata
10 & 8 18 42 & +70 41 21 & 73 & 4149& 17.68\\
12 & 8 13 49 & +70 43 33 & 22 & 2065 & 18.65 \\ 
14 & 8 18 56 & +70 43 27 & 27 & 3907 & 17.47 \\
16 & 8 18 59 & +70 43 51 & 13 & 948 & 19.00 \\
20 & 8 19 01 & +70 42 44 & 8 & 383 & 20.19 \\
21 & 8 19 04 & +70 41 16 & 115 & 39153 & 14.72\\
22 & 8 19 06 & +70 43 08 & 32 & 6443 & 16.65\\
23 & 8 19 06 & +70 43 30 & 39 & 9477 & 16.22 \\
24 & 8 19 08 & +70 44 03 & 10 & 579 & 19.37 \\
28 & 8 19 13 & +70 43 31 & 19 & 1934 & 17.93\\
29 & 8 19 14 & +70 42 21& 27 & 4517 &17.58 \\
30 & 8 19 14 & +70 44 20 & 69 & 20614 & 15.77\\
31 & 8 19 15 & +70 43 49 & 18 & 1625 & 18.37\\
32 & 8 19 16 & +70 41 49 & 37 & 3782 & 17.83\\
33 & 8 19 17 & +70 44 56 & 25 & 2140 & 18.34 \\
36 & 8 19 19 & +70 43 14 & 37 & 5945 & 16.98 \\
37 & 8 19 20 & +70 45 17 & 20 & 1011 & 19.12 \\
39 & 8 19 23 & +70 44 43 & 23 & 1971 & 18.24\\
42 & 8 19 25 & +70 45 12 & 22 & 1383 & 18.75\\
43 & 8 19 27 & +70 42 06 & 16 & 549 & 18.75\\
44 & 8 19 27 & +70 43 31& 48 & 7566 & 16.82 \\
45 & 8 19 27 & +70 45 19 & 32 & 2344 & 18.08 \\
48 & 8 19 36 & +70 44 54 & 50 & 4695 & 16.93\\
\enddata
\tablecomments{Numbers, coordinates, and diameters for each hole taken from \citet{puc92}.  Hole coordinates have been converted to the J2000 epoch for ease of comparison.}
\label{tab1}
\end{deluxetable}
\clearpage

\begin{deluxetable}{cccccc}
\tablecolumns{6}
\tabletypesize{\footnotesize}
\small
\tablewidth{0pt}
\tablecaption{Photometric Properties of THINGS \hi\ Holes}
\tablehead{
    \colhead{Hole} &
     \colhead{$\alpha$} &
      \colhead{$\delta$} &
      \colhead{Diameter} &
      \colhead{No. of Stars} &
        \colhead{Integrated} \\
      
       \colhead{Number} &
     \colhead{(J2000)} &
      \colhead{(J2000)} &
      \colhead{(arcsec)}  &
       \colhead{in CMD} &
         \colhead{m$_{F555W}$} \\
       
        \colhead{(1)} &
     \colhead{(2)} &
      \colhead{(3)} &
      \colhead{(4)}  &
       \colhead{(5)} &
       \colhead{(6)}  \\
      
}
\startdata
4 & 8 18 40 & +70 41 36 & 44 & 1733 & 18.81  \\
6 & 8 18 49 & +70 43 40 & 29 & 3781 & 17.98 \\
8 & 8 18 52 & +70 42 46 & 21 & 2636 & 17.26 \\
9 & 8 18 53 & +70 44 27 & 17 & 1067 & 19.35 \\
12 & 8 18 58 & +70 42 56 & 24 & 4067 & 16.75 \\
14 & 8 19 00 & +70 42 26 & 19 & 2289  & 18.13 \\
16 & 8 19 04 & +70 43 28 & 46 & 16146 & 15.67 \\
17 & 8 19 05 & +70 41 24 & 128 & 61244 & 14.37 \\
19 & 8 19 13 & +70 44 24 & 67 & 25053 & 15.66\\
20 & 8 19 15 & +70 45 25 & 20 & 1644 & 18.44 \\
21 & 8 19 15 & +70 45 01 & 23 & 2312 & 18.40 \\
23 & 8 19 18 & +70 43 21 & 45 & 11644 & 16.24 \\
26 & 8 19 24 & +70 43 29 & 29 & 4253 & 17.54 \\
27 & 8 19 26 & +70 42 14 & 19 & 1126 & 18.89  \\
29 & 8 19 27 & +70 45 29 & 29 & 2202 & 18.28 \\
30 & 8 19 28 & +70 44 38 & 19 & 1590 & 18.62 \\
31 & 8 19 29 & +70 43 38 & 30 & 3190 & 18.05 \\
34 & 8 19 35 & +70 45 05 & 40 & 3673 & 17.33 \\
39 & 8 19 44 & +70 43 55 & 29 & 640 & 19.31 \\

\enddata
\tablecomments{\hi\ holes from THINGS \citep{wal08} that are within the HST/ACS field of view.  Numbers, coordinates, and diameters for each \hi\ hole are from \citet{bag09}}
\label{tab2}
\end{deluxetable}
\clearpage

\begin{landscape}
\begin{deluxetable}{cccccccc}
\tablecolumns{8}
\tabletypesize{\scriptsize}
\small
\tablewidth{0pt}
\tablecaption{Derived Properties of Puche et al. \hi\ Holes}
\tablehead{
    \colhead{Hole} &
     \colhead{Diameter} &
      \colhead{DV} &
      \colhead{Kinematic Age}&
    \colhead{E$_{Hole}$ (P92)} &
        \colhead{E$_{Hole}$ (new)} &
      \colhead{E$_{SFKA}$} &
      \colhead{Putative} \\
      
       \colhead{Number} &
     \colhead{(pc)} &
      \colhead{\kms} &
      \colhead{(Myr)}  &
        \colhead{(10$^{50}$ erg)}&
           \colhead{(10$^{50}$ erg)}&
       \colhead{(10$^{50}$ erg)}  & 
       \colhead{Efficiency$_{KA}$} \\

        \colhead{(1)} &
     \colhead{(2)} &
      \colhead{(3)} &
      \colhead{(4)}  &
       \colhead{(5)} &
        \colhead{(6)}  &
        \colhead{(7)} &
         \colhead{(8)} \\
      
}
\startdata
10 & 1171 & 8.8 & 65 & 500 & 364  & 2400  & 15\%\\
12 & 349 & 7.7 & 22 & 18.0 & 16  & 130  & 12\% \\
14 & 421 & 3.9 & 53 & 15.0  & 19 & 2700  & 0.7\% \\
16 & 199 & 1.0 & 97 & 0.30  & 0.5 & 1000  & $<$0.1\% \\
20 & 125 & 6.5? & 9.4 & 0.4  & 2.0 & 10  & 20\% \\
21 & 1820 & 6.6 & 135 & 2800  & 7700 & 60000  & 13\%  \\
22 & 508 & 6.5 & 38 & 48.0 & 230 &  1600  & 15\%  \\
23 & 619 & 4.5 & 67 & 60 &  211 & 3300  & 6.4\%  \\
24 & 162 & 7.5? & 11 & 1.66 & 5.0 &  20 & 27\%  \\
28 & 302 & 7.2? & 21 & 11.0 & 83 & 220  & 37\% \\
29 & 469 & 3.9 & 59 & 21.4 & 190 & 1300 & 15\%\\
30 & 1083 & 8.1 & 65 & 737 &  2200 & 5700  & 39\% \\
31 & 289 & 2.5? & 56 & 2.20 & 13 & 900  & 1.4\% \\
32 & 583 & 4.8 & 59 & 48.4 &  420 & 2600  & 16\%  \\
33 & 388 & 3.5 & 54 & 10.2 &  17 & 780  & 2.2\% \\
36 & 579 & 8.3 & 34 & 119 &  990 & 1800  & 55\%  \\
37 & 312 & 11.9 & 13 & 28.8 &  34 & 45  & 74\% \\
39 & 359 & 3.5 & 50 & 8.00 &  16 & 1500  & 1.1\% \\
42 & 355 & 10.1 & 17 & 34.0 &  39 & 75  & 52\%  \\
43 & 255 & 4.8 & 26 & 3.73 &  25 & 90 & 28\%  \\
44 & 752 & 7.9  & 47 & 217 &  1200 & 7300 & 17\% \\
45 & 596 & 6.6  & 37 & 53.5 &  51 & 930  & 6\%   \\
48 & 786 &14.1 & 27 & 649 &  560 & 450  & 120\%  \\
\enddata
\tablecomments{(1) Hole numbers from \citet{puc92}; (2) Hole diameters calculated assuming a distance of 3.4 Mpc; (3) Expansion velocities from \citet{puc92}; (4) Kinematic age assumes the diameter from column 2 and expansion velocity from column 3; (5) Energy from \citet{puc92} was adjusted for a distance of 3.39 Mpc;  (6) $E_{Hole}$ computed using Equation 1 and the diameters and expansion velocities in this table, along with the values of n$_{0}$ computed in \S4.1; (7) Total energy over the inferred kinematic age of the \hi\ hole computed using STARBURST99; (8) The ratio of E$_{Hole}$ to E$_{kinematic}$.}
\label{tab3}
\end{deluxetable}
\end{landscape}
\clearpage

\begin{deluxetable}{ccccccccc}
\tablecolumns{11}
\tabletypesize{\scriptsize}
\small
\tablewidth{0pt}
\tablecaption{Derived Properties of THINGS \hi\ Holes}
\tablehead{
    \colhead{Hole} &
     \colhead{Diameter} &
      \colhead{DV} &
      \colhead{Kinematic Age}&
    \colhead{E$_{Hole}$ (P92)} &
        \colhead{E$_{Hole}$ (new)} &
      \colhead{E$_{SFKA}$} &
      \colhead{Putative} &
       \colhead {Type} \\
      
       \colhead{Number} &
     \colhead{(pc)} &
      \colhead{\kms} &
      \colhead{(Myr)}  &
        \colhead{(10$^{50}$ erg)}&
         \colhead{(10$^{50}$ erg)}&
       \colhead{(10$^{50}$ erg)}  & 
       \colhead{Efficiency$_{KA}$} & 
       \colhead{}\\

        \colhead{(1)} &
     \colhead{(2)} &
      \colhead{(3)} &
      \colhead{(4)}  &
       \colhead{(5)} &
        \colhead{(6)}  &
        \colhead{(7)} &
        \colhead{(8)} &
        \colhead{(9)}\\
      
}
\startdata
4 & 731 &  7 &  51 &  100 & 170 & 500 & 34\%  &1\\
6 & 486 &  7 &  34 &  50 & 180 & 600 & 31\% & 1\\
8 & 345 &  10 &  17 &  32 & 140 & 500 & 28\%  & 3\\
9 & 279 &  10 &  14 &  13 & 54 & 15 & 360\%  & 3\\
12 & 403 &  11 &  18 &  63 & 450 & 330 & 140\%  & 3\\
14 & 318 &  15 &  10 &  40 & 310 & 20 & 1600\% &  2\\
16 & 761 &  7 &  53 &  200 & 2100 & 4600 & 45\% &  1\\
17 & 2110 &  7 &  147 &  3200 & 18000 & 100000 & 18\% &  1\\
19 & 1107 &  7 &  77 &  500 & 3200 & 9500 & 34\% &  1\\
20 & 329 &  12 &  13 &  25 & 56 & 100 & 56\% & 3\\
21 & 378 &  9 &  21 &  20 & 88 & 300 & 29\% &  3\\
23 & 738 &  16 &  23 &  630 & 2900 & 2400 & 120\% &  2\\
26 & 483 &  7 &  34 &  50 & 133 & 2200 &  6\% &  1\\
27 & 312 &  7 &  22 &  20 & 25 & 190 & 13\% &  1\\
29 & 479 &  7 &  33 &  32 & 43 & 900 & 5\% &  1\\
30 & 318 &  13 &  12 &  32 & 40 & 150 & 26\% &  2\\
31 & 491 &  7 &  34 &  50 &  83 & 1000 & 8\% &  1\\
34 & 657 &  7 &  46 &  80 & 79 & 730 & 11\% &  1\\
39 & 474 &  18 &  13 &  160 & 66 & 50 & 130\% &  2\\

\enddata
\tablecomments{(1) $-$ (5) Hole number, size, expansion, velocity, and $E_{Hole}$ from the catalog of \citet{bag09}; (6) Total energy over the inferred kinematic age of the \hi\ hole computed using STARBURST99; hole 6 had no appreciable SF within its kinematic age; (6) $E_{Hole}$ computed using Equation 1 and the diameters and expansion velocities in this table, along with the values of n$_{0}$ computed in \S 4.1;  (7) Total energy over the inferred kinematic age of the \hi\ hole computed using STARBURST99;  (8) Feeback efficiency: The ratio of $E_{Hole}$ to $E_{SFKA}$; (8) The ratio of $SFR_{KA}$ to the SFR averaged over the past 500 Myr. (9) Hole type as classified by \citet{bag09}, where type 1 holes are blown out and type 2 or 3 holes are still appreciably expanding.}
\label{tab4}
\end{deluxetable}
\clearpage

\begin{deluxetable}{cccccc}
\tablecolumns{6}
\tabletypesize{\footnotesize}
\small
\tablewidth{0pt}
\tablecaption{Photometric Properties of Control Fields}
\tablehead{
    \colhead{Hole} &
     \colhead{$\alpha$} &
      \colhead{$\delta$} &
      \colhead{Aperture Diameter} &
      \colhead{No. of Stars} &
        \colhead{Integrated} \\
      
       \colhead{Number} &
     \colhead{(J2000)} &
      \colhead{(J2000)} &
      \colhead{(arcsec)}  &
       \colhead{in CMD} &
         \colhead{m$_{F555W}$} \\
       
        \colhead{(1)} &
     \colhead{(2)} &
      \colhead{(3)} &
      \colhead{(4)}  &
       \colhead{(5)} &
       \colhead{(6)}   \\
      
}
\startdata

c1 & 8 19 46 & +70 44 29 & 30 & 839 & 19.38 \\
c2 & 8 19 39 & +70 43 55 & 30 & 1309 & 19.20 \\
c3 & 8 19 28 & +70 44 09 & 30 & 4089 & 17.64 \\
c4 & 8 19 35 & +70 43 08 & 30 & 1549 & 18.78 \\
c5 & 8 19 32 & +70 42 40 & 30 & 1863 & 18.08 \\
c6 & 8 19 21 & +70 42 28 & 30 & 3598 & 17.96 \\
c7 & 8 19 12 & +70 42 55 & 30 & 5183 & 17.40 \\
c8 & 8 18 51 & +70 42 05 & 30 & 1988 & 18.72 \\
c9 & 8 18 45 & +70 42 39 & 30 & 2147 & 17.74 \\
\enddata
\tablecomments{Control fields shown in cyan in Figure \ref{map2}.  Corresponding CMDs in Figure \ref{ccmd1} and SFHs in Figure \ref{csfh1}.}
\label{tab5}
\end{deluxetable}
\clearpage

\begin{deluxetable}{ccccccc}
\tablecolumns{7}
\tabletypesize{\footnotesize}
\small
\tablewidth{0pt}
\tablecaption{Photometric Properties of \hi\ Holes used by \citet{rho99}}
\tablehead{
    \colhead{Hole} &
     \colhead{$\alpha$} &
      \colhead{$\delta$} &
      \colhead{Aperture Diameter} &
      \colhead{No. of Stars} &
      \colhead{Hole} &
        \colhead{Integrated} \\
      
       \colhead{Number} &
     \colhead{(J2000)} &
      \colhead{(J2000)} &
      \colhead{(arcsec)}  &
       \colhead{in CMD} &
         \colhead{Type}\ &
         \colhead{m$_{F555W}$} \\
       
        \colhead{(1)} &
     \colhead{(2)} &
      \colhead{(3)} &
      \colhead{(4)}  &
       \colhead{(5)} &
       \colhead{(6)}  &
       \colhead{(7)}  \\
      
}
\startdata
10 & 8 18 40 & +70 41 39 & 30.00 & 893 & a & 19.75\\
12 & 8 18 50 & +70 43 41 & 20.00 & 1989 & b & 18.63 \\ 
14 & 8 18 55 & +70 43 34 & 20.00 & 2810 & c & 18.12\\
16 & 8 18 58 & +70 43 01 & 20.00 & 2740 & d & 17.12 \\
20 & 8 18 59 & +70 42 52 & 20.00 & 2639 & c & 17.44 \\
21 & 8 19 02 & +70 41 21 & 30.00 & 4077 & d & 17.18 \\
22 & 8 19 04 & +70 43 16 & 20.00 & 3064 & d & 17.24 \\
23 & 8 19 04 & +70 43 40 & 20.00 & 3051& c & 17.85\\
24 & 8 18 07 & +70 44 14 & 20.00 & 2825 & c & 18.08 \\
28 & 8 19 12 & +70 43 52 & 20.00 & 2526 & c & 18.01\\
30 & 8 19 13 & +70 44 32 & 20.00 & 2062 & c & 18.59 \\
31 & 8 19 14 & +70 44 04 & 20.00 & 2374 & c & 18.09 \\
32 & 8 19 12 & +70 41 54 & 24.00 & 2615 & c & 18.31 \\
33 & 8 19 16 & +70 45 10 & 20.00 & 1628 & c & 18.81\\
36 & 8 19 19 & +70 43 21 & 24.00 & 3324 & d & 17.65\\
37 & 8 19 19 & +70 45 32 & 20.00 & 1205 & b & 19.04 \\
39 & 8 19 22 & +70 44 51 & 20.00 & 1893 & c & 18.29\\
42 & 8 19 21 & +70 45 21 & 20.00 & 1336 & b & 18.76\\
43 & 8 19 26 & +70 42 16 & 16.00 & 805 & e & 19.40\\
44 & 8 19 24 & +70 43 33 & 30.00 & 4229 & d & 17.60 \\ 
45 & 8 19 25 & +70 45 37 & 20.00 & 1023 & c & 19.04\\
48 & 8 19 35 & +70 45 06 & 30.00 & 1892 & d & 18.14\\
\enddata
\tablecomments{(1) $-$ (4) Numbers, coordinates, and diameters for each aperture taken from \citet{rho99}, which are based on the \hi\ hole catalog of \citet{puc92}.  Coordinates have been converted to J2000 for ease of comparison; (6) Classifications from \citet{rho99}: a $-$ empty hole, b $-$ faint foreground star, c $-$ galaxian background, d $-$ possible star cluster, e $-$ possible photoionization region; (7) integrated magnitudes calculated from the ACS CMDs (Figures \ref{rho_cmd1} \& \ref{rho_cmd2}).}
\label{rho_tab}
\end{deluxetable}
\clearpage

\newpage

\newpage

\newpage

\newpage

\newpage

\newpage

\newpage

\newpage

\newpage

\newpage

\clearpage

\newpage

\newpage

\newpage

\newpage

\newpage

\newpage

\newpage

\newpage

\newpage

\newpage

\newpage


\clearpage

\newpage

\newpage

\newpage

\newpage

\clearpage

\newpage

\newpage

\newpage

\end{document}